%% file: fatalic.tex
\documentclass[a4paper,11pt]{article}
\pdfoutput=1
\usepackage{jinstpub}
\usepackage{fatalic-defs}
\usepackage{lineno}
\usepackage{multirow}
\title{FATALIC: A novel CMOS front-end readout ASIC for the ATLAS Tile Calorimeter}

\abstract{
  The present article introduces a novel ASIC architecture, designed in the context of the ATLAS Tile Calorimeter upgrade 
  program for the High-Luminosity phase of the Large Hadron Collider at CERN. The architecture is based on radiation-tolerant 
  \SI{130}{\nm} \acrlong{cmos} technology, embedding both analog and digital processing of detector signals. A detailed 
  description of the ASIC is given in terms of motivation, design characteristics, simulated and measured performance. 
  Experimental studies, based on 24 prototype chips under real particle beam conditions are also presented in order to 
  demonstrate the potential of the architecture as a reliable front-end readout electronic solution.
}

\author[1]{S.~Angelidakis}
\author[]{W.M.~Barbe}
\author[]{R.~Bonnefoy}
\author[]{H.~Chanal}
\author[]{C.~Fayard}
\author[1]{R.~Madar}
\author[]{S.~Manen}
\author[]{M.-L.~Mercier}
\author[]{E.~Nibigira}
\author[]{D.~Pallin}
\author[]{N.~Pillet}
\author[]{L.~Royer}
\author[]{A.~Soulier}
\author[]{R.~Vanda\"ele}
\author[]{F.~Vazeille}
\affiliation[]{Laboratoire de Physique de Clermont-Ferrand, CNRS/IN2P3, Universit\'e Clermont Auvergne,\\4 avenue Blaise Pascal, Aubi\`ere, France}

\note{Corresponding author.}
\emailAdd{Stylianos.Angelidakis@cern.ch}
\emailAdd{romain.madar@clermont.in2p3.fr}
\begin{document}
\maketitle
\clearpage

\input{TeX/part1_intro.tex}

\input{TeX/part2_asic.tex}

\input{TeX/part3_cards.tex}

\input{TeX/part4_reco.tex}

\input{TeX/part5_tbeam.tex}

\section{Conclusion}
\label{sec:conclusion}
The above sections conclude the presentation of \gls{fatalic} and demonstrate the full potential of the \SI{130}{nm} CMOS 
technology, realised as a \gls{fe} readout electronic architecture for the strenuous conditions of the HL-LHC. The fast 
channels offer excellent performance for the processing of detector signals in the input charge range up to \SI{1.2}{nC}, 
with a noise of \SI{6.1}{fC} (\SI{8}{fC}) and linearity better than 1.5\% (0.3\%), up to approximately \SI{850}{pC},
according to measurements (simulation). The disagreement between measurement and simulation is not attributed to \gls{fatalic} 
itself but rather to imperfections of the present All-in-One cards to support the slow-channel of \gls{fatalic}, as discussed
in Section\,\ref{subsec:SimAndPerf}. The performance of the fast channels was also probed with particle beams, in which 
\gls{fatalic} was used to measure the energy of electrons, muons and pions, reproducing the nominal estimation of the Tile 
EM scale constant as well as the average energy loss per unit distance of muons traversing the TileCal.

The slow channel also exhibits the expected performance and was successfully used to process low amplitude currents in 
calibration scans using the \gls{cs} system, in order to correct the \gls{pmt} gains in the respective readout channels. 
At the same time, however, it exposes the limitations imposed by the CMOS technology, along with possible directions for 
improvement. The main limitation is the large, approximately \SI{7}{nA} $\sfrac{1}{f}$ noise, introduced by the input 
stage, which does not comply with the specification ($<$\SI{1}{nA}) defined for new the Tile \gls{fe} electronics. The 
adopted, current-driven architecture is not offered for further reduction of the noise, which would therefore require 
migration to a different (bi-CMOS) technology or relocation of the slow channel outside \gls{fatalic}.
%-------------------------------------------------------------------------------

\clearpage
%\acknowledgments

\bibliography{fatalic}

\end{document}

%% file: TeX/part1_intro.tex
%%%%%%%%%%%%%%%%%%%%%%%%%%%%%%%%%%%%%%%%%%%%%%
\section{Introduction}
\label{sec:intro}
%%%%%%%%%%%%%%%%%%%%%%%%%%%%%%%%%%%%%%%%%%%%%%

The \gls{lhc}~\cite{Evans:2008zzb} at CERN is scheduled to undergo a decisive upgrade~\cite{Apollinari:2015bam}, 
which will extend its physics potential well beyond the initial design goal~\cite{Jakobs:2011zz,Schmidt:2016jra}. 
The first phase (Phase-I) of the upgrade will take place during the long technical shutdown of the \gls{lhc}, in 
2019-2020, to improve the injector and the collider collimation system. In the second phase (Phase-II), scheduled 
for 2024-2026, the \gls{lhc} will be further enhanced with new technologies, including 11-12\,\si{T} triplet magnets 
and compact crab cavities with ultra-precise phase control. The final High Luminosity LHC (HL-LHC) is expected to 
begin operation in 2026, delivering proton-proton ($pp$) collisions with a maximum instantaneous luminosity of 
\SI{7.5e34}{cm^{-2}.s^{-1}} (7.5 times higher than the initial design). In terms of total integrated luminosity, 
a goal of 3000-4000 $\si{fb^{-1}}$ is defined.

In order to perform in the high intensity radiation environment of the HL-LHC, where up to 200 inelastic $pp$ collisions 
(pile-up events) per \SI{25}{ns} bunch crossing are expected, the ATLAS detector~\cite{PERF-2007-01} is
scheduled for upgrade of its sub-detector systems as well as the Trigger and Data AcQuisition (TDAQ) 
strategies~\cite{CERN-LHCC-2015-020}. In this context, the Phase-II upgrade program foresees the replacement of 
the readout electronics of the \gls{TileCal}~\cite{TCAL-2010-01,Collaboration:2285583}, the central hadronic 
calorimeter of ATLAS, with new architectures that will be able to deliver reliable measurements during the 
HL-LHC operation. Among the different designs proposed and evaluated for the \gls{fe} readout electronics, 
the Laboratoire de Physique de Clermont-Ferrand (LPC) presented the solution of an ASIC that embeds both
the analog processing and digitisation of the detector signal, the \gls{fatalic}. The following sections 
intend to provide a detailed description of the \gls{fatalic} motivation and design, and present simulation 
and experimental measurements to demonstrate its potential as a \gls{fe} readout system.

%%%%%%%%%%%%%%%%%%%%%%%%%%%%%%%%%%%%%%%%%%%%%%
\section{The ATLAS Tile Calorimeter}
\label{sec:TileCal}
%%%%%%%%%%%%%%%%%%%%%%%%%%%%%%%%%%%%%%%%%%%%%%

\begin{figure}[t]
  \begin{center}
    \subfloat[]{\includegraphics[width=0.64\textwidth]{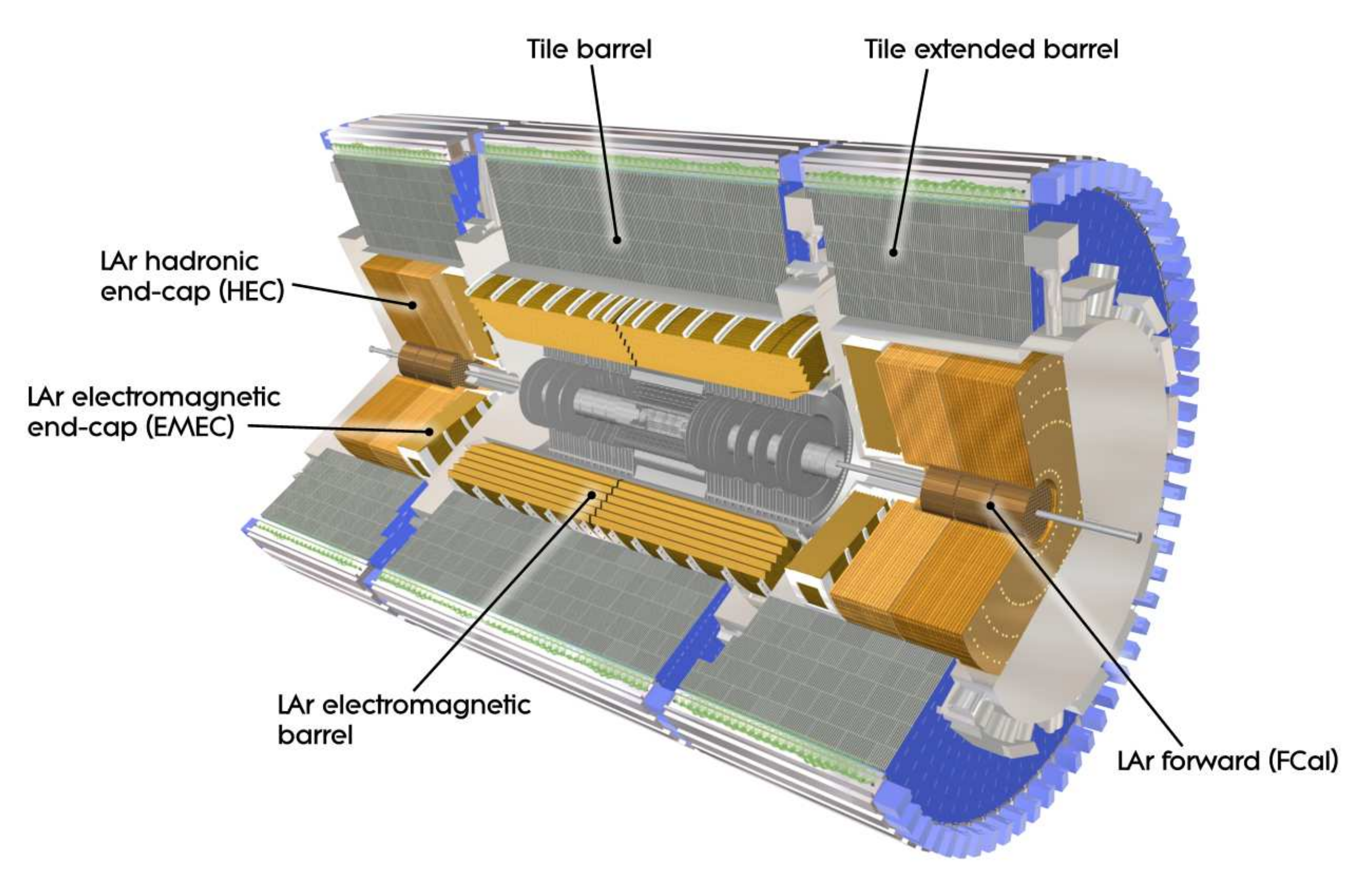}\label{fig:tile_a}}
    \subfloat[]{\includegraphics[width=0.36\textwidth]{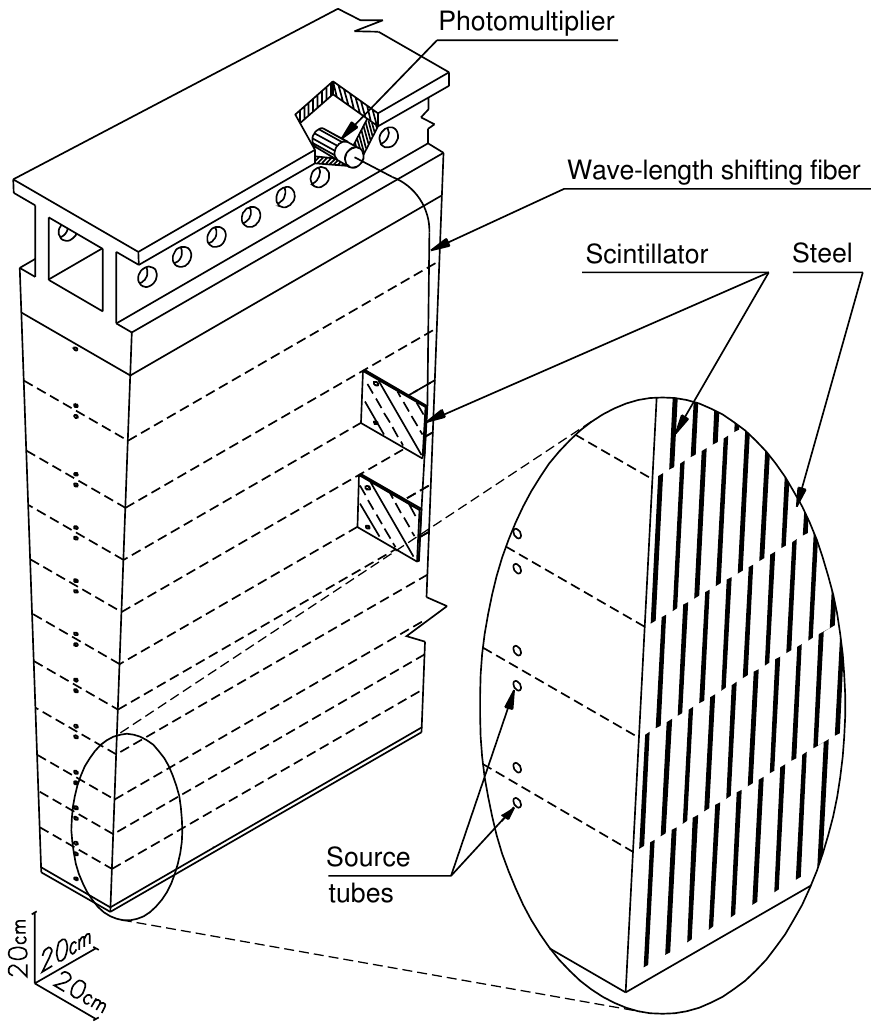}\label{fig:tile_b}}\\
    \subfloat[]{\includegraphics[width=1.00\textwidth]{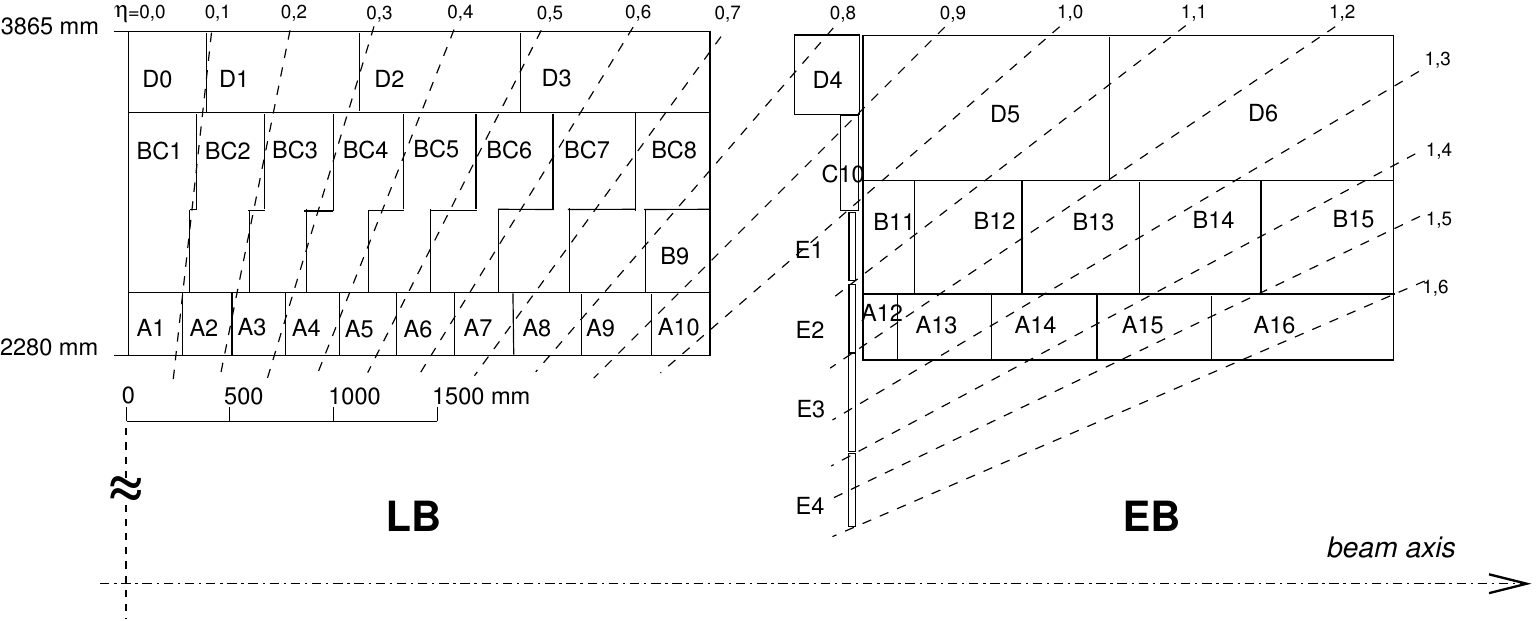}\label{fig:tile_c}}
    \caption{(a) The calorimeter system of ATLAS. (b) Schematic of a wedge module. (c) Tile cell mapping.
    \label{fig:general-context:tilcal-views}}
  \end{center}
\end{figure}

\gls{TileCal} is a sampling calorimeter, constructed of steel plates as absorber and scintillating tiles as active 
medium, and it is important for the measurement of jet- and missing-energy, jet substructure, electron isolation and 
triggering (including muon information). The position of \gls{TileCal} in the ATLAS calorimeter complex can be seen 
in figure\,\ref{fig:tile_a}. It is divided into four cylinders, two of which form the central Long-Barrel (LB) while 
the other two constitute Extended-Barrel (EB) partitions, covering the pseudorapidity\footnote{
%%%%-----------------------------------------------------------------------
ATLAS uses a right-handed coordinate system, centered at the nominal interaction point. 
The $x$-axis points towards the center of the LHC ring, the $y$-axis points upwards and 
the $z$-axis points along the beampipe. Cylindrical coordinates ($r$,$\phi$) are used in 
the transverse plane, $\phi$ being the azimuthal angle around the $z$-axis. The pseudorapidity 
is defined in terms of the polar angle $\theta$ as $\eta = -\ln(\tan\frac{\theta}{2})$.}
%%%%-----------------------------------------------------------------------
range $|\eta|<1.7$. Each Tile cylinder is made of $64$ wedge modules (figure\,\ref{fig:tile_b}) in the azimuthal
coordinate. The \glspl{pmt} with the associated \gls{fe} readout electronics and high voltage distribution cards 
are inserted at the outer radius of each module, hosted by a train of two \SI{1.4}{m} long ``drawers'' which form
a ``super-drawer'' (one super-drawer can host up to 48 \glspl{pmt}).

The scintillation light is collected from the two opposite sides of the tiles by \gls{wls} fibres, which are bundled,
to define readout cells, and coupled to a pair of \glspl{pmt}. The current pulse induced at the \gls{pmt} anodes, with 
a typical width of \SI{10}{\ns}, is therefore proportional to the energy deposited in the respective cell and is fed to 
the \gls{fe} electronics to be amplified, shaped and digitised at the \SI{40}{MHz} \gls{lhc} clock rate. This arrangement 
of the readout employs a total of 9852 \glspl{pmt} and segments \gls{TileCal} into three radial layers (figure\,\ref{fig:tile_c}), 
classified as "A", "BC" and "D", with depth 1.5, 4.1, and 1.8 interaction lengths, respectively,
at $\eta=0$. Each A-BC-D cell grouping in the same $\eta$-direction forms a projective tower currently used for the fast 
trigger. Additional scintillators (E cells) are installed in the region between the LB and the EB, mainly to measure energy 
escaping the electromagnetic calorimeter, while at higher $|\eta|$, Minimum Bias Trigger Scintillators (MBTS) are used to 
monitor minimum-bias event rates. The E and MBTS scintillators have been used to monitor the luminosity during van der Meer 
scans\footnote{
%------------------------------------------------------------
Van der Meer scans enable the measurement of the absolute luminosity of a particle collider by 
sweeping the beams transversely across each other. Being responsible for luminosity measurements 
in ATLAS, \gls{TileCal} must be able to measure very low Minimum Bias event rates.}
%------------------------------------------------------------
~\cite{Balagura:2011yw} and are also useful in electron identification.

To ensure stable measurements, \gls{TileCal} incorporates three calibration systems. A \gls{cis}~\cite{Tang:2013vya} 
is used to monitor the response of the \gls{fe} electronics to known injected charge, but also to derive the conversion 
factors between the electronic responses and the input charge. Next, the \gls{cs} system~\cite{Starchenko:2002ju,Shalanda:2003rq} 
uses a radioactive $^{137}$Cs $\gamma$-source, hydraulically circulated through a system of tubes that traverses every row 
of scintillating tiles. The illumination of the tiles with \SI{0.662}{MeV} $\gamma$ produces a uniform, low current signal 
at the \gls{pmt} anodes that allows inter-calibration of the detection chains (scintillators, \gls{wls} fibres, \glspl{pmt}, 
\gls{fe} electronics). After initial adjustment of the \gls{pmt} gains, the \gls{cs} system is used to measure the small
variations among the read-out responses, which are used to derive the necessary calibration coefficients with respect to
a unique reference value. Lastly, the laser system~\cite{system:2016tae} injects light pulses to the \glspl{pmt} for 
frequent monitoring of the gains between \gls{cs} scans.

%---------------------------------
\section{Upgrade of the Tile readout}
\label{sec:PhysicsSpecs}
%---------------------------------

In the current Tile readout scheme, the 256 super-drawers are interfaced to back-end Read-Out Drivers (RODs) through
256 optical links (plus another 256 links for redundancy) with a bandwidth of \SI{800}{Mbps} per link. Each super-drawer
stores the digitised samples from the read-out Tile cells in pipeline memories, while analog trigger signals, corresponding 
to each Tile tower, are transmitted to the off-detector trigger pre-processor. These trigger signals are formed by summation
of the \gls{pmt} outputs, in dedicated analog boards, with rough adjustment in time but without calibration to account for 
gain variations. Upon trigger acceptance, the data of each super-drawer are forwarded to the RODs.

For the HL-LHC, the TDAQ strategy\,\cite{,Collaboration:2285583} foresees a fully digital calorimeter trigger with 
higher granularity and precision. Each super-drawer is replaced by four independent mini-drawers, with half the length 
of the current drawers to allow easier access to the electronics and improve the reliability of the cooling circuits. 
At the \SI{40}{MHz} rate, each mini-drawer transmits the entire set of digitised data to an off-detector PreProcessor 
(PPr)\,\cite{Carrio:2014hqa} over two \SI{9.6}{Gbps} optical links (the full readout scheme employs 2048 optical links 
plus another 2048 links for redundancy). The PPr stores the data in pipeline memories and, in parallel, transmits calibrated 
trigger primitives (energy measured in grouped or individual Tile cells) to the trigger system. Upon receipt of a trigger 
acceptance signal, the reconstructed energy, the time and a quality factor for each Tile cell are forwarded to the data 
network for event aggregation and storage.

In the new Tile readout system, the present reliable but outdated Tile \gls{fe} electronics will be replaced by a new
architecture that will be able to endure the harsh radiation conditions anticipated at the HL-LHC (up to \SI{50}{krad} 
Total Ionising Dose, as presented in Section\,\ref{subsec:rad}) and handle the expected dynamic range of the input signal. 
The lowest expected signal from physics events is defined by the minimum ionisation Landau peak of muons traversing an A 
cell (smallest Tile cell) at normal incidence, with a most probable value of \SI{350}{MeV}~\cite{Adragna:2009zz}. Considering 
the typical electromagnetic (EM) scale constant of \SI{1.05}{pC/GeV} and the $e/\mu$ response ratio of $0.91$, the respective 
charge delivered by a single \gls{pmt} is approximately \SI{200}{fC}. On the other hand, the largest expected signal is that of 
energetic jets, depositing up to \SI{1.3}{TeV} in EM scale (\SI{1.5}{TeV} in hadronic scale) in a single Tile cell. To 
cover this range, the maximum charge requirement for one readout channel is set to \SI{800}{pC}. There is however interest 
in extending the range up to \SI{1.2}{nC} to be able to measure higher energy jets, expected in rare and possibly new 
physics events. In addition to the processing of physics signals, each \gls{fe} electronic card must include a separate 
channel for large time-constant integration of low amplitude currents. This channel is needed for calibration scans using 
the \gls{cs} system ($\SI{50}{nA}-\SI{100}{nA}$) but also for the monitoring of minimum-bias event rates and the 
instantaneous luminosity ($\SI{20}{pA}-\SI{10}{\mu A}$).

%% file: TeX/part2_asic.tex
%%%%%%%%%%%%%%%%%%%%%%%%%%%%%%%%%%%%%%%%%%%%%%%%%%%%%%%%%%%%%%%%%%%%%%
\section{The FATALIC Architecture}
\label{sec:strategy}
%%%%%%%%%%%%%%%%%%%%%%%%%%%%%%%%%%%%%%%%%%%%%%%%%%%%%%%%%%%%%%%%%%%%%%

\begin{figure}[b]
\centering
  \includegraphics[height=0.3\textheight]{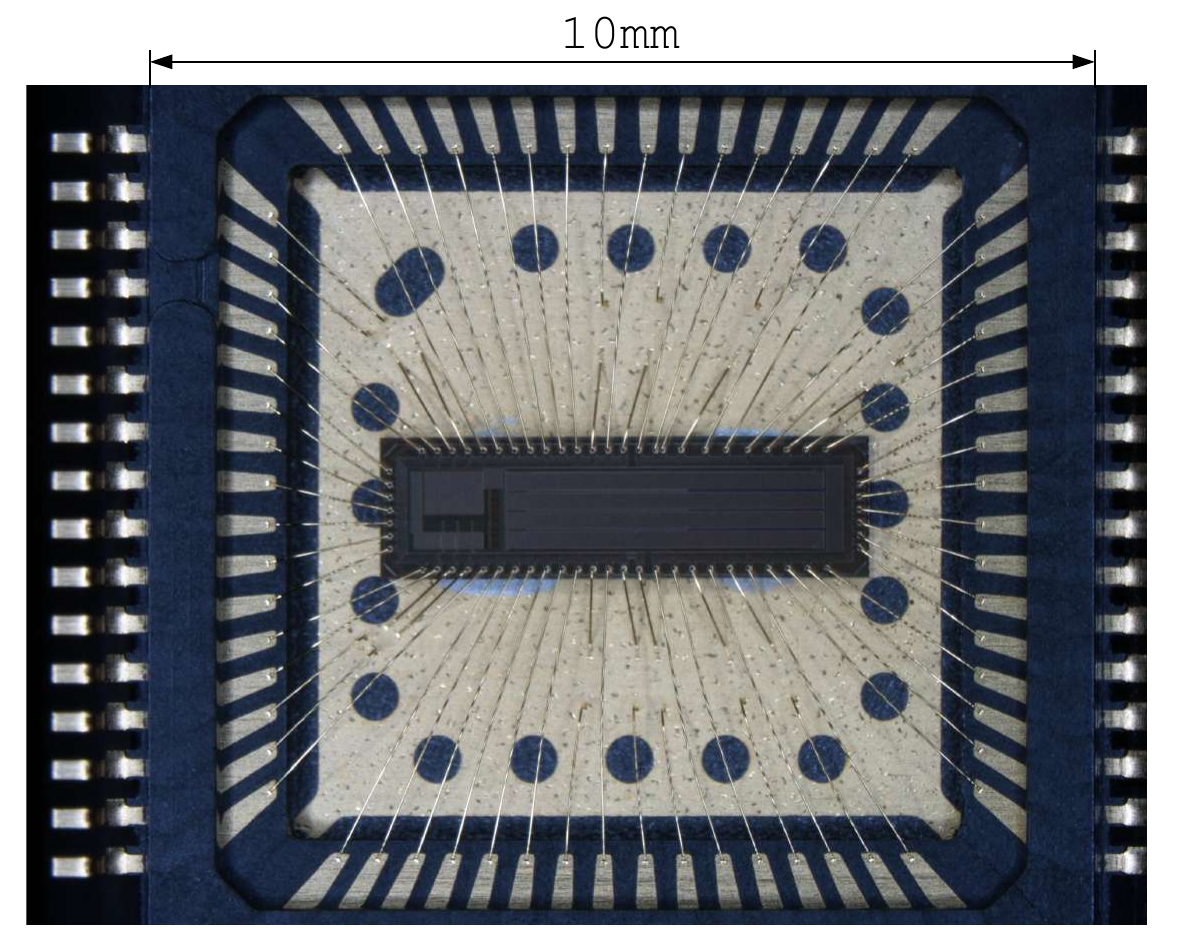}
  \caption{\gls{fatalic} chip, wire-bonded inside an LQFP 64-pin package.\label{fig:fatalic}}
\end{figure}

\gls{fatalic} is an ASIC, based on \SI{130}{nm} GlobalFoundries (GF) Complementary Metal-Oxide-Semiconductor (CMOS) technology, 
designed to replace the current, discrete \gls{fe} readout electronics of the \gls{TileCal}. The proposal of an ASIC, embedding 
both analog signal processing and digitisation, is motivated by its significant advantages in terms of simplicity, radiation
tolerance, power consumption and production cost for a large number of chips. On the other hand, since the operating low voltage 
(\SI{1.6}{V}) is dictated by the technology, \gls{fatalic} must rely on a current-driven, rather than a voltage-driven architecture 
to handle the large input dynamic range. Hence, an input stage employing current conveyors is implemented to adjust the 
impedance and distribute the signal to the different channels. Figure\,\ref{fig:fatalic} shows a \gls{fatalic} chip, wire-bonded 
inside a 64-pin LQFP package.

\begin{figure}[tb]
  \centering
  \includegraphics[width=0.8\textwidth]{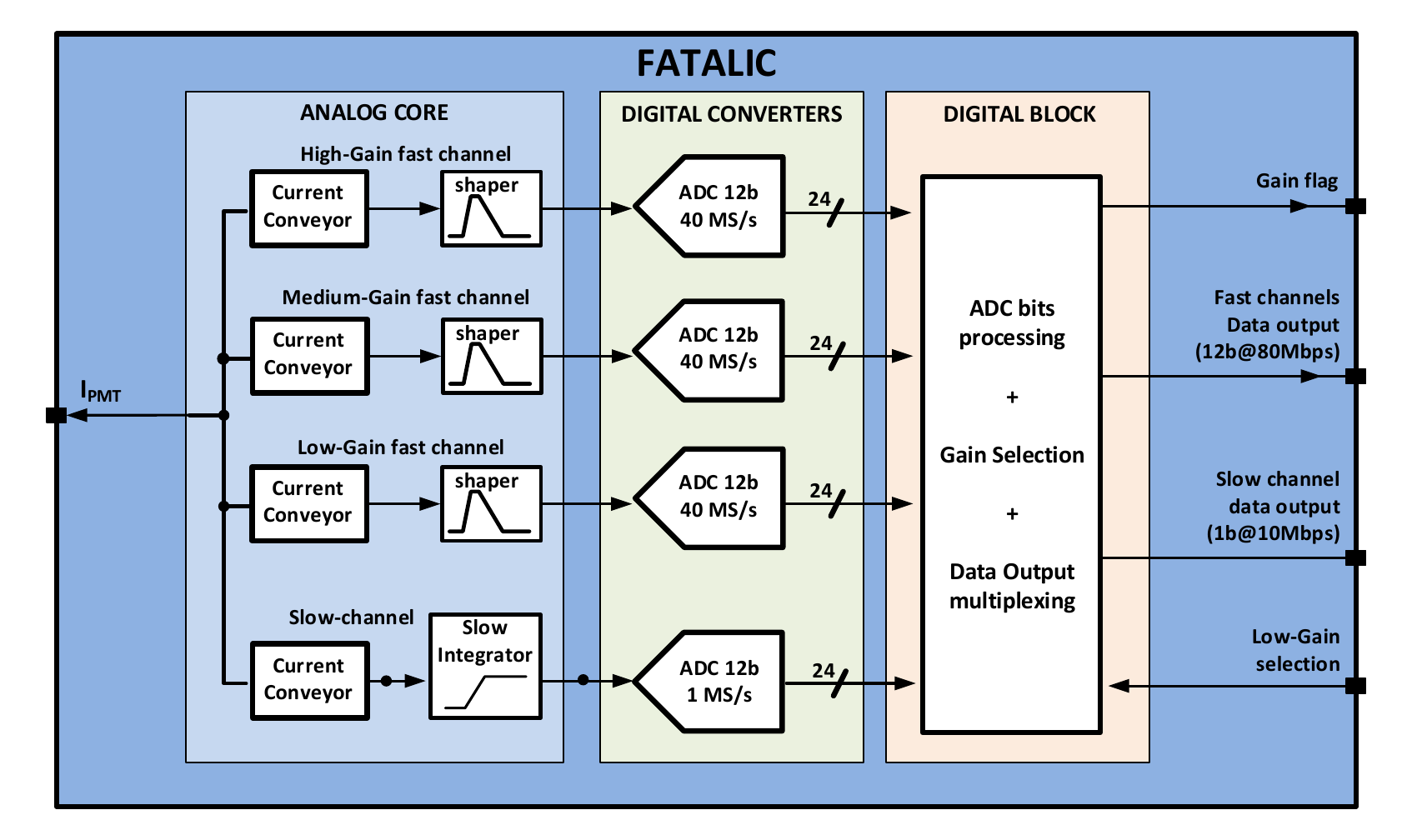}
  \caption{Block diagram of \gls{fatalic}.}
  \label{fig:fatalic5}
\end{figure}

The specifications of \gls{fatalic} are listed in table\,\ref{tab:f5_specs}, while a block diagram summarising the architecture is given 
in figure\,\ref{fig:fatalic5}. In order to handle the large input range (up to \SI{1.2}{nC}), \gls{fatalic} provides three fast channels to
process the \gls{pmt} signal at the \SI{40}{MHz} LHC clock rate, with relative amplification $\times$1 (low gain), $\times$8 (medium gain)
and $\times$64 (high gain). This three-gain scheme is chosen, instead of using two gains, in order to improve the energy resolution (at the
Tile cell level) across the input range, as shown by the related study in Section\,\ref{sec:2gain}. In parallel, the signal is routed to an 
additional, slow channel for integration over a large ($\SI{100}{\mu s}$) time constant.

\begin{table}[t]
	\begin{center}
		\begin{tabular}{l c}
			\toprule
			Technology & 130\,nm CMOS GF \\
			Number of channels per ASIC & 1 \\
			Polarity & negative \\
			Fast channel dynamic range in charge & \SI{25}{fC} - \SI{1.2}{nC}\\
			Fast channel dynamic range in current peak& \SI{1.25}{\mu A} - \SI{60}{mA}\\
			Rise time of the current peak & \SI{4}{ns}\\
			Fall time of the current peak & \SI{36}{ns} \\
			Fast channel noise (rms) & <\SI{12.5}{fC} \\
			Slow channel dynamic range in average current & \SI{0.5}{nA} - \SI{1}{\mu A} \\
			Slow channel Noise (rms) & 0.25\,nA \\
			Power consumption & $\sim$\SI{200}{mW} \\
			Power supply      & 1.6\,V \\
			Output            & 12-bit words \\
			\bottomrule
		\end{tabular}
		\caption{Specifications of \gls{fatalic}.\label{tab:f5_specs}}
	\end{center}
\vspace*{-\baselineskip}
\end{table}

%---------------------------------
\subsection{Fast channels}
\begin{figure}[!t]
  \centering
  \includegraphics[width=0.6\textwidth]{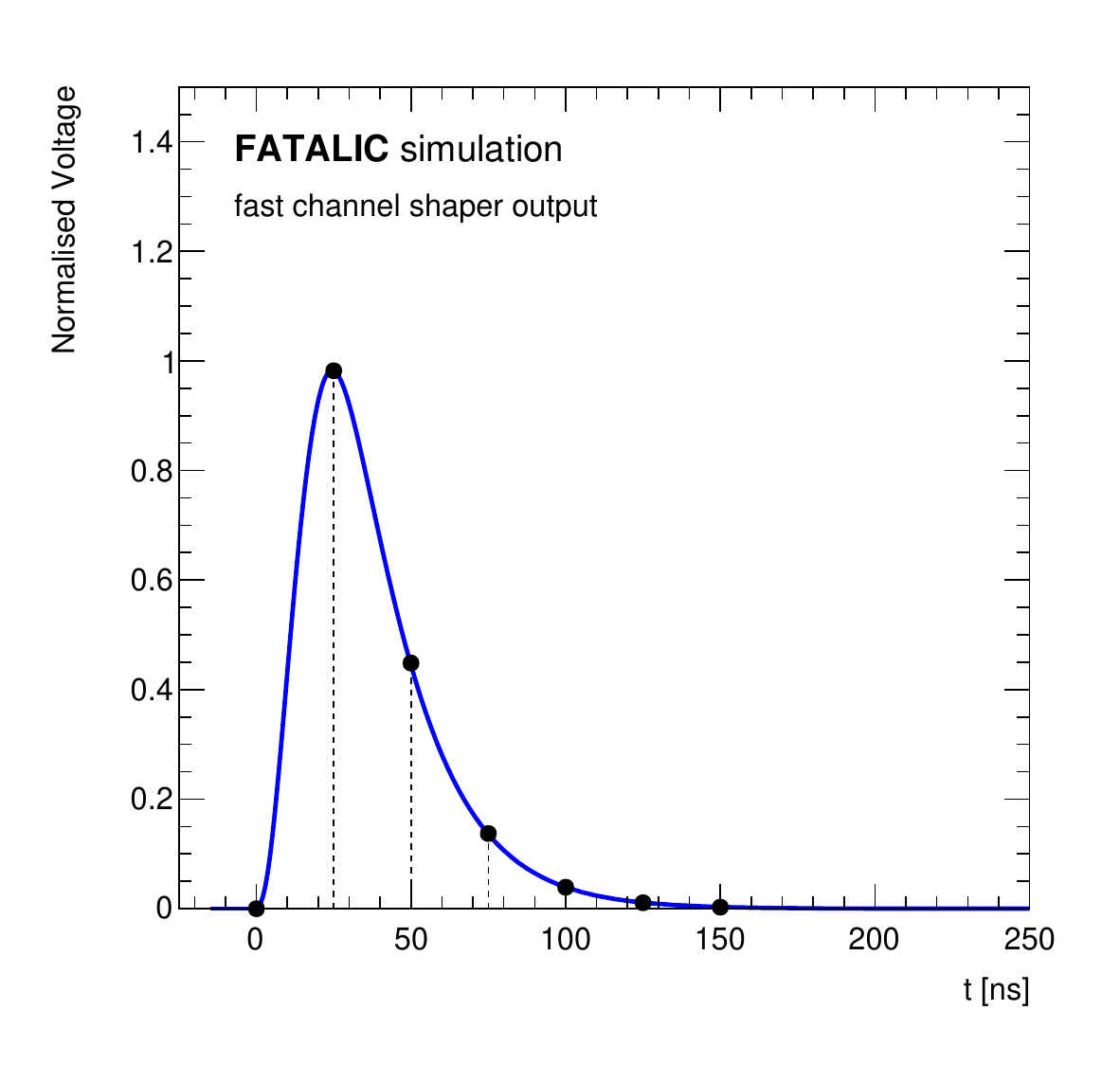}
  \caption{Typical analog output pulse (simulation) from the fast channel shapers. At the \SI{40}{MHz} clock rate, the 2$^\text{nd}$ sample coincides with 
  the pulse peak.\label{fig:pulse}}
\end{figure}

%---------------------------------
In the fast channels the signal is read by three current conveyors with different input impedances, which define the respective gain ratios. 
Current integration and current-to-voltage conversion follow at the transimpedance-shaping stage by three identical shapers, composed by a 
differential amplifier with RC feedback loop. A \SI{25}{ns} time constant is used to produce an asymmetric output pulse with \SI{25}{ns} 
peaking time, as shown in figure\,\ref{fig:pulse}. The digitisation is carried out in the ASIC, by 12-bit \SI{40}{MS/s} \glspl{adc}, to avoid 
degradation of the signal during transmission to an external converter. This establishes an effective 18-bit dynamic range and defines the
range of each channel: $\SI{25}{fC}$-$\SI{20}{pC}$ (high gain), $\SI{200}{fC}$-$\SI{164}{pC}$ (medium gain) and $\SI{1.6}{pC}$-$\SI{1.3}{nC}$ (low gain).

%---------------------------------
\subsection{Slow channel}
%---------------------------------
The slow channel is designed to integrate low amplitude currents in the range from \SI{0.5}{nA} to $\SI{1}{\mu A}$ with minimum contamination
from $\sfrac{1}{f}$ noise, induced by the input stage. A current conveyor with low input impedance is therefore used to drive approximately 87\% 
of the \gls{pmt} signal to the slow channel, while current integration and current-to-voltage conversion are carried out by a differential 
amplifier with large time-constant ($\SI{100}{\mu s}$) RC feedback. The integrated signal is then sampled by a 12-bit \SI{833}{kS/s} \gls{adc}. 
Finally, to minimise the contamination from white noise in the measurement of such low amplitude currents, the data are obtained after averaging
over a time interval of \SI{10}{ms}, as described in Section\,\ref{sec:cards}.

%---------------------------------
\subsection{Outputs}
\label{sec:output}
%---------------------------------

The 12-bit fast channel samples are read-out from 12 respective output pins. In order to comply with the bandwidth of the
uplink to the back-end electronics (see Section\,\ref{sec:cards}), \gls{fatalic} is restricted to provide the data of only 
two of the three channels; the 12-bit data of the medium-gain channel are always delivered upon the rising edge of the \SI{40}{MHz} 
clock, while the data of either the low- or the high-gain channel (alternative gain) are delivered upon the falling edge. 
The selection of the alternative gain is made by the embedded digital block, depending on the saturation of the most sensitive 
channel (dynamic gain switch), and is declared by an output flag. The dynamic gain switch can however be forced to provide 
only low gain readout by a dedicated input bit. On the other hand, slow channel data are delivered serially, at \SI{10}{Mbps} 
(\SI{833}{kHz} for the 12 bits), through a single readout pin.

\clearpage

%%%%%%%%%%%%%%%%%%%%%%%%%%%%%%%%%%%%%%%%%%%%%%%%%%%%%%%%%%%%%%%%%%%%%%
\section{Development of the integrated circuit}
\label{sec:ASICdev}
%%%%%%%%%%%%%%%%%%%%%%%%%%%%%%%%%%%%%%%%%%%%%%%%%%%%%%%%%%%%%%%%%%%%%%

%~~~~~~~~~~~~~~~~~~~~~~~~~~~~~~~~~~~~~~~~~~~~~~~~~
\subsection{Current conveyor system}
\label{subsec:Conveyor}
%~~~~~~~~~~~~~~~~~~~~~~~~~~~~~~~~~~~~~~~~~~~~~~~~~

\begin{figure}[b]
	\begin{center}
		\includegraphics[width=\textwidth]{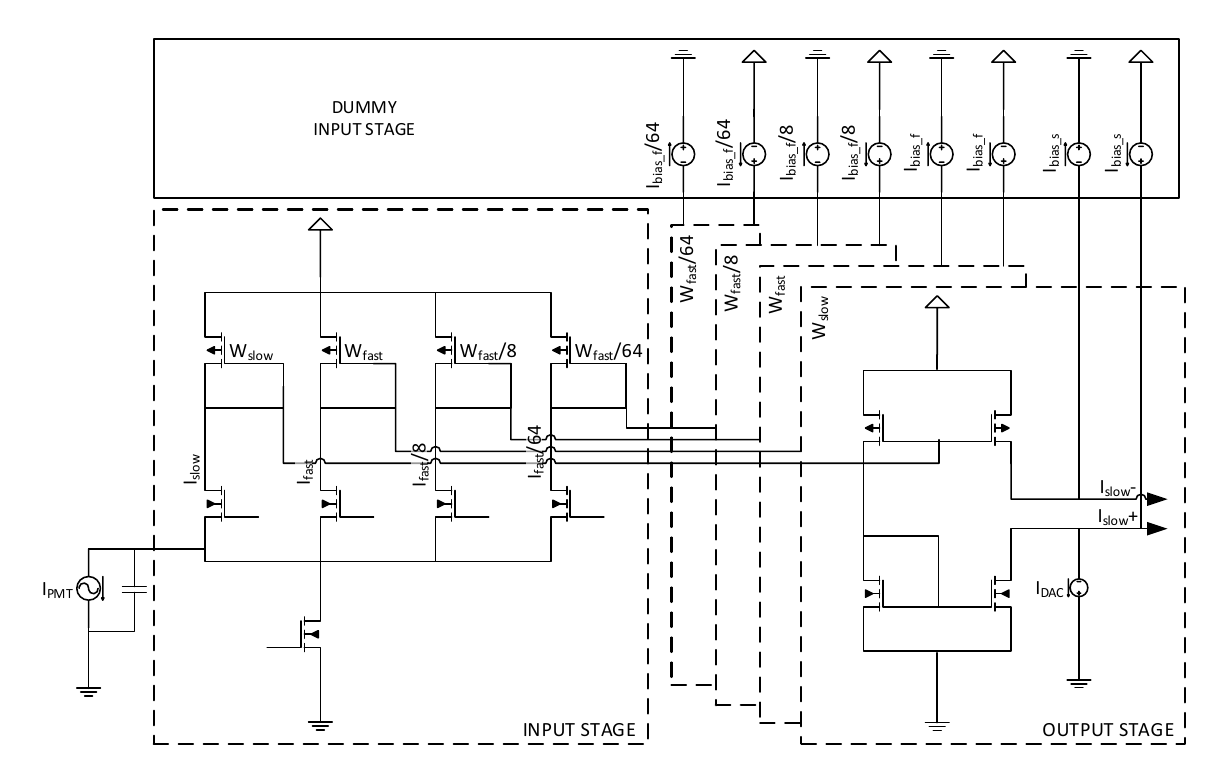}
		\caption{The current conveyor system, consisting of the input stage, the four output stages and a replica 
        (dummy) of the input stage in order to subtract the biasing current from the differential output.}
		\label{iconv}
	\end{center}
\end{figure}

The current conveyor system, schematically depicted in figure~\ref{iconv}, consists of one input stage and four output stages, 
which provide differential signal to each of the \gls{fatalic} channels. At the input stage, a current mirror sets the biasing 
current to the nominal value of \SI{0.5}{mA}. The \gls{pmt} signal is read at the source of four common-gate NMOS\footnote{
%------------------------------------------
Negative channel Metal Oxide Semiconductor}
%------------------------------------------
with the same length (L) but with different widths: $\mathrm{W_{fast}/1}$ (high gain), $\mathrm{W_{fast}}/8$ (medium gain), $\mathrm{W_{fast}}/64$ 
(low gain) and $\mathrm{W_{slow}=8\,W_{fast}}$ (slow channel). A PMOS\footnote{
%------------------------------------------
Positive channel Metal Oxide Semiconductor}
%------------------------------------------
current mirror is then used to replicate the current from the drain of each NMOS onto the respective output stage, while a replica of 
the input stage (dummy) is implemented to return just the biasing current to be subtracted from the final output.

\begin{figure}[t]
	\begin{center}
		\includegraphics[width=0.6\textwidth]{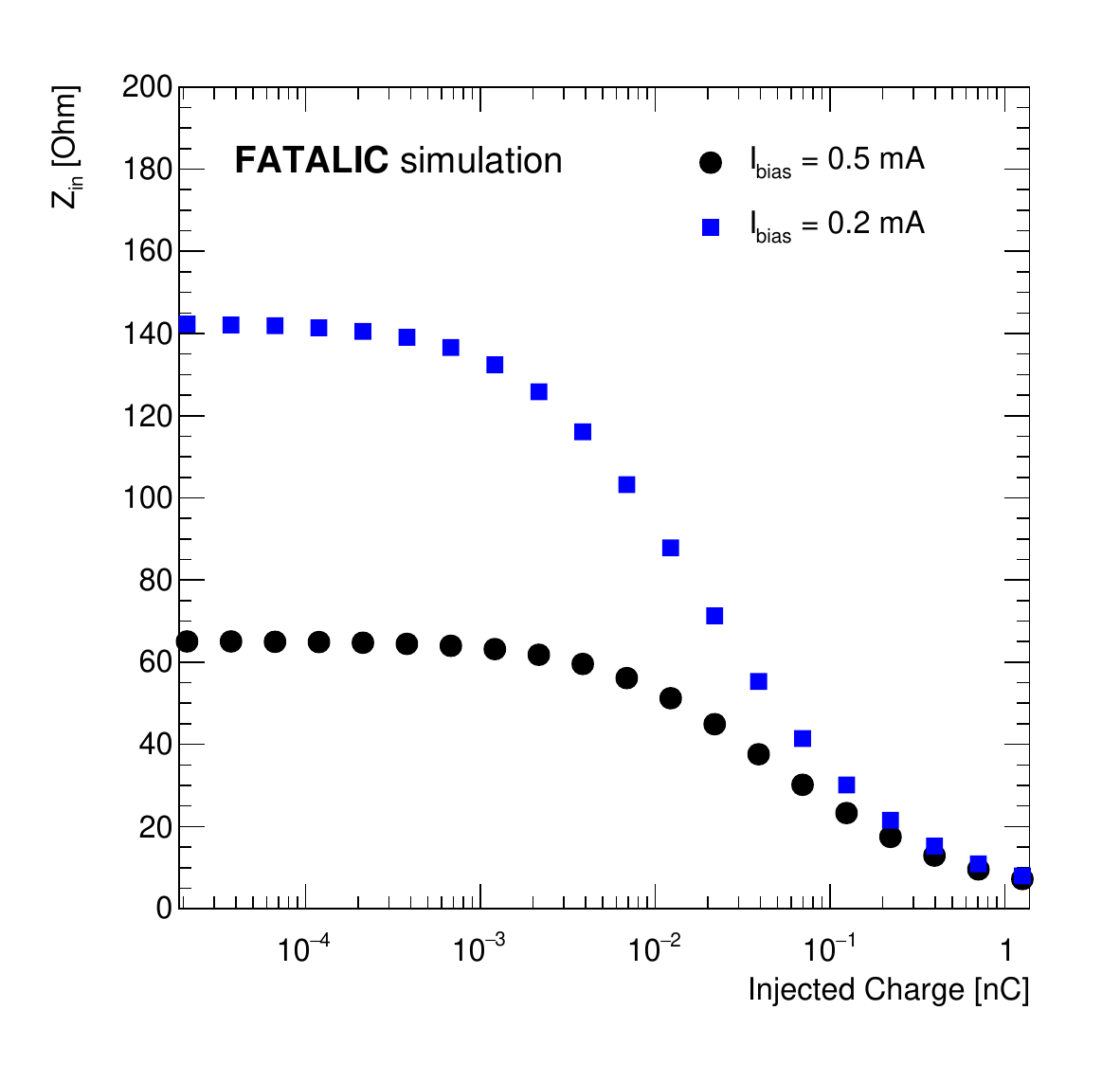}
		\caption{Input impedance as a function of the input charge for the nominal biasing current, $\Ibias=0.5$\,mA, 
        and for a test value of $\Ibias=0.2$\,mA.}
		\label{fig:Zin_vs_Idc}
	\end{center}
\end{figure}

In terms of noise, the performance of such current-conveyor is modest compared to a charge preamplifier, but well adequate, 
given the large dynamic range of input stage. The input impedance is kept below $\SI{70}{\Omega}$ for the entire input range, 
as presented in figure\,\ref{fig:Zin_vs_Idc}. Finally, to account for the open-loop configuration of the input stage, which 
induces large dispersion among the pedestal values, a tuning system with a \gls{dac} is also added to the above design.

%~~~~~~~~~~~~~~~~~~~~~~~~~~~~~~~~~~~~~~~~~~~~~~~~~
\subsection{Signal shaping}
\label{subsec:Shaping}
%~~~~~~~~~~~~~~~~~~~~~~~~~~~~~~~~~~~~~~~~~~~~~~~~~

\begin{figure}[t]
  \centering
  \subfloat[][]{\includegraphics[width=0.6\textwidth]{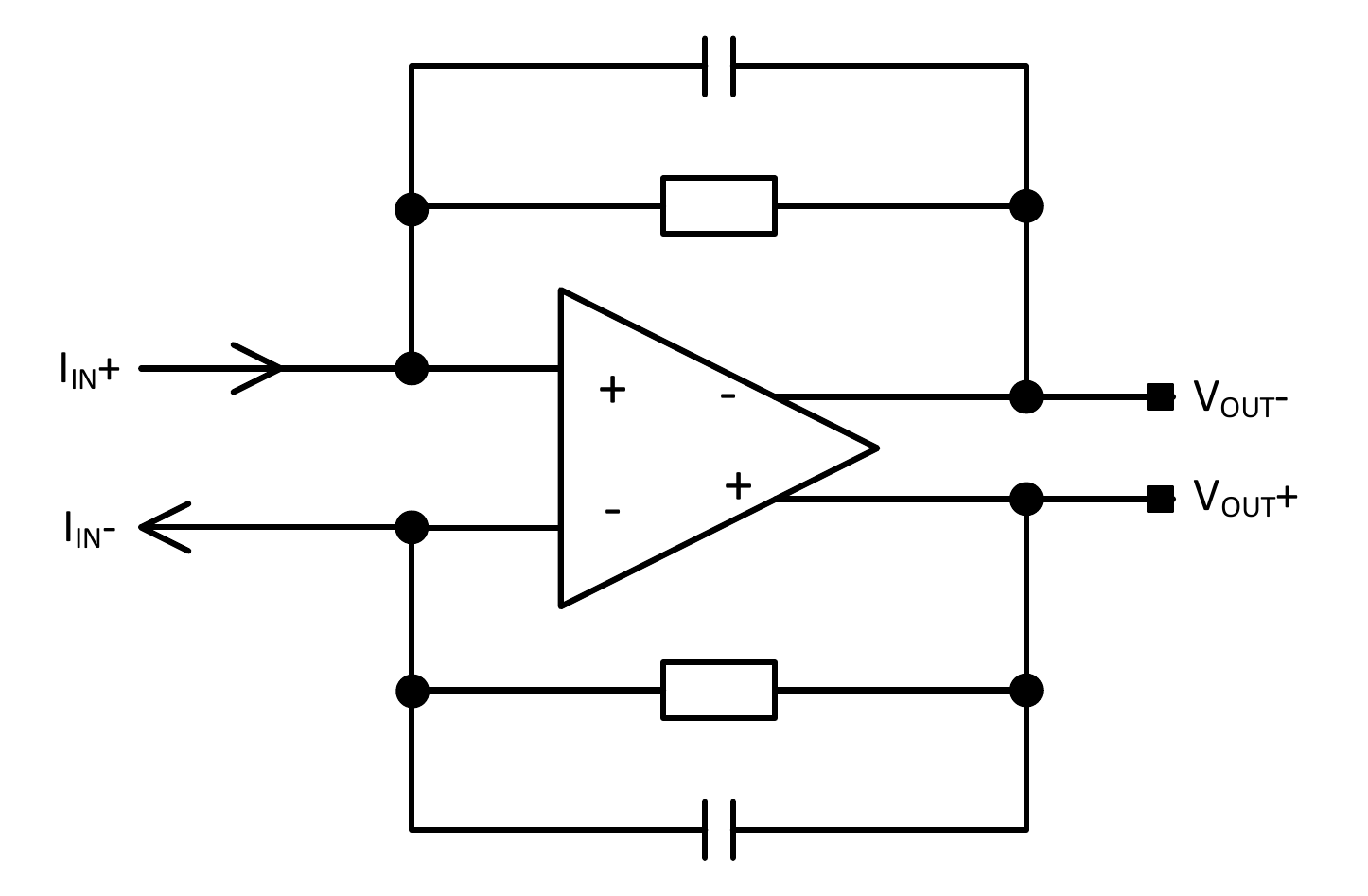}\label{fig:amp_a}}\\
  \subfloat[][]{\includegraphics[width=0.8\textwidth]{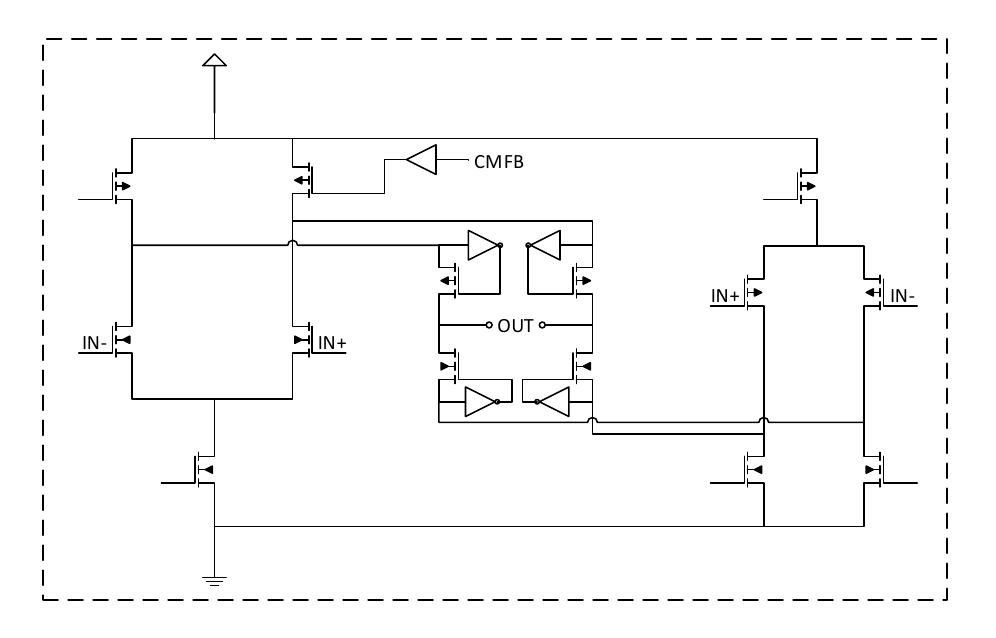}\label{fig:amp_b}}
  \caption{(a) Schematic of the shaper, consisting of a differential amplifier with RC feedback loop. 
           (b) Architecture of the differential amplifier.\label{fig:amp}}
\end{figure}

The shaper (figure\,\ref{fig:amp_a}) is a differential amplifier with RC feedback loop, with a time constant of \SI{25}{ns} 
($R=\SI{5}{k\Omega}$, $C=\SI{5}{pF}$) in the fast channels and $\SI{100}{\mu s}$ ($R=\SI{500}{k\Omega}$, $C=\SI{200}{pF}$) 
in the slow channel. The amplifier is based on a folded-cascode boosted architecture, depicted in figure\,\ref{fig:amp_b}, 
with two differential input pairs, one NMOS and one PMOS, and Common Mode FeedBack (CMFB) to control the output common mode. 
Its open-loop characteristics are listed in table\,\ref{tab:opamp}. 

\begin{table}[h]
	\centering
	\begin{tabular}{l c}
		\toprule
		Open-loop gain       & \SI{84}{dB}\\
		-3db bandwidth       & \SI{38}{kHz}\\
		Unity-gain bandwidth & \SI{650}{MHz}\\
		Slew rate            & $\SI{220}{V/\mu s}$\\
		Phase margin         & 85$^\circ$\\
		Supply voltage       & \SI{1.6}{V}\\
		Consumption          & \SI{2.3}{mA}\\
		\bottomrule
	\end{tabular}
    \caption{\label{tab:opamp}Open-loop characteristics of the shaper with the \gls{adc} input capacitive load of \SI{1.6}{pF}.}
\end{table}	

%~~~~~~~~~~~~~~~~~~~~~~~~~~~~~~~~~~~~~~~~~~~~~~~~~
\subsection{Signal digitisation}
\label{subsec:ADC}
%~~~~~~~~~~~~~~~~~~~~~~~~~~~~~~~~~~~~~~~~~~~~~~~~~

The shaper output is sampled by three 12-bit \SI{40}{MS/s} \glspl{adc} in the fast channels and one 12-bit \SI{833}{kS/s} 
\gls{adc} in the slow channel. The \gls{adc} design by LPC is based on the pipeline architecture
%, which is well adequate for the particular resolution and sampling rate, 
with 1.5-bit-per-stage resolution.
%. is chosen to avoid degradation of the performance due to the comparator offsets. 
Among the various efficient \gls{adc} architectures developed and improved over the last ten years, the pipelined \gls{adc} 
has been adapted for high resolution, speed and dynamic range with relatively low power consumption and low component count. 
In CMOS technologies, resolutions in the range of 10-14 bits with a sampling frequency up to \SI{100}{MS/s} are typically 
achieved with power consumption lower than \SI{100}{mW}.

\begin{figure}[tb]
	\centering
	\includegraphics[width=0.9\textwidth]{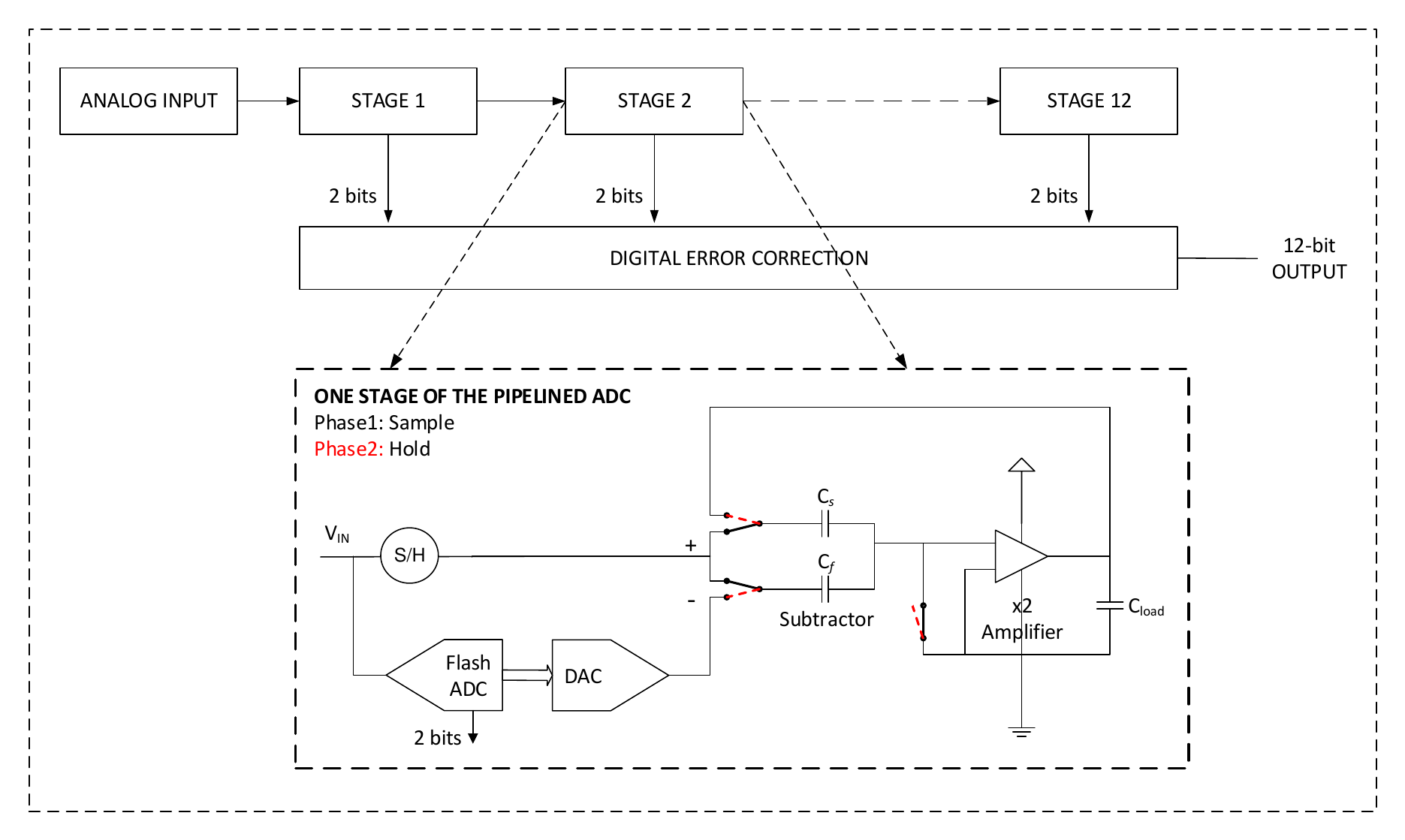}
	\caption{Block diagram of the 12-bit pipelined ADC with two output bits per stage.}
	\label{fig:adc}
\end{figure}

Figure\,\ref{fig:adc} presents the block diagram of the pipelined ADC, with two output bits per stage, displaying the architecture
of one of the stages. Each stage receives a differential input voltage, in the range $\pm$\SI{500}{mV}, which is read by a 
sample-and-hold (S/H) and a 2-bit flash \gls{adc}. The flash \gls{adc} compares the input to two threshold voltages ($\pm$\SI{125}{mV}) 
and outputs a 2-bit word, while a \gls{dac} and a residue amplifier provide the input to the next stage, as presented in 
table\,\ref{tab:adc_stage}. Although a 2-bit word is delivered, the effective resolution is 1.5-bit since the combination \texttt{11} is 
avoided. This bit redundancy limits the degradation of the Integral Non-Linearity\footnote{
%------------------------------------------------------------
The INL is defined as the deviation, in Least-Significant-Bits (LSBs), of the output code from the ideal transfer-function.}
%------------------------------------------------------------
(INL) due to variations of the residue amplifier gain or the comparator offsets (the INL is unaffected by variations of the offset 
voltage for up to $\pm$12.5\% of the full-scale input voltage). The complete architecture includes 12 cascading stages and is 
followed by the digital correction block (see Section\,\ref{subsec:DigitalBlock}), which delivers the final digital code. The 
estimated power consumption of the \gls{adc} is \SI{48}{mW}.

\begin{table}[h]
  \centering
  \begin{tabular}{rcl c c c}
  \toprule
                      & $V_\mathrm{in}^i$  &  & $[b_2b_1]$  & DAC          & $V_\mathrm{in}^{i+1}$\\
  \midrule
  $ V_\mathrm{ref}/4$ &-& $ V_\mathrm{ref}$   & [10] & $V_\mathrm{ref}/2$  & $2V_\mathrm{in}^i - V_\mathrm{ref}$\\
  $-V_\mathrm{ref}/4$ &-& $ V_\mathrm{ref}/4$ & [01] & 0                   & $2V_\mathrm{in}^i$\\
  $-V_\mathrm{ref}$   &-& $-V_\mathrm{ref}/4$ & [00] & $-V_\mathrm{ref}/2$ & $2V_\mathrm{in}^i + V_\mathrm{ref}$\\
  \bottomrule
  \end{tabular}
  \caption{Each stage $i$ of the pipeline converts the input voltage $V_\mathrm{in}^{i}$, in the dynamic range $\pm V_\mathrm{ref}$, 
           into a 2-bit word $[b_2b_1]$ and feeds the amplified residual $V_\mathrm{in}^{i+1}=2(V_\mathrm{in}^i-\mathrm{DAC})$ to the 
           next stage.\label{tab:adc_stage}}
\end{table}

%~~~~~~~~~~~~~~~~~~~~~~~~~~~~~~~~~~~~~~~~~~~~~~~~~
\paragraph{The comparator.}

\begin{figure}[t]
  \centering
  \includegraphics[width=0.9\textwidth]{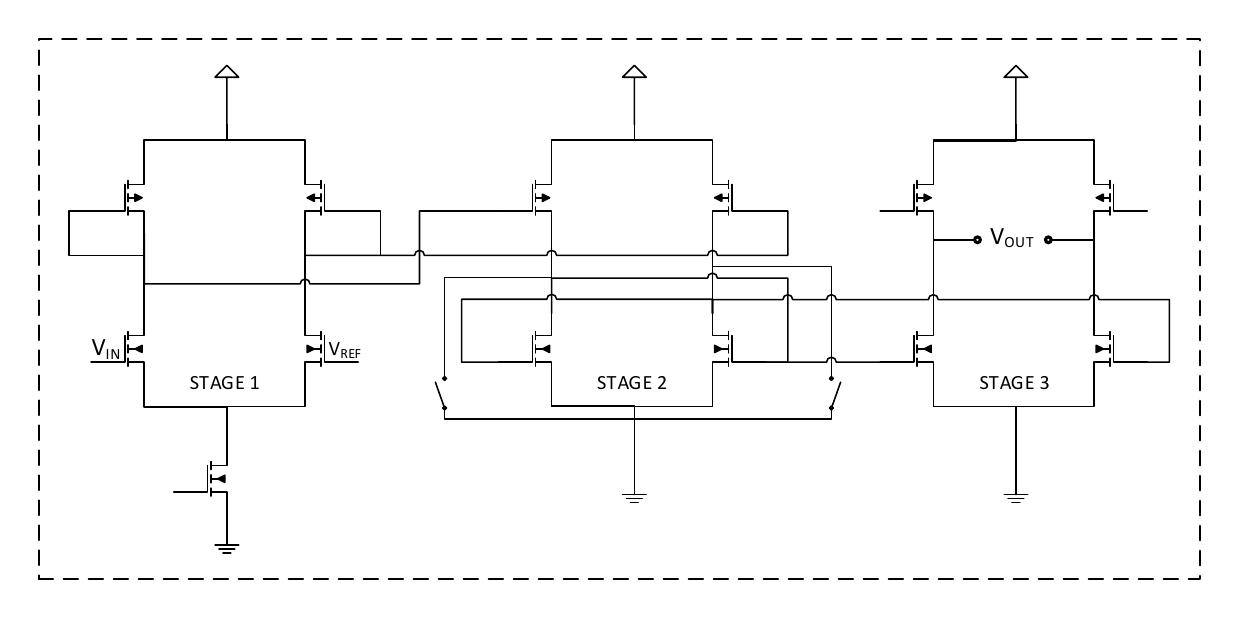}
  \caption{Architecture of the comparator (single-ended representation for simplicity).}
  \label{fig:comp_adc}
\end{figure}

The architecture of the comparator is presented in figure\,\ref{fig:comp_adc}. The transconductance input stage is fully differential, 
comparing the differential input signal to the differential threshold voltage, while isolating the input from kick-back noise, 
induced by the switching of the subsequent latched stage. The latched stage performs the comparison when the switch-transistors are 
OFF and a reset when they are ON. The state of the comparator, when it is latched, is memorised thanks to the bistable third stage. Lastly, 
two NOT gates perform the final digital shaping of the output signal. The main characteristics of the comparator are given in 
table\,\ref{tab:comp}.

%The latch state of the comparator is memorised by a bi-stable third stage.

%~~~~~~~~~~~~~~~~~~~~~~~~~~~~~~~~~~~~~~~~~~~~~~~~~
\paragraph{The \gls{dac}.}

The \gls{dac} employs a set of switches, controlled by the comparators, to select the differential reference voltage to 
be applied on the feedback capacitor of the residue amplifier. Given the $\pm$\SI{500}{mV} input dynamic range, the 
reference voltages are set to $\pm$\SI{250}{mV}.

\begin{table}[h]
\centering
	\vspace{0.5cm}
	\begin{tabular}{l c}
	\toprule
	Clocking frequency & 40\,MHz\\
	Sensitivity        & 50\,$\mu$V\\
	Power supply       & 1.6\,V\\
	Power consumption  & 2.1\,mW\\
	\bottomrule
	\end{tabular}
	\caption{\label{tab:comp}Characteristics of the comparator.}
\end{table}	

%~~~~~~~~~~~~~~~~~~~~~~~~~~~~~~~~~~~~~~~~~~~~~~~~~
\paragraph{The gain-2 residue amplifier.}

The residue amplifier, displayed in figure\,\ref{fig:adc}, is a differential amplifier with capacitive feedback. Capacitive, 
rather than resistive feedback is used for better component matching, which is crucial for the accuracy of the amplification 
and, therefore, the linearity of the \gls{adc}. A capacitance of \SI{800}{fF} is sufficient to minimise both the thermal 
noise ($kT/C$) and the component mismatch ($\sim$$1/\sqrt{C}$). On the other hand, the design requires a small die surface 
and low supply current for dynamic performance. To match both the sampling ($C_\mathrm{S}$) and feedback ($C_\mathrm{F}$) 
capacitance to \SI{800}{fF}, with an accuracy better than 0.1\%, an array of four \SI{1600}{fF} MIM capacitor unit cells is 
drawn in common-centroid layout, while dummy switches are added to counterbalance parasitic capacitors introduced by the 
reset switches. Lastly, the timing sequence, controlling the sample, hold/amplification and reset phases of each stage, has 
been carefully defined to prevent charge disruption.

%\begin{figure}[bt]
%\centering
%\includegraphics[width=0.8\textwidth]{figures/asic/etage_adc}
%\caption{One stage of the pipelined ADC (single-ended representation for simplicity).}
%\label{fig:ADC_stage}
%\end{figure}

%~~~~~~~~~~~~~~~~~~~~~~~~~~~~~~~~~~~~~~~~~~~~~~~~~
\subsection{Digital block}
\label{subsec:DigitalBlock}
%~~~~~~~~~~~~~~~~~~~~~~~~~~~~~~~~~~~~~~~~~~~~~~~~~

\begin{figure}[tb]
\centering
\includegraphics[width=0.9\textwidth]{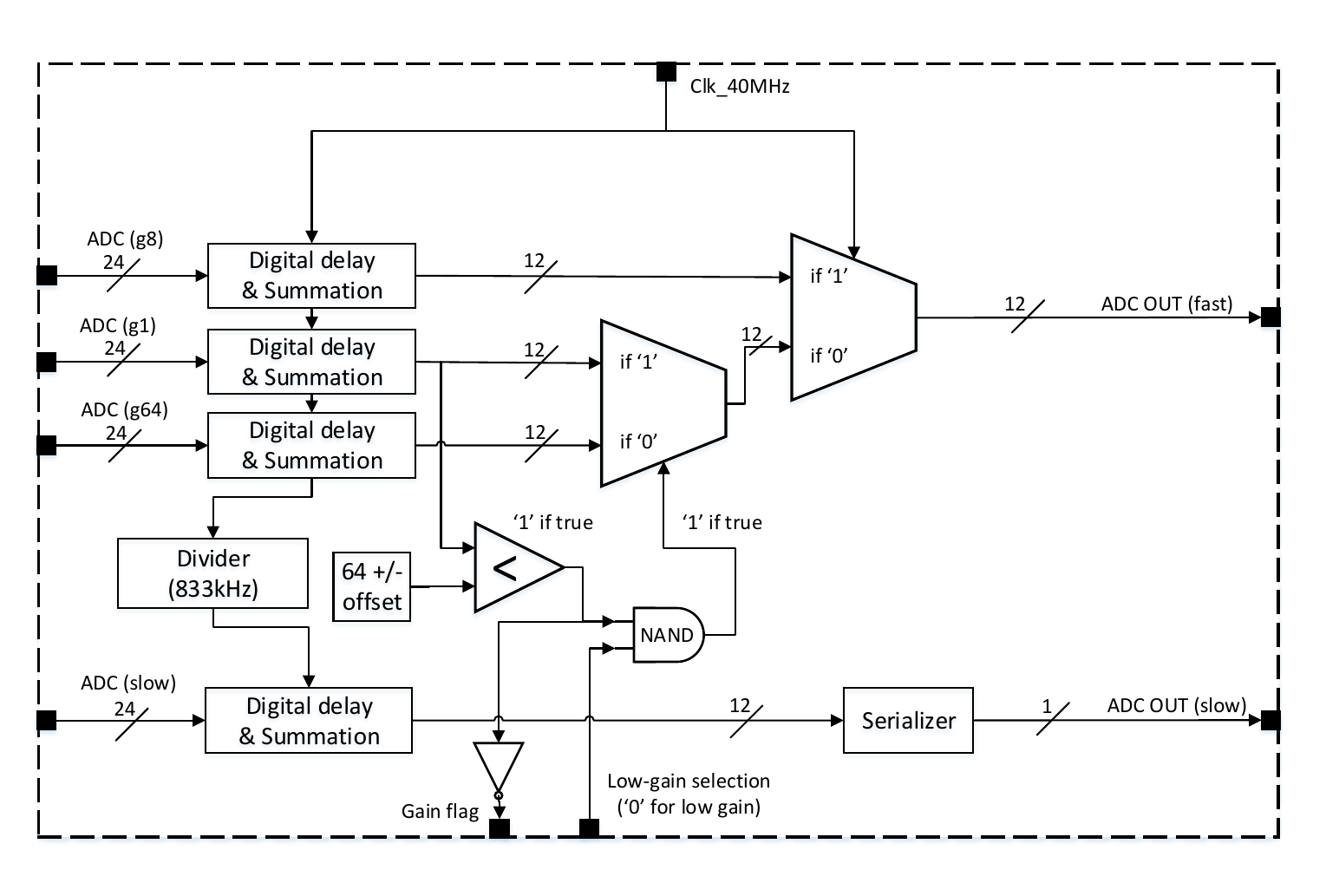}
\caption{Architecture of the digital block.}
\label{bloc_digital2}
\end{figure}

Each \gls{adc} delivers twelve 2-bit words at the \SI{40}{MHz} clock rate. The result of the complete conversion is therefore 
a 24-bit word with 12 redundant bits. The digital block (figure\,\ref{bloc_digital2}) synchronises the outputs using shift 
registers to compensate for the delay due to the position of each stage in the pipeline. The total latency of this process is
equal to eight clock cycles (\SI{200}{ns}) and it is fixed for every power cycle. Digital summation (with carry-save) of the
most-significant bit of each stage with the least-significant bit of the previous stage is then performed to the final 12-bit 
code every \SI{25}{ns} for the fast channels and $\SI{1.2}{\mu s}$) for the slow channel. The second function of the digital block 
is to select the alternative gain output between the high-gain and the low-gain channel data (see Section\,\ref{sec:output}). 
This selection is made according to the output of the low-gain channel; if it is equal or higher (lower) than \SI{600}{ADC} counts, 
then the low-gain (high-gain) channel data are delivered.

%~~~~~~~~~~~~~~~~~~~~~~~~~~~~~~~~~~~~~~~~~~~~~~~~~
\subsection{Floorplan}
%~~~~~~~~~~~~~~~~~~~~~~~~~~~~~~~~~~~~~~~~~~~~~~~~~

\begin{figure}[tb]
\bigskip
\centering
  \includegraphics[width=0.9\textwidth]{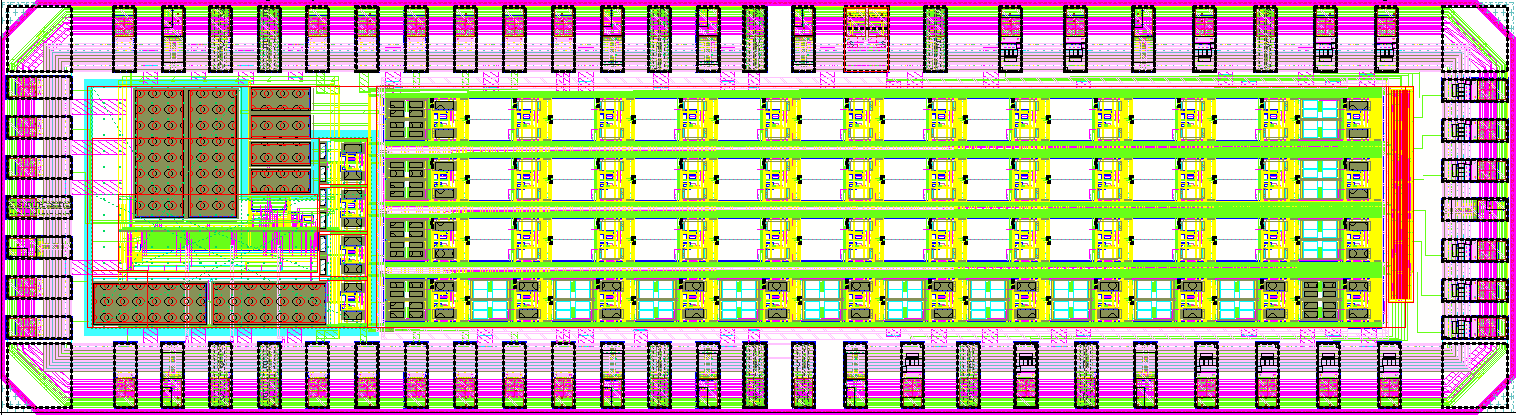}
  \caption{The floorplan of \gls{fatalic}. Two main regions are distinguished; the region hosting the analog blocks
  (left) and the region hosting the \glspl{adc} and the digital block.\label{fig:fatalic_floor}}
\end{figure}

The floorplan of \gls{fatalic} (figure\,\ref{fig:fatalic_floor}) is optimised for minimum surface, while preserving signal 
integrity. Two main regions are distinguished; the region hosting the input stage and the shapers (analog block), and 
the region hosting the \glspl{adc} and the digital block. The two regions are isolated by a high impedance (BFMOAT) layer 
to reduce the coupling. The analog power and reference voltage rails are also decoupled by embedded large capacitors. The 
total surface of the chip measures $\SI{7.5}{mm^2}$, while the core area is limited to $\SI{2.3}{mm^2}$.

%~~~~~~~~~~~~~~~~~~~~~~~~~~~~~~~~~~~~~~~~~~~~~~~~~
\subsection{Design verification}
\label{subsec:Verify}
%~~~~~~~~~~~~~~~~~~~~~~~~~~~~~~~~~~~~~~~~~~~~~~~~~

All standard verification checks have been performed for the design of \gls{fatalic}. The analog part was tested with Monte Carlo simulations,
using the Virtuoso Analog Design Environment by Cadence. The yield with process and mismatch is evaluated to 90\%, while the offset 
of the input stage, which is an open loop configuration, is tunable to maintain a high yield and was assessed with mismatch simulations.
Corners simulations did not highlight sensitivity of the ASIC to the temperature or the power supply. For the digital part of the ASIC, 
simulations were performed with SOC Encounter, while for the full ASIC, mixed-mode simulations were performed with the Virtuoso Analog
Mixed-Signal (AMS) Designer. Finally, the good performance of the ASIC is confirmed with all of the 24 prototype chips that were produced.

%% file: TeX/part3_cards.tex
%%%%%%%%%%%%%%%%%%%%%%%%%%%%%%%%%%%%%%%%%%%%%%%%%%%%%%%%%%%%%%%%%%%%%%
\section{Associated cards}
\label{sec:cards}
%%%%%%%%%%%%%%%%%%%%%%%%%%%%%%%%%%%%%%%%%%%%%%%%%%%%%%%%%%%%%%%%%%%%%%

\begin{figure}[t]
  \begin{center}
    \includegraphics[width=1.0\textwidth]{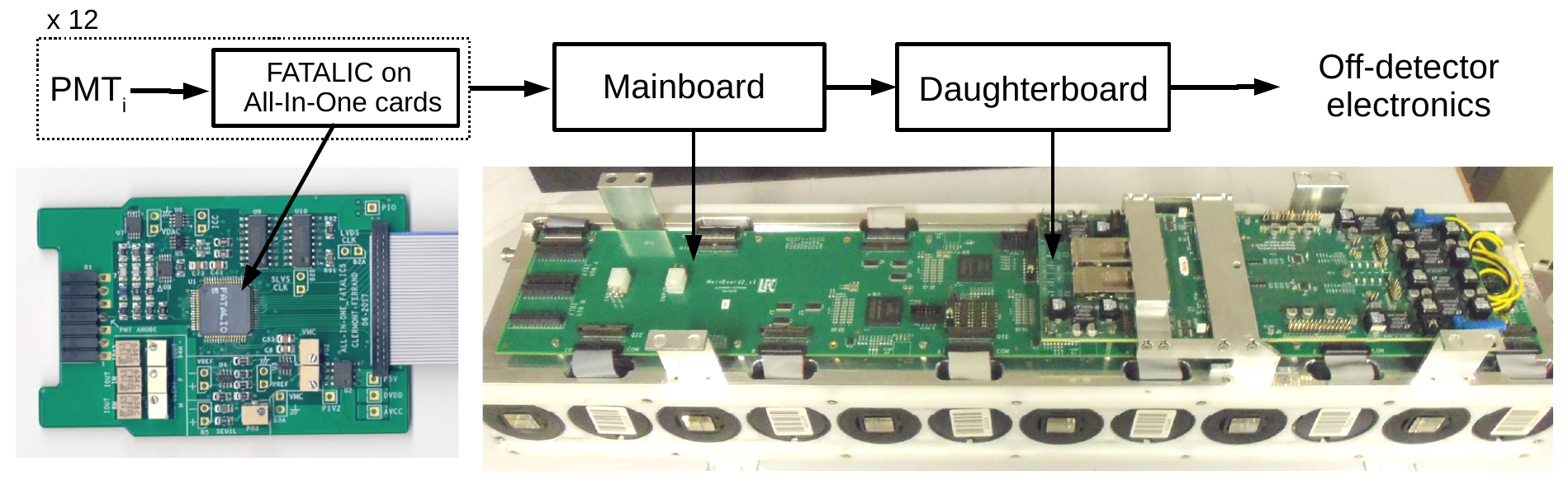}
    \caption{Schematic view of the readout chain. Left: the All-In-One card supporting FATALIC. Right: a 
    fully equipped mini-drawer housing 12 \glspl{pmt} along with the associated readout electronic cards.
    \label{fig:cards:overview}}
  \end{center}
\end{figure}

Figure\,\ref{fig:cards:overview} displays a fully equipped mini-drawer, hosting 12 \glspl{pmt}, 12 ``All-in-One'' cards, each of 
which contains one \gls{fatalic} chip and the dedicated \gls{cis}, and the Mainboard, which controls the All-in-One cards and 
transmits the data to the Daughterboard. Both the All-in-One cards and the Mainboard have been designed by LPC for the purposes 
of \gls{fatalic}. The Daughterboard, also shown in figure\,\ref{fig:cards:overview}, is the on-detector interface to the back-end 
electronics. It is divided into two independent sides, each of which establishes a \SI{9.6}{Gb/s} uplink to deliver the data of 
six readout channels to the off-detector PPr, while a \SI{4.8}{Gb/s} downlink is used to transmit the \gls{lhc} clock as well as 
control and configuration commands.

%---------------------------------
\subsection{All-in-one card}
%---------------------------------

The All-in-One card contains \gls{fatalic} and the \gls{cis}. It is based on a 6-layer Printed Circuit Board (PCB) with 
dimensions $\SI{7.0}{cm}\times\SI{4.7}{cm}$. On one side, a 7-pin connector attaches the card to the high-voltage divider, 
at the basis of the \gls{pmt}, while on the opposite side a 40-conductor ribbon cable establishes communication with the 
Mainboard. Finally, nine on-board potentiometers adjust the low voltage supply and the pedestal for each channel.

\begin{figure}[p]
  \begin{center}
    \includegraphics[width=0.9\textwidth]{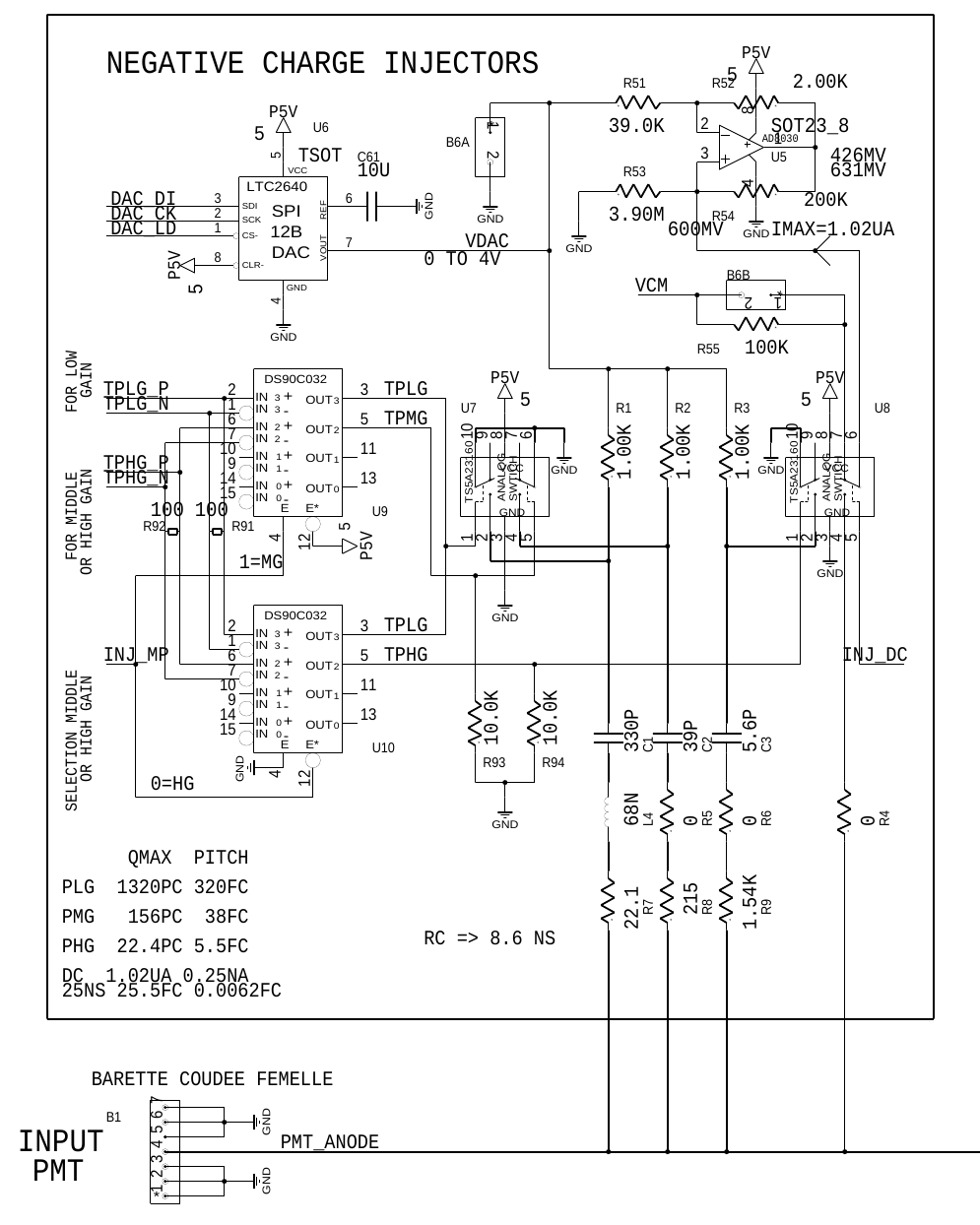}
    \caption{Schematic diagram of the \gls{cis}.\label{fig:cards:cis}}
  \end{center}
\end{figure}

A schematic diagram of the \gls{cis} is given in figure\,\ref{fig:cards:cis}. The charge injection is driven by a 12-bit 
\gls{dac} with maximum output \SI{4.095}{Volts}. The \gls{dac} charges one of the three available
capacitors, \SI{5.6}{pF}, \SI{39}{pF} or \SI{330}{pF}, for the scanning of the high-, medium- or low-gain dynamic 
range, respectively. The connected capacitance is defined through analog switches, controlled by two timing signals
from the Mainboard. The same timing signals control the charge/discharge (through appropriately adjusted resistances) 
cycles reproducing the \gls{pmt} pulse shape. Finally, the \gls{dac} is also connected to a Howland DC current source
to allow calibration of the slow channel.

%---------------------------------
\subsection{Mainboard}
%---------------------------------

The twelve All-in-One cards of one mini-drawer are controlled by the $\SI{28}{cm}\times\SI{10}{cm}$ Mainboard, which 
serialises the data for transmission to the Daughterboard and distributes clocks and commands from the Daughterboard 
to the \gls{fe} electronics. It also distributes the \SI{10}{V} low-voltage supply, by means of point-of-load regulators, 
to power both the All-in-One cards and the Daughterboard. The connection to the Daughterboard is established with a 
400-pin FPGA Mezzanine Connector (FMC).

The Mainboard is divided into two sections, each of which receives a separate \SI{10}{V} supply and carries two Altera 
Cyclone4 FPGAs. Each FPGA controls three All-in-One cards and communicates with the Daughterboard through a Serial Peripheral 
Interface (SPI) port. The data of each of the two output fast channels are transmitted to the Daughterboard in Low-Voltage 
Differential Signaling (LVDS) with a ``2-Lane Output, 16-bit Serialisation'' at \SI{320}{Mbps}. The necessary synchronisation 
clocks (\SI{160}{MHz} and \SI{40}{MHz}) are also generated by the FPGA, in common for the three cards. On the other hand, 
slow channel data are digitally summed in the FPGA over a time interval of \SI{10}{ms} and delivered to the Daughterboard 
through a standard Inter-Integrated Circuit (I$^2$C) bus.

%\begin{figure}[t]
%  \begin{center}
%    \includegraphics[width=0.8\textwidth]{figures/cards/serialization.jpg}
%    \caption{Serialisation scheme.\label{fig:mbserialisation}}
%  \end{center}
%\end{figure}

%% file: TeX/part4_reco.tex
%%%%%%%%%%%%%%%%%%%%%%%%%%%%%%%%%%
\section{Simulation and performance}
\label{subsec:SimAndPerf}
%%%%%%%%%%%%%%%%%%%%%%%%%%%%%%%%%%

The following paragraphs describe the performance of \gls{fatalic} in terms of noise, linearity and radiation
tolerance, based on simulation and experimental measurements with 24 prototype chips. For the simulation of the 
ASIC, the Cadence Virtuoso Analog Mixed-Signal Design Environment is used. Regarding the experimental measurements, 
it must be noted that the All-in-One cards, used to accomodate \gls{fatalic} for the purposes of these studies,
are not adequate to support the slow channel of \gls{fatalic}. Specifically, due to the particularly large gain 
of the slow channel, the resistance range of the respective on-board potentiometer is not sufficient for the 
adjustment of the pedestal within the input dynamic range. As a workaround, it was decided to reduce the biasing 
current in the input stage, from the nominal value of \SI{0.5}{mA} to \SI{0.2}{mA}. This however affected the 
performance of the fast channels, causing dynamic amplification of the signal depending on the amplitude, with 
significant impact on the linearity. This effect would normally be eliminated by equipping the All-in-One cards 
with potentiometers of larger resistance range or by implementing slow control functionalities in \gls{fatalic}.

%---------------------------------
\subsection{Noise measurements}
\label{subsec:ped}
%---------------------------------

\begin{figure}[!bt]
\centering
\subfloat[][]{\includegraphics[width=0.48\textwidth]{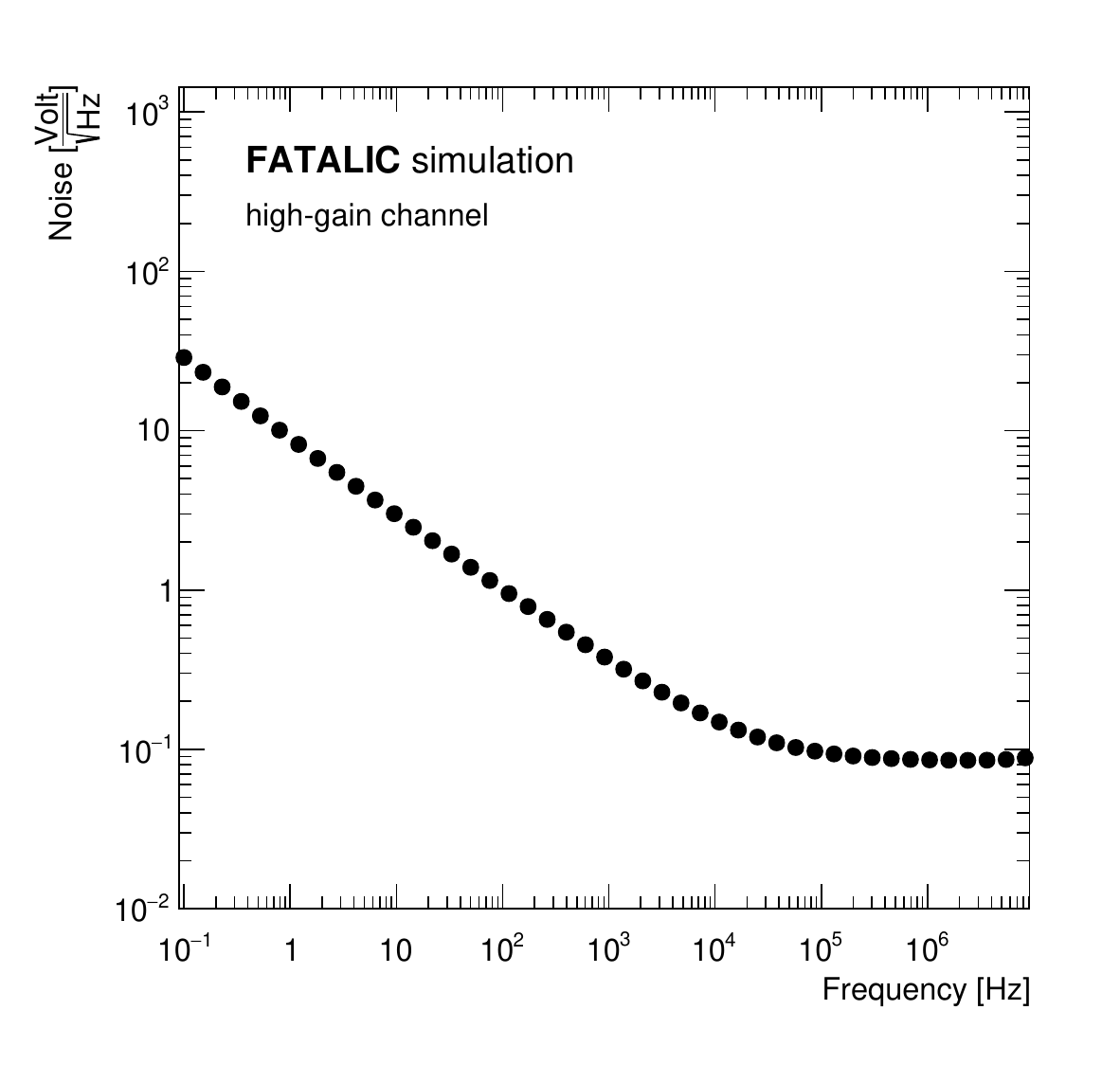}\label{fig:gnoise_a}}
\subfloat[][]{\includegraphics[width=0.48\textwidth]{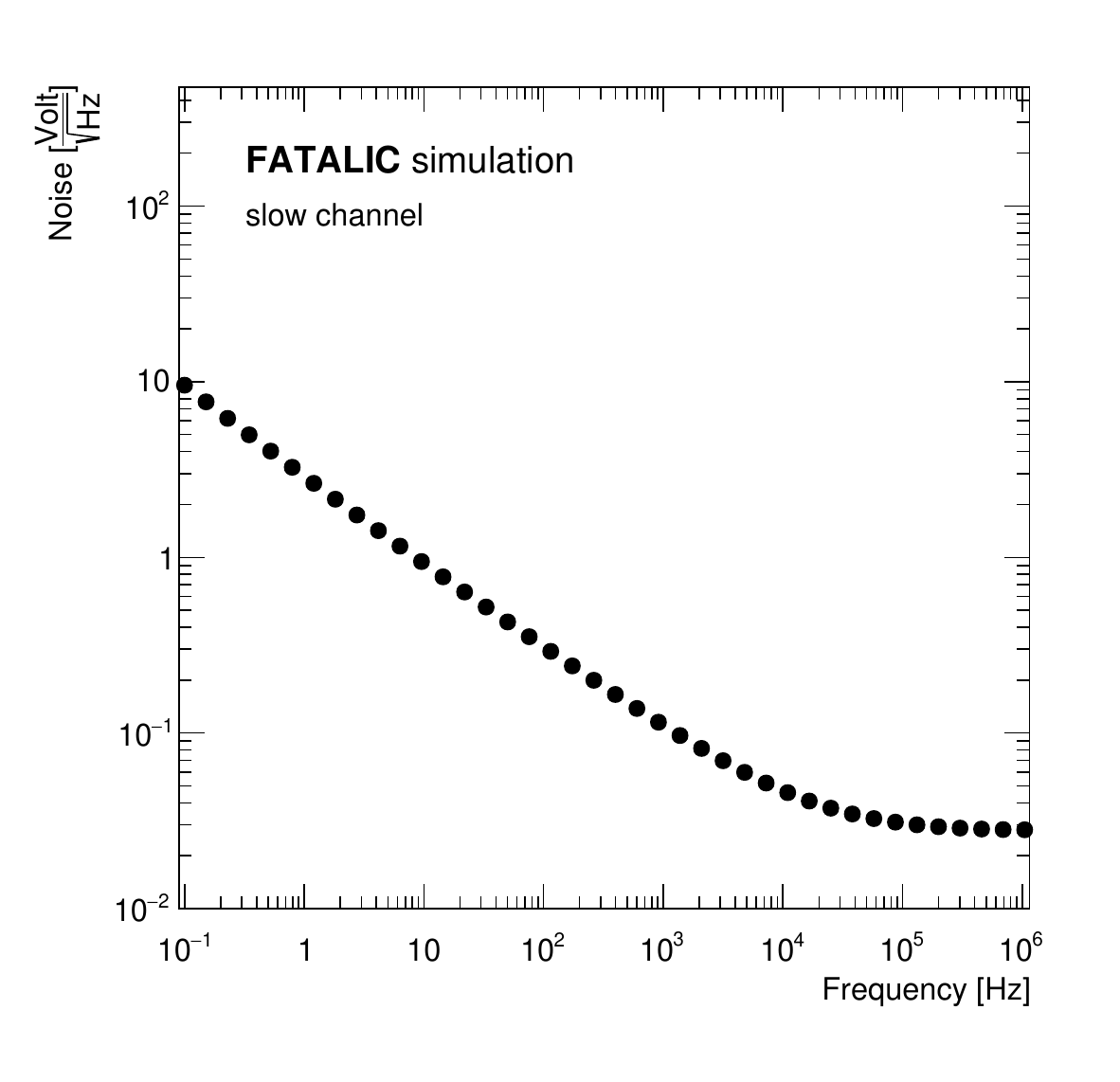}\label{fig:gnoise_b}}
\caption{Simulation of the noise spectral density in the (a) fast channels (high gain) and (b) slow channel.}
\label{fig:gnoise}
\end{figure}

The dominant noise introduced in the fast channels is white noise from the input stage, expected from simulation 
to be approximately \SI{8}{fC} in the high-gain channel. On the other hand, the dominant noise in the slow channel 
is $\sfrac{1}{f}$ noise from the input stage, estimated at the level of \SI{7}{nA}. This substantially exceeds the
specification of \SI{0.25}{nA} (see table\,\ref{tab:f5_specs}) initially defined for \gls{fatalic} because the 
lowest region of the noise frequency spectrum was not properly included in the simulations used to define the design.
%due to overlooking the lower region of the noise frequency spectrum.
The spectral density of the noise for the fast and slow channels is given in figure\,\ref{fig:gnoise}.

\begin{figure}[!t]
  \centering
  %\subfloat[][]{\includegraphics[width=0.4\textwidth]{figures/development/Pedestal_CH2_HighGain}\label{fig:ped_sum_a}}
  %\subfloat[][]{\includegraphics[width=0.4\textwidth]{figures/development/CorrMatrix}\label{fig:ped_sum_b}}\\
  \subfloat[][]{\includegraphics[width=0.48\textwidth]{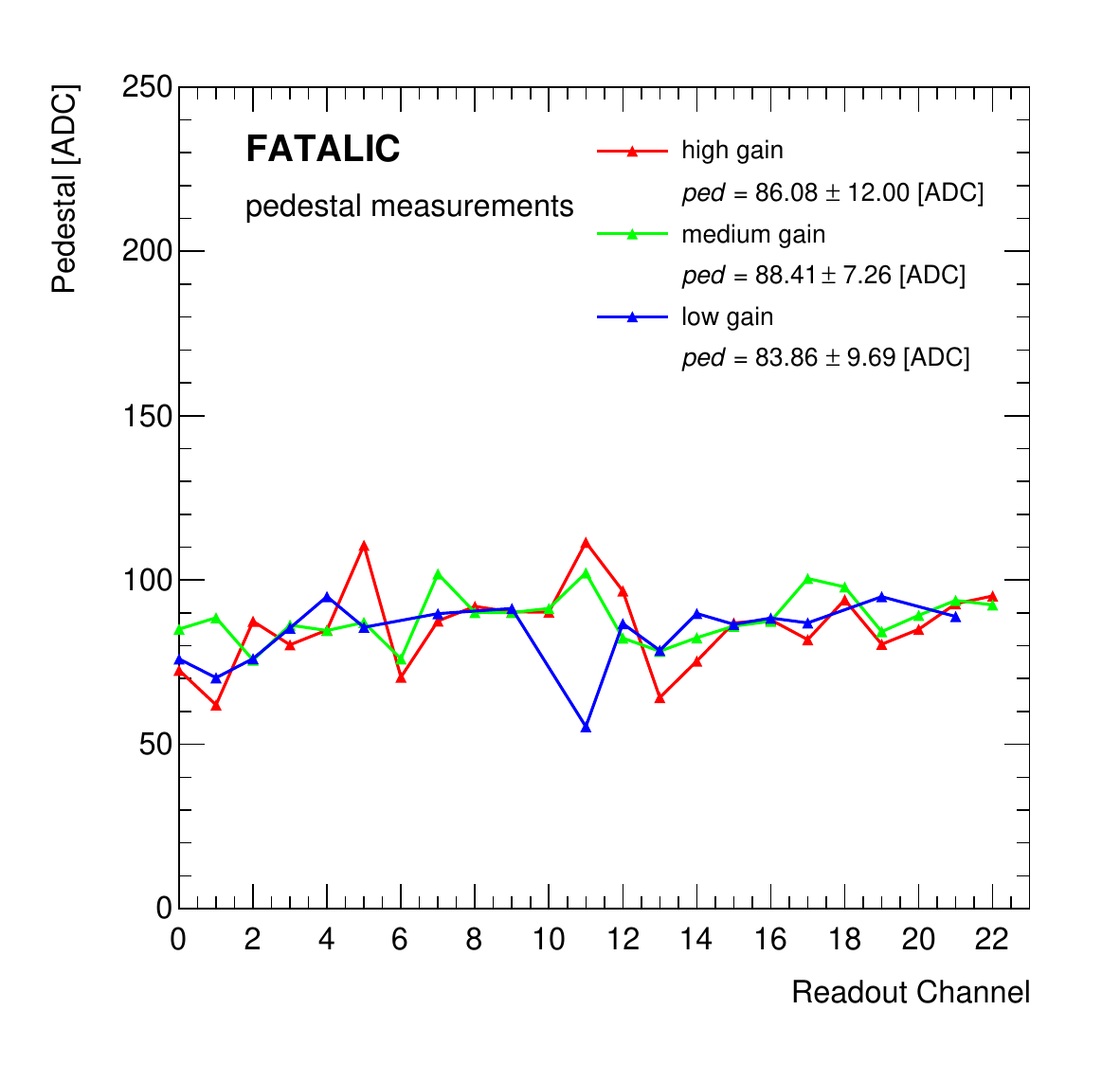}\label{fig:ped_sum_c}}
  \subfloat[][]{\includegraphics[width=0.48\textwidth]{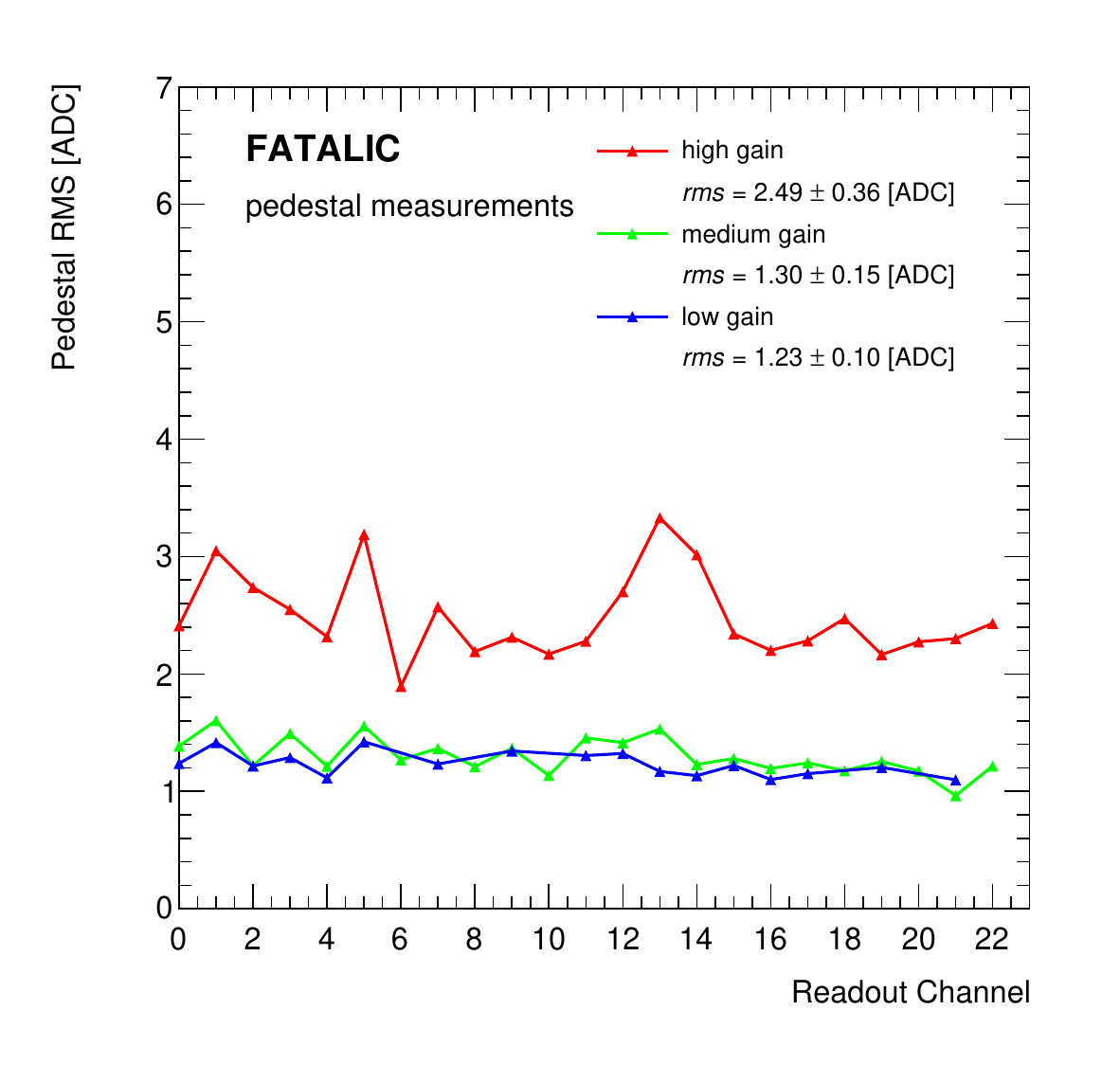}\label{fig:ped_sum_d}}\\
  \subfloat[][]{\includegraphics[width=0.48\textwidth]{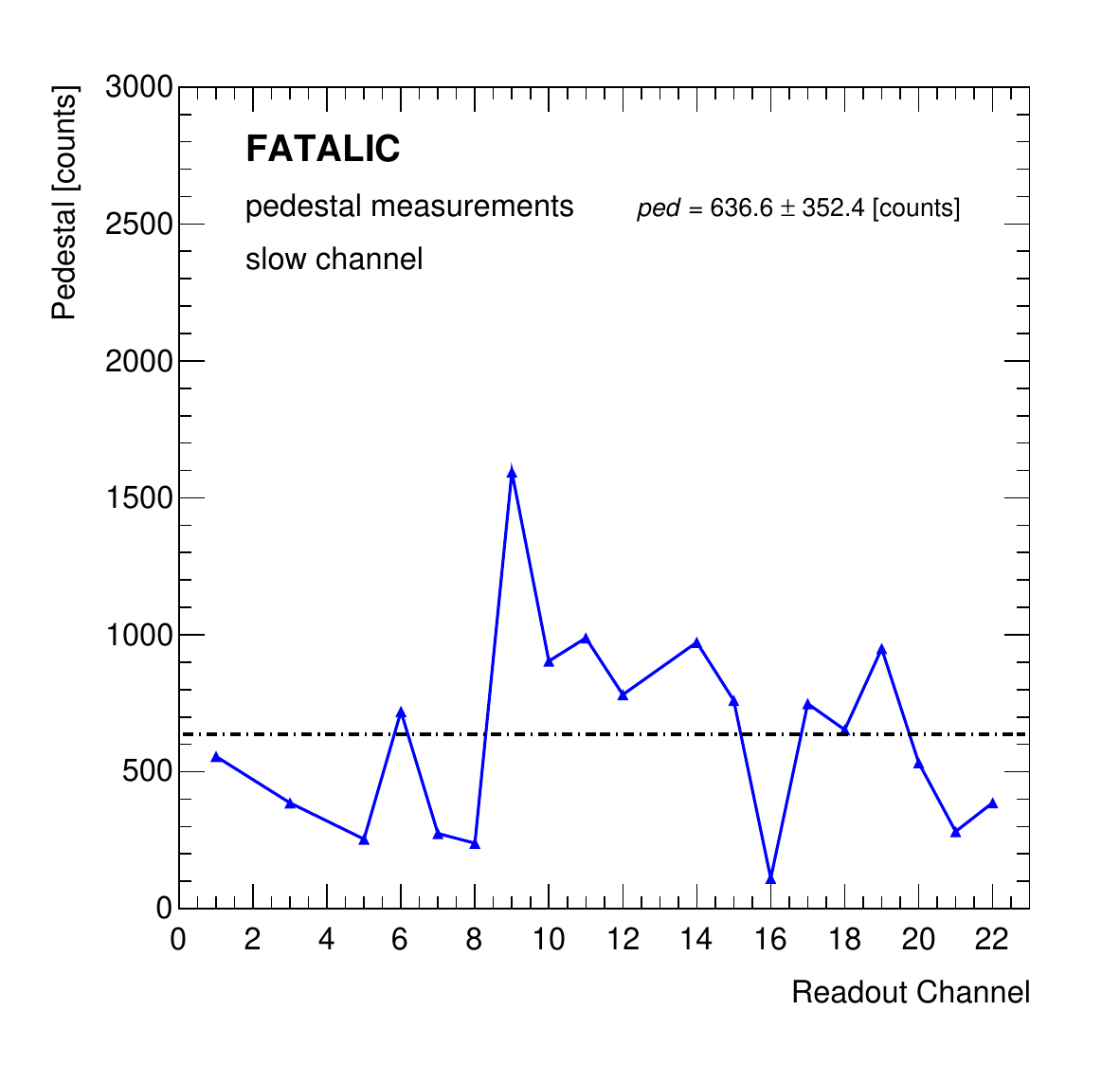}\label{fig:ped_sum_e}}
  \subfloat[][]{\includegraphics[width=0.48\textwidth]{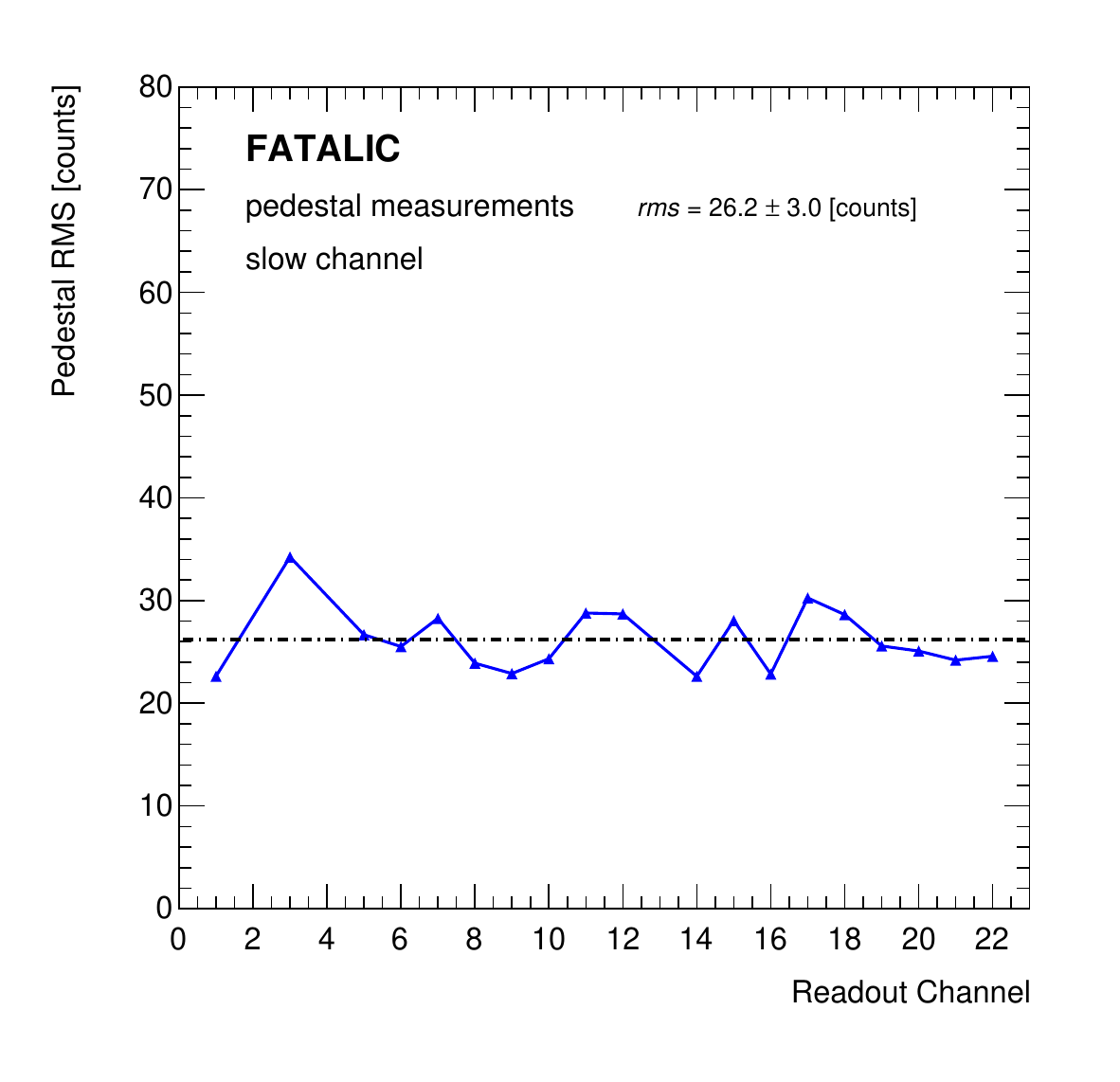}\label{fig:ped_sum_f}}
  \caption{Pedestal measurements with 24 prototype \gls{fatalic} chips. (a) Mean and (b) standard deviation 
           of the pedestal in the high-, medium- and low-gain channels. (c) Mean and (d) standard deviation 
           of the pedestal in the slow channels.\label{fig:ped_sum}}
\end{figure}

Experimental measurements of the noise were taken with the All-in-One cards connected to \glspl{pmt} under high voltage, 
without signal. Figure\,\ref{fig:ped_sum_c},\,\ref{fig:ped_sum_d} present the mean and standard deviation of the pedestal
distribution in the fast channels, obtained by gaussian fit. The noise, estimated from the standard deviation, averages 
to $(2.49\pm 0.36)$\,ADC counts in the high-gain (dominated by white noise from the input stage), $(1.30\pm 0.15)$\,ADC 
counts in the medium-gain and $(1.23\pm 0.10)$\,ADC counts in the low-gain channel (dominated by noise from the \gls{adc}). 
Using the fC/ADC conversion factors of Section~\ref{subsec:linearity}, the above numbers can also be translated into units 
of input charge, namely $(6.1\pm 0.9)$\,fC, $(26.5\pm 3.1)$\,fC and $(260.0\pm 21.1)$\,fC, respectively. The small 
difference between the measured noise in the high gain channel and the simulation could be explained by the reduction 
of the biasing current, described above. Finally, the results for the case of the slow channel are presented 
in figure\,\ref{fig:ped_sum_e},\,\ref{fig:ped_sum_f}. In this case the average noise is found $(26.2\pm 0.7)$\,counts which, 
considering the design ratio of 0.25\,nA/count, corresponds to $(6.6\pm 0.8)$\,nA, in good agreement to the simulation.

%---------------------------------
\subsection{Linearity measurements}
\label{subsec:linearity}
%---------------------------------

\begin{figure}[!tbh]
  \centering
  \subfloat[][]{\includegraphics[width=0.48\textwidth]{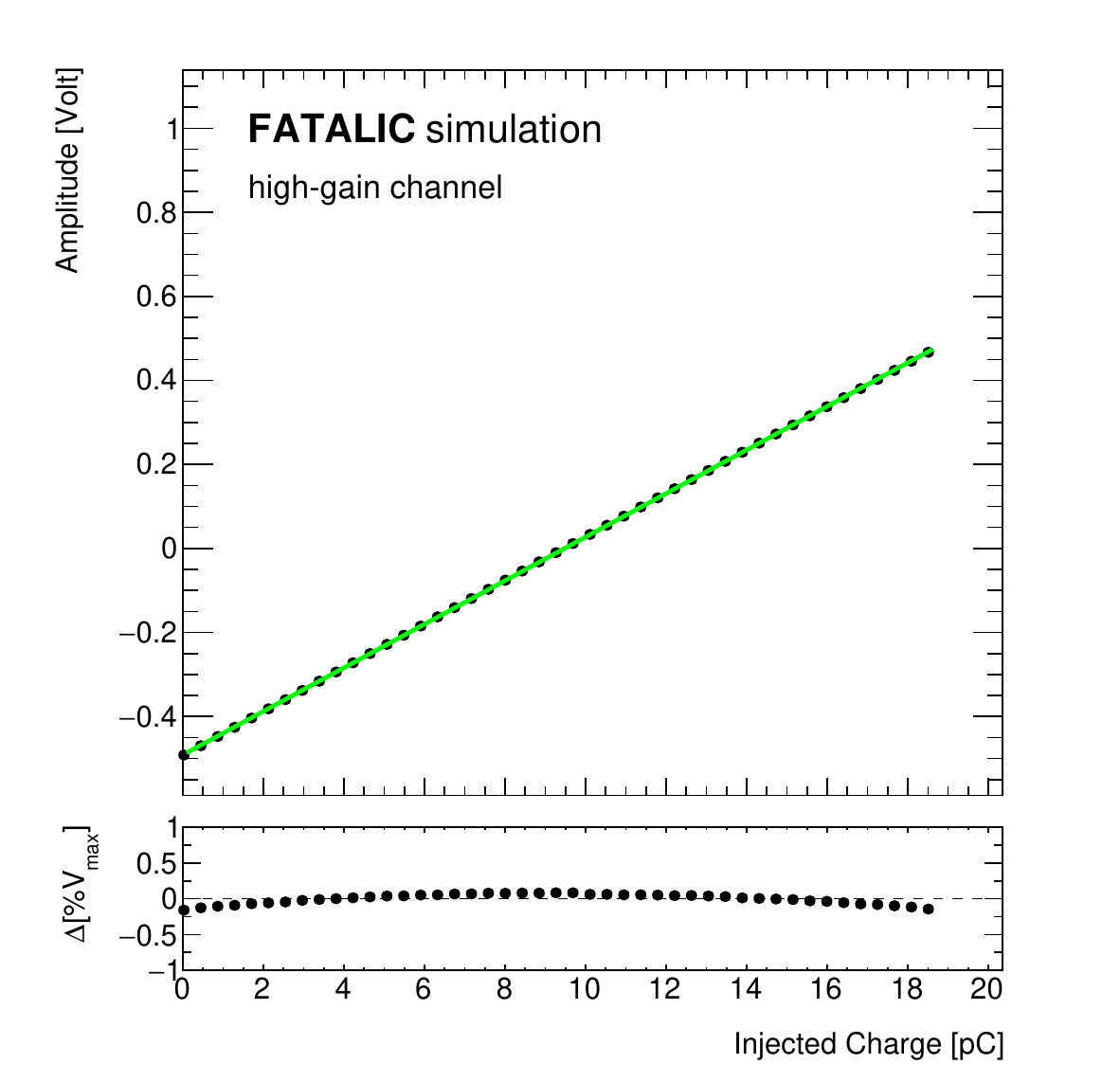}\label{fig:glin_a}}
  \subfloat[][]{\includegraphics[width=0.48\textwidth]{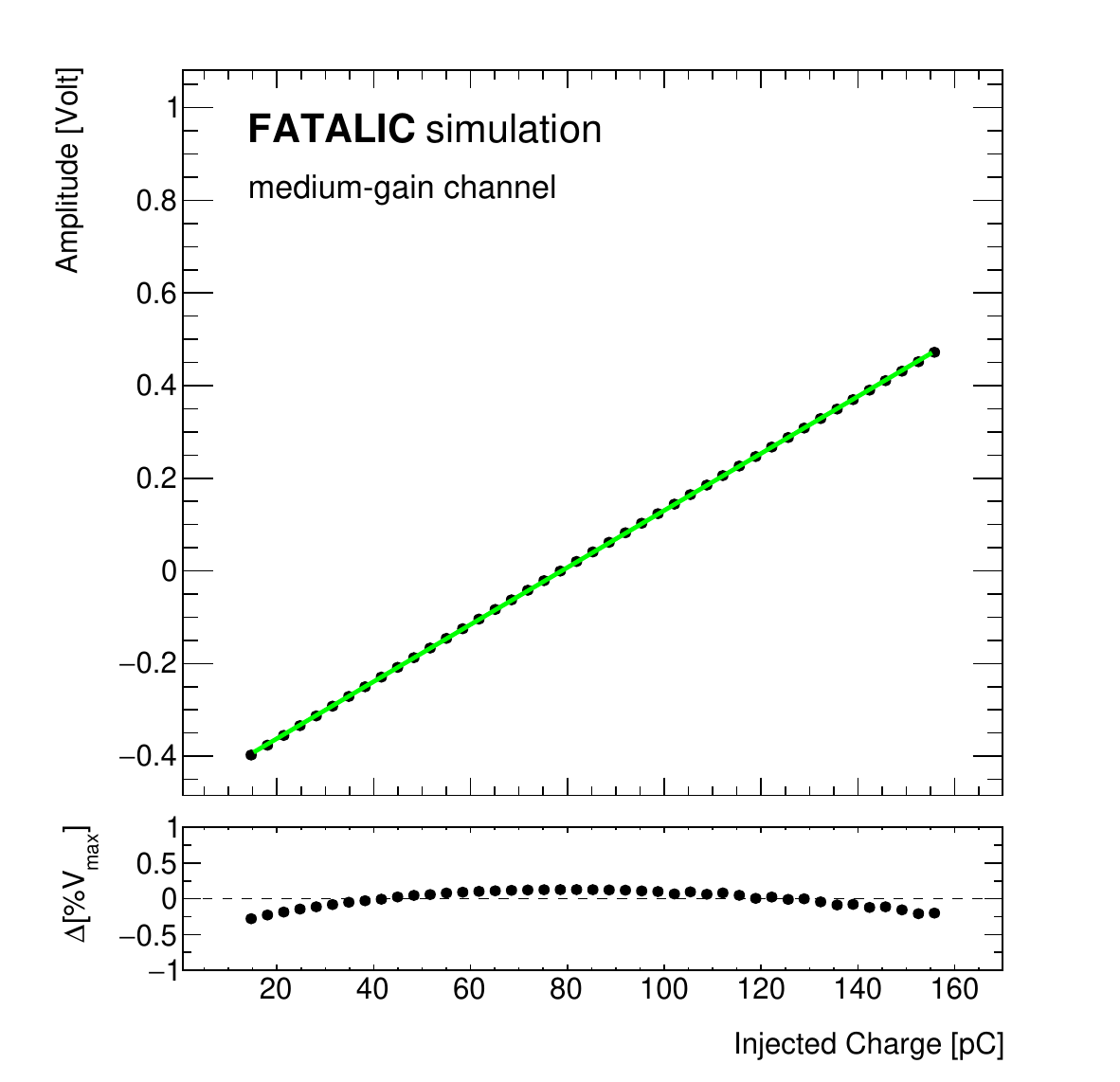}\label{fig:glin_b}}\\
  \subfloat[][]{\includegraphics[width=0.48\textwidth]{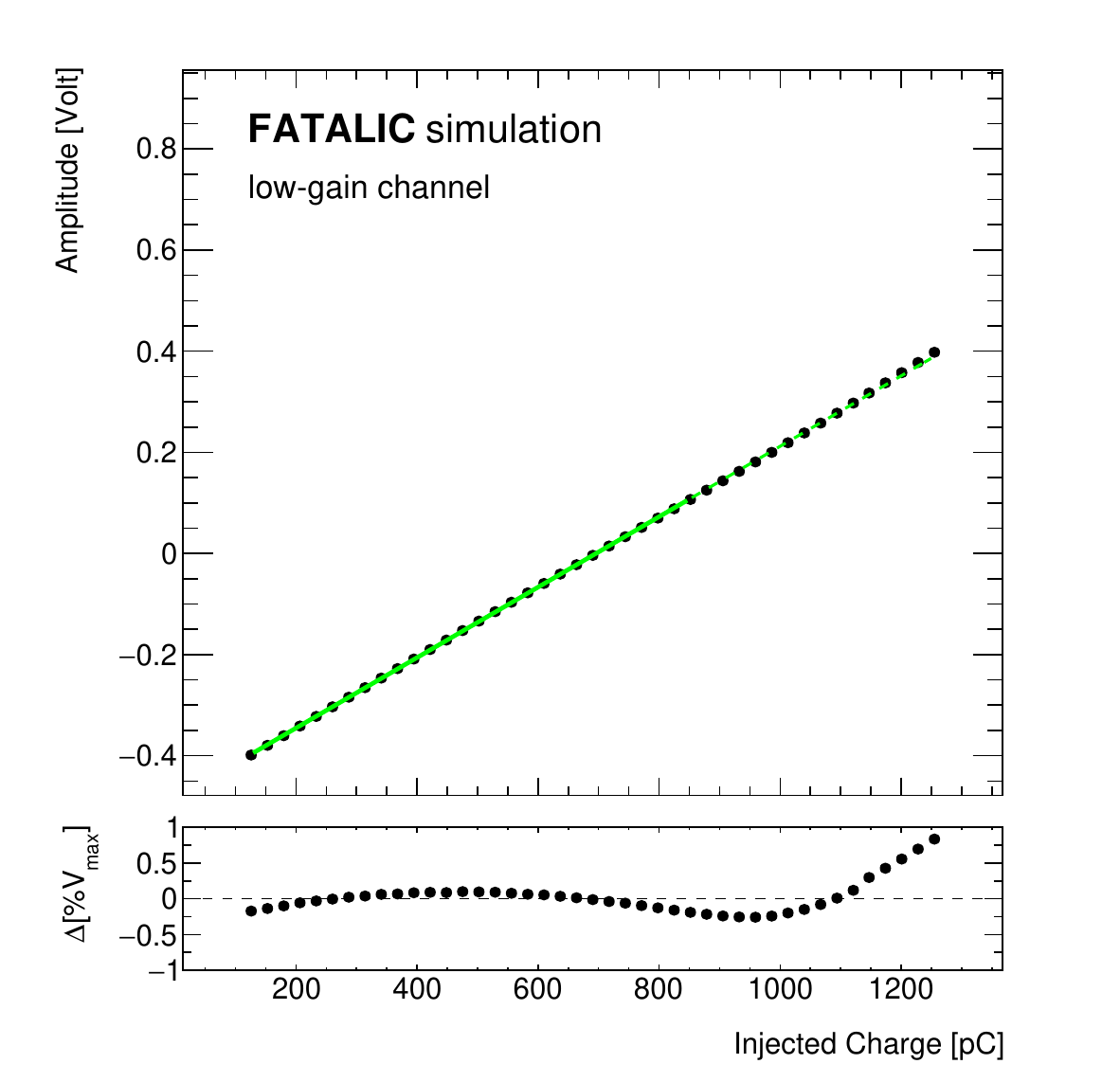}\label{fig:glin_c}}
  \subfloat[][]{\includegraphics[width=0.48\textwidth]{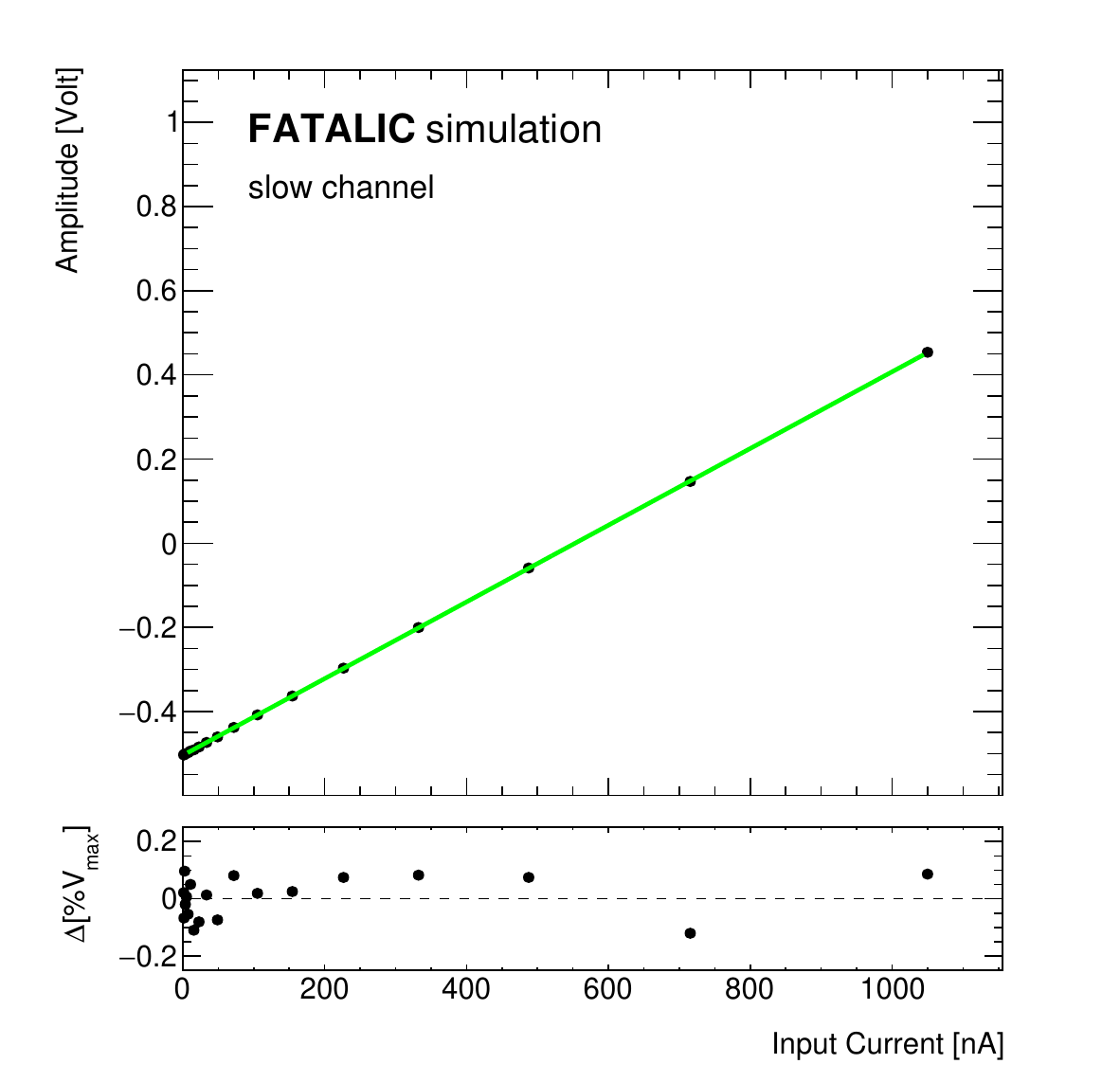}\label{fig:glin_d}}
  \caption{Simulation results, showing the deviation from linearity in the (a) high-gain, (b) medium-gain, 
           (c) low-gain and (d) slow channel.\label{fig:glin}}
\end{figure}

Figure\,\ref{fig:glin_a}, \ref{fig:glin_b} and \ref{fig:glin_c} present the linearity, obtained by simulation, of 
the analog pulse peak amplitude (in millivolt) with the input charge. The non-linearity, defined as the deviation 
from a linear fit over the maximum channel response (\SI{1}{V}), is estimated to be below 0.3\% in the input range 
up to \SI{850}{pC}, while at higher charge values it increases, reaching approximately 0.6\% at \SI{1.2}{nC}. In 
the case of the slow channel, the deviation from linearity is expected to be less than 0.2\% over the entire input 
dynamic range, as shown in figure~\ref{fig:glin_d}.

\begin{figure}[tb]
  \centering
  \subfloat[][]{\includegraphics[width=0.48\textwidth]{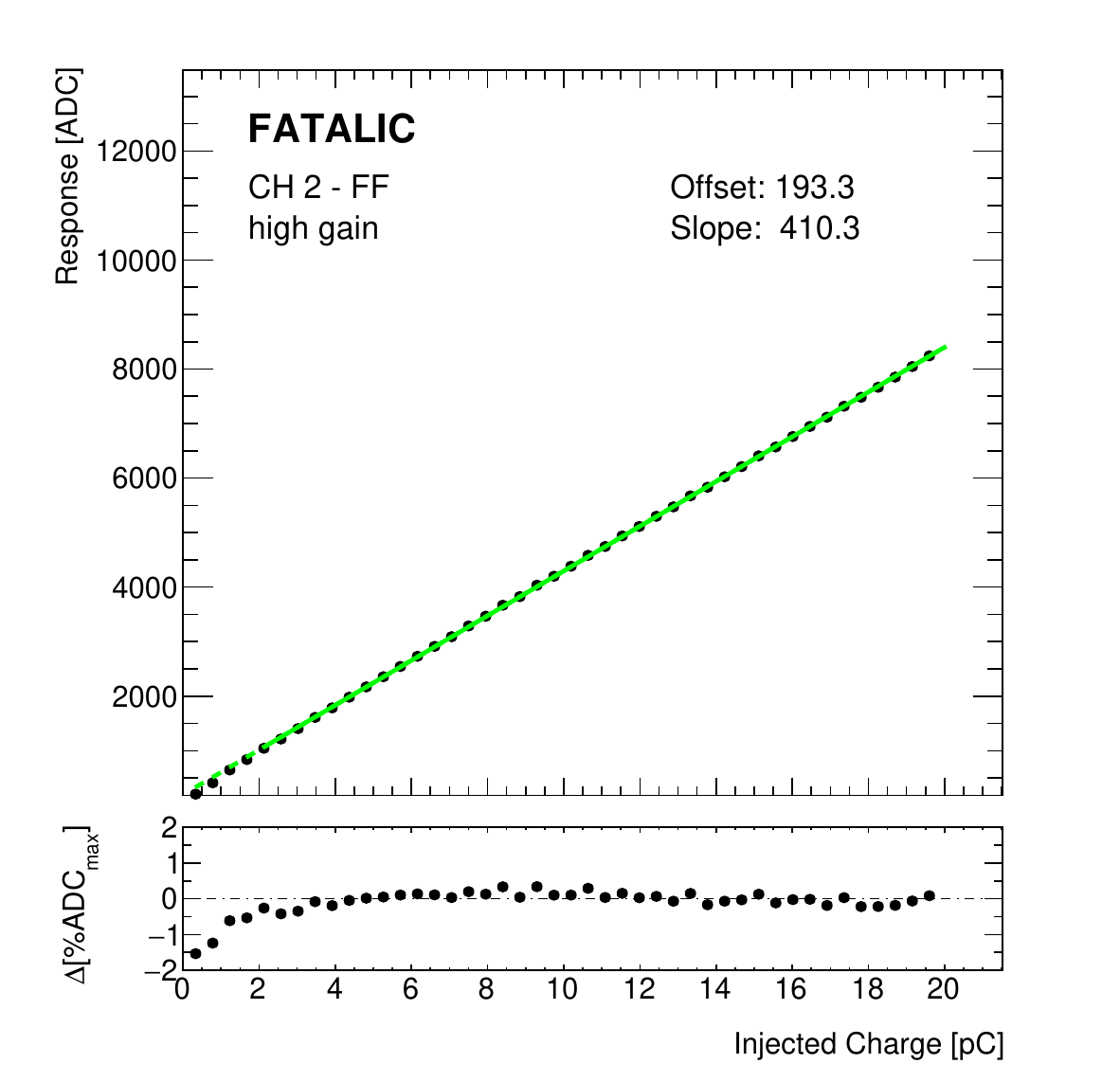}\label{fig:lin_ex_a}}
  \subfloat[][]{\includegraphics[width=0.48\textwidth]{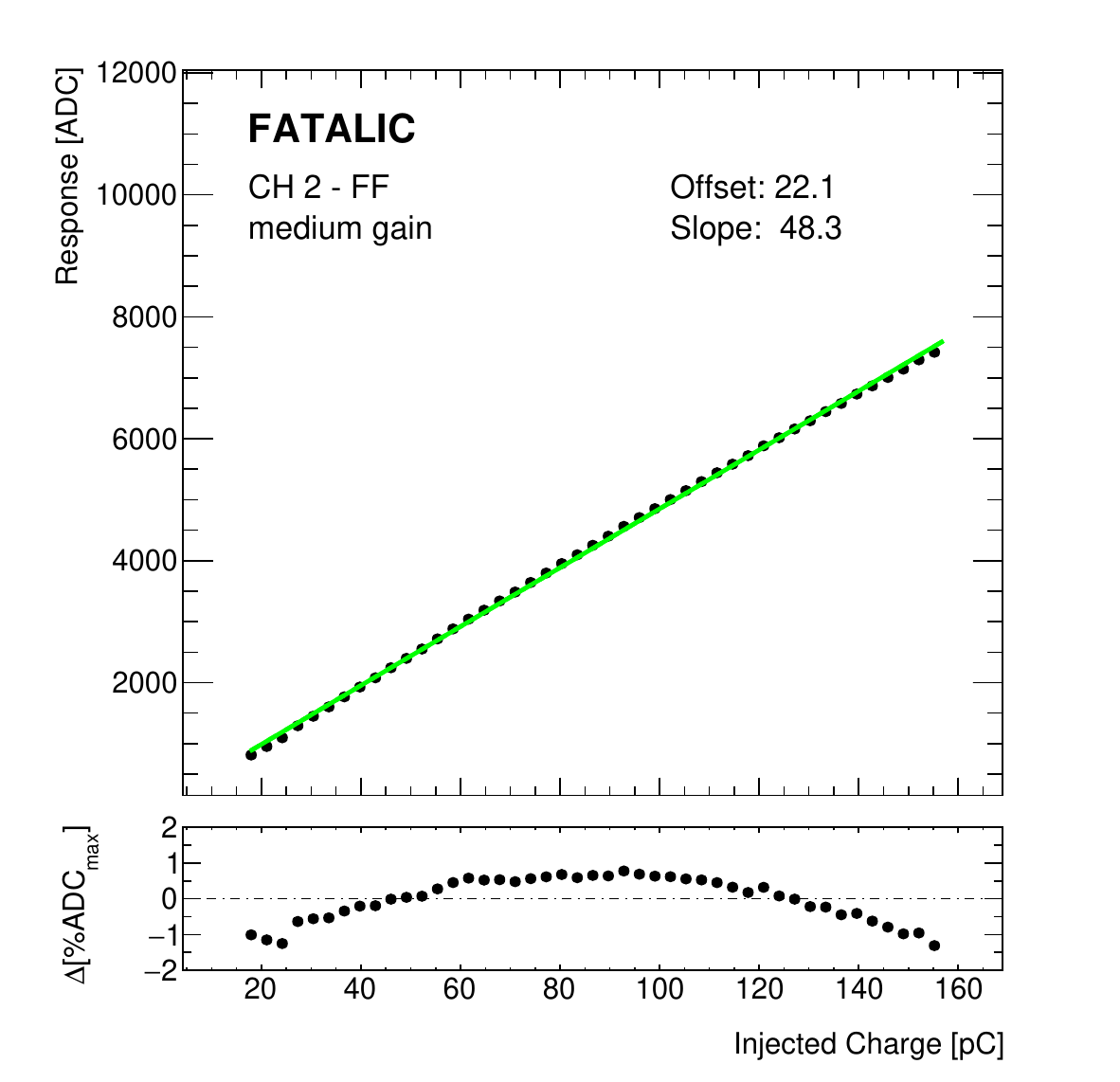}\label{fig:lin_ex_b}}\\
  \subfloat[][]{\includegraphics[width=0.48\textwidth]{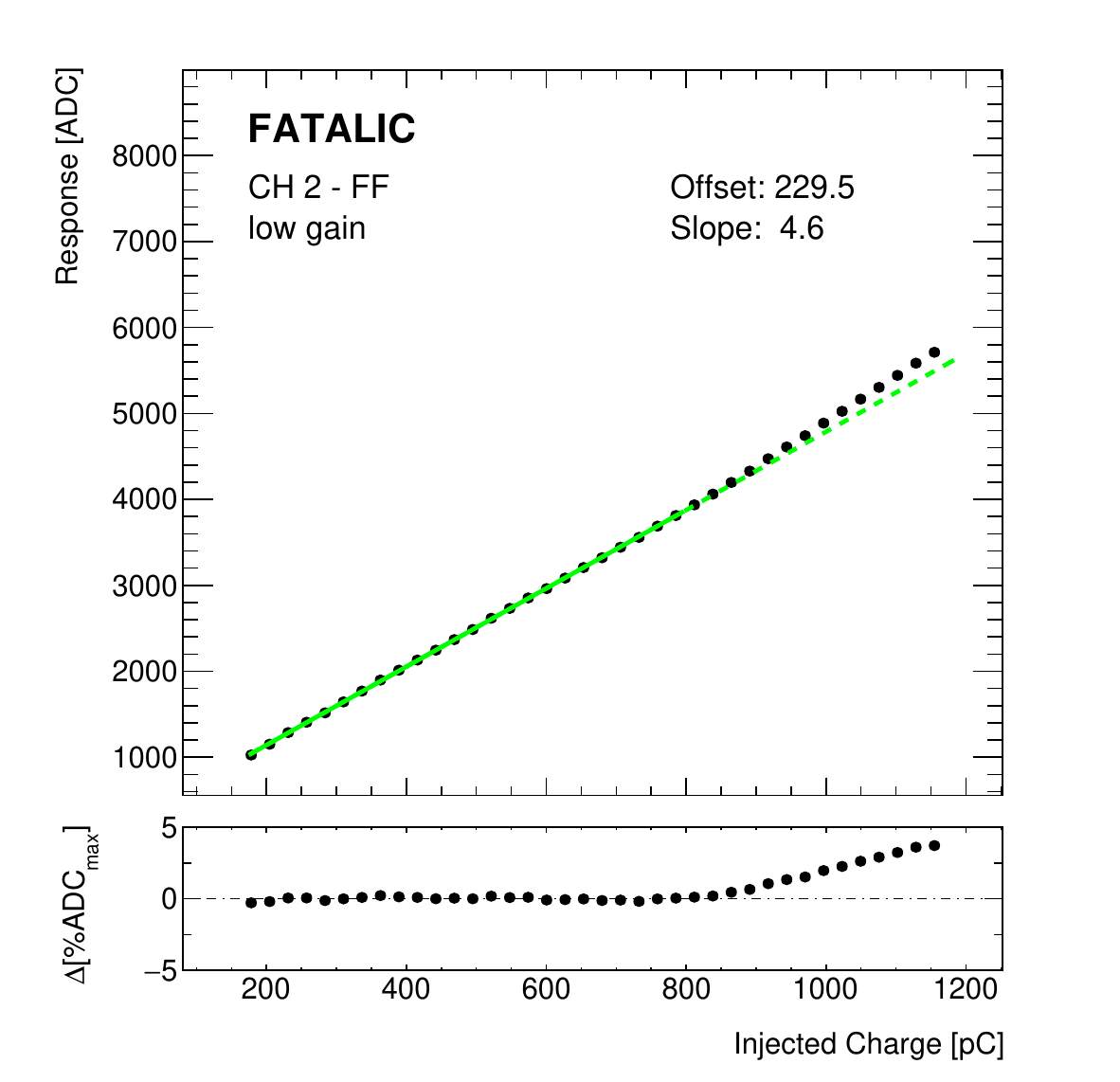}\label{fig:lin_ex_c}}
  \subfloat[][]{\includegraphics[width=0.48\textwidth]{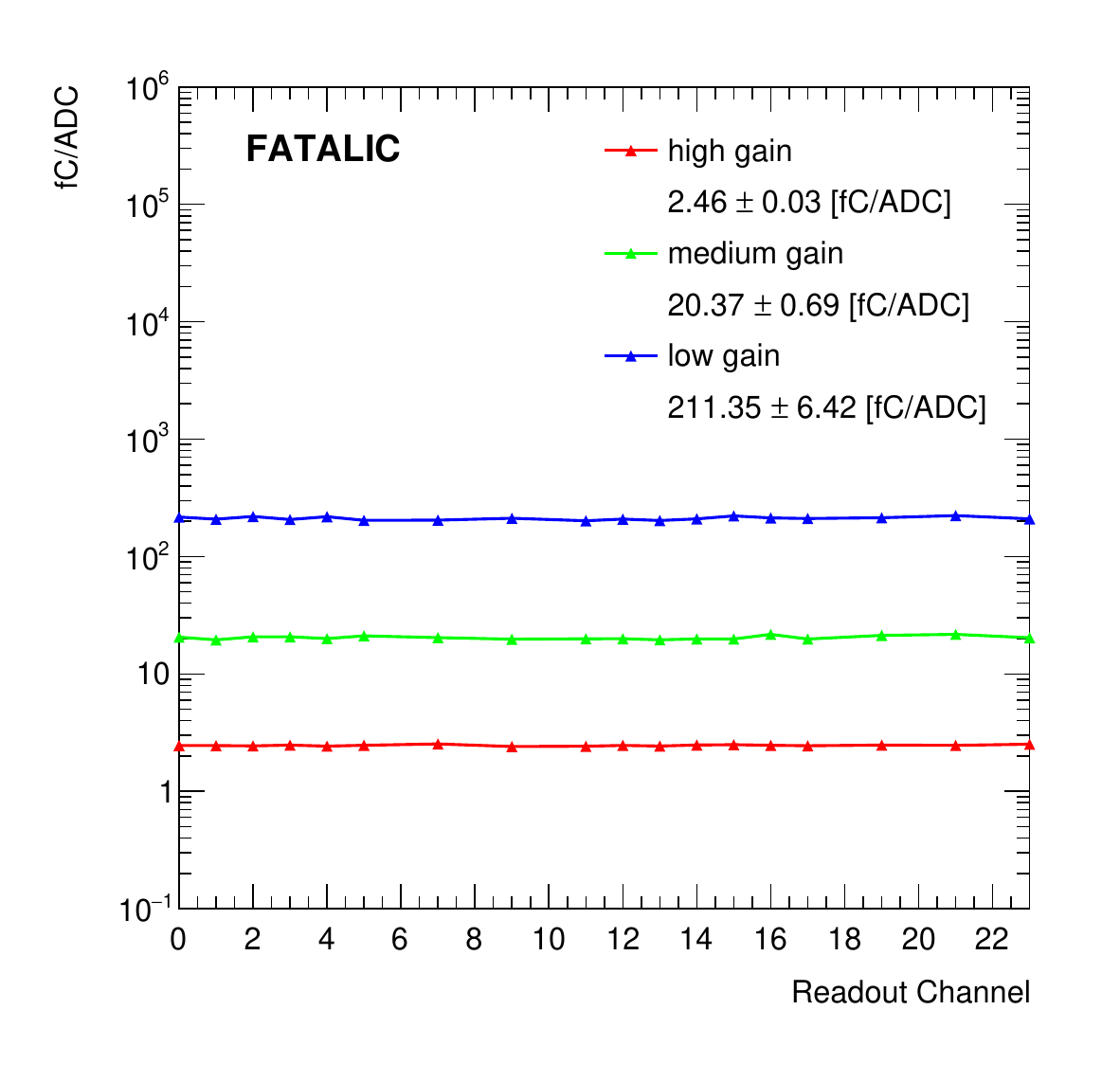}\label{fig:lin_ex_d}}
  \caption{ Linearity measurements using the \gls{cis}. Example showing the response as a function of the injected charge 
            for one \gls{fatalic} chip in the (a) high, (b) medium and (c) low gain channel. (d) Summary of the measured 
            fC/ADC conversion factors from all tested chips.
            \label{fig:lin_ex}}
\end{figure}

Experimental studies of the linearity are carried out using the on-board \gls{cis}. Since the linearity of the CIS has 
not been verified, the response in this case is obtained from the sum of the selected digitised samples, which is less 
sensitive to the shape of the injected pulse. Figure\,\ref{fig:lin_ex_a}, \ref{fig:lin_ex_b} and \ref{fig:lin_ex_c} present 
an example of the measured response as a function of the injected charge, for one chip. In the high-gain channel, the 
maximum deviation from linearity (average from all the tested prototype chips, relative to the maximum response)  is 
measured to be $(0.3\pm 0.1)\%$ above \SI{2}{pC}, increasing to $(1.3\pm 0.2)\%$ for lower charge values. In the 
medium-gain channel, the deviation is $(1.4\pm 0.6)\%$, while in the low-gain channel it is $(0.4\pm 0.1)\%$ below 
\SI{800}{pC}, reaching $(3.5\pm 0.7)\%$ at \SI{1.2}{nC}. Finally, the fC/ADC conversion factors, summarised in 
figure\,\ref{fig:lin_ex_d}, are extracted from the slope of the linear interpolation and average to $(2.46 \pm 0.03)$\,fC/ADC 
for the high-gain, $(20.4 \pm 0.7)$\,fC/ADC for the medium-gain and $(211.3 \pm 6.4)$\,fC/ADC for the low-gain channel.

Two effects have been confirmed to contribute significantly to the observed difference between the measured linearity
and the expectation, both regarding the present All-in-One card. The first one is the workaround, described above, of 
reducing the biasing current in the input stage from its nominal value in order to enable the operation of the slow 
channel. The second contribution comes from imperfections in the \gls{cis}, which have significant impact on the 
injected pulse shape (while in simulation \gls{fatalic} is always injected with the same \gls{pmt} pulse shape). Both 
of these imperfections would be corrected in a subsequent version of the All-in-One card.

%---------------------------------
\subsection{Performance of the analog and digital block}
\label{subsec:blockperf}
%---------------------------------

\begin{figure}[h]
\centering
  \subfloat[][]{\includegraphics[width=0.48\textwidth]{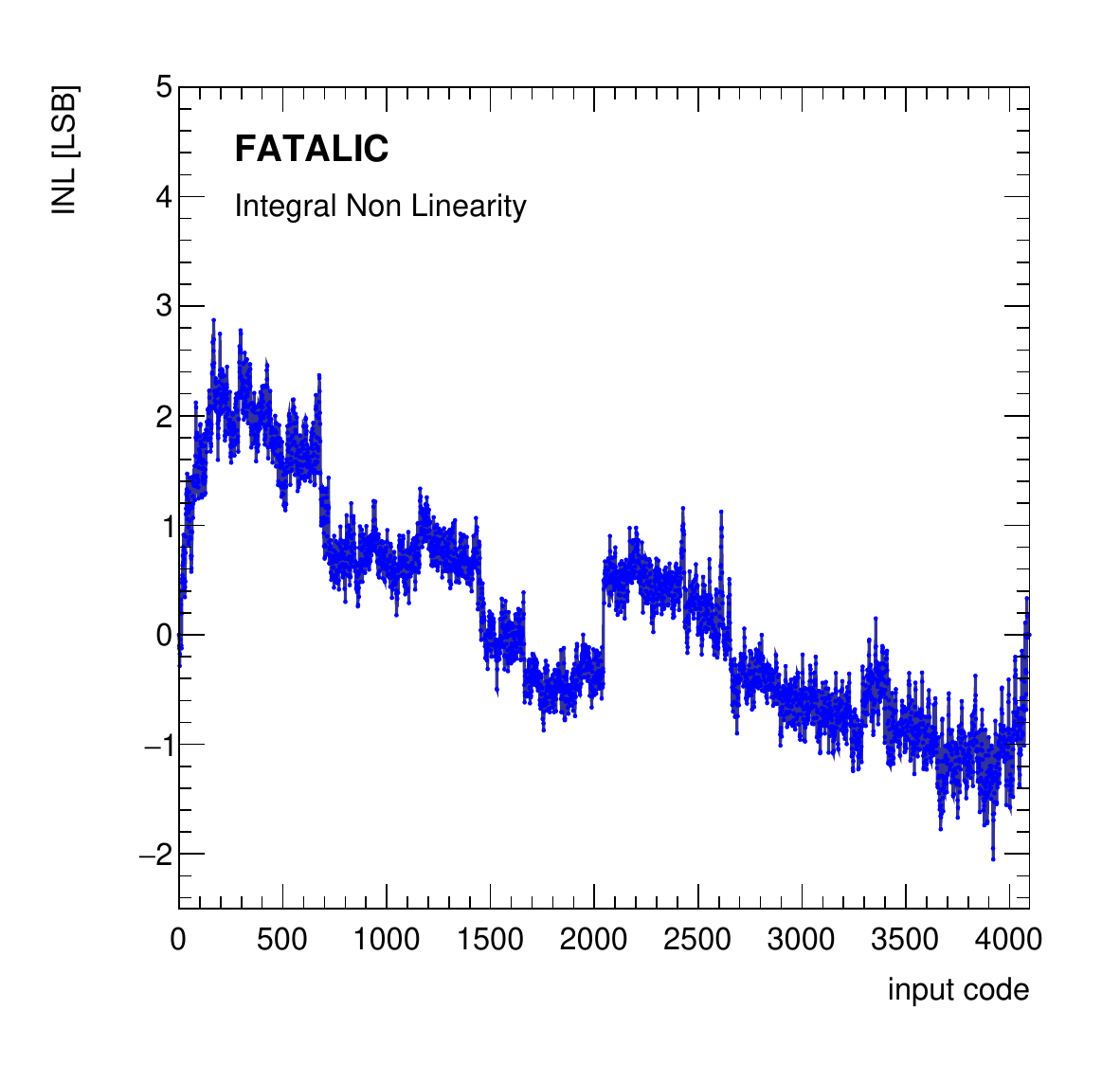}}
  \subfloat[][]{\includegraphics[width=0.48\textwidth]{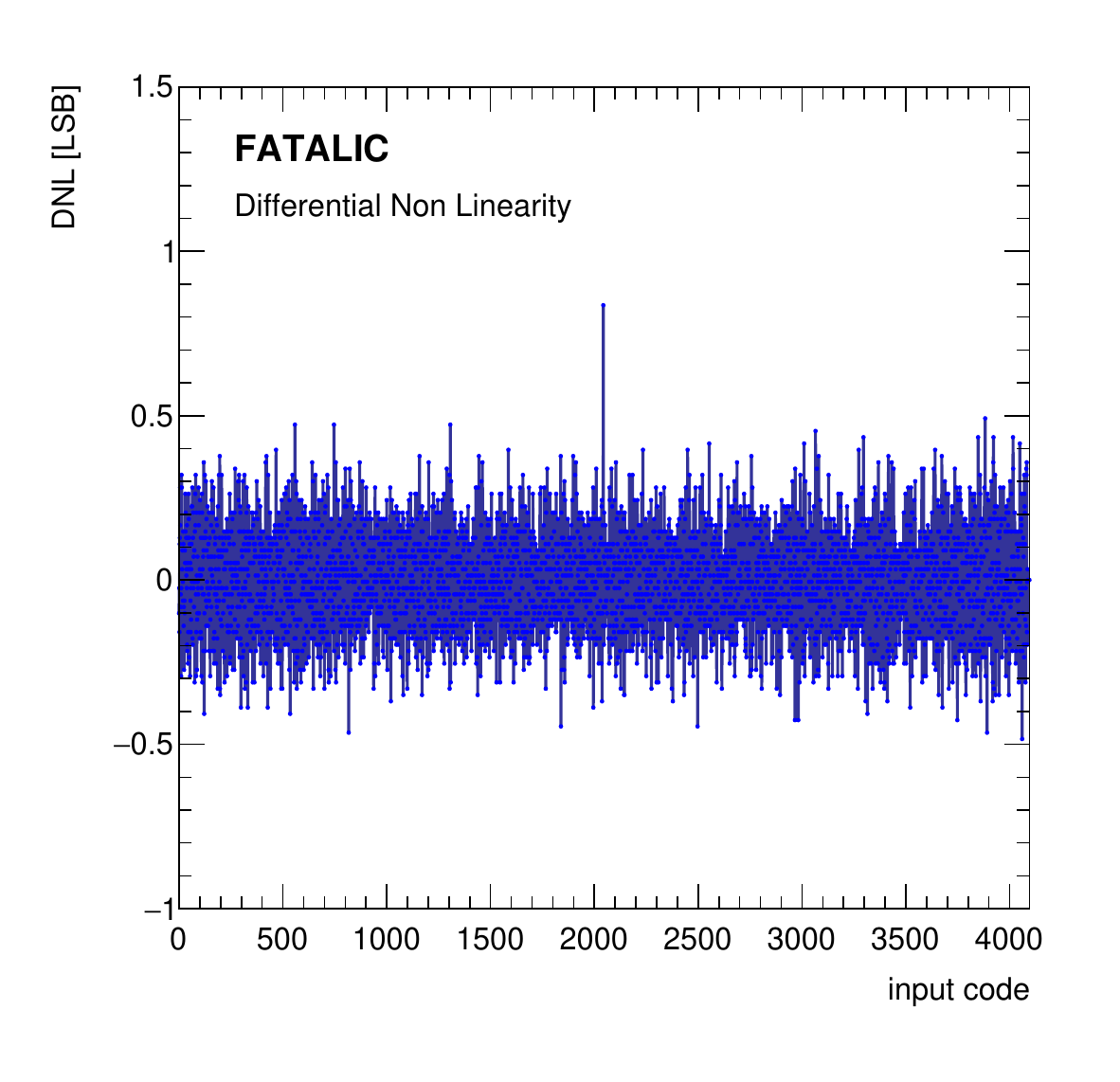}}
  \caption{(a) INL and (b) DNL error of the pipelined \glspl{adc}.\label{fig:adc_inldnl}}
\end{figure}

Dedicated studies have been carried out in order to assess the performance of the analog and digital block of 
\gls{fatalic} separately. For these studies, one chip of a previous version of \gls{fatalic} was used, which 
provided a dedicated output of the analog block and also contained an additional \gls{adc}, independent of the 
analog block. In order to measure the INL error, a differential slow ramping signal is delivered at the input 
of the \gls{adc}, which is clocked at the nominal frequency of \SI{40}{MHz}, and the observed output is compared 
to the ideal transfer function across the entire dynamic range. Similarly, the Differential Non-Linearity (DNL) 
error is estimated from the difference between the actual step width and the ideal step of 1\,LSB. The results 
are shown in figure\,\ref{fig:adc_inldnl}. The INL error is found to be within $\pm$3\,LSB, while the DNL error 
is less than $\pm$1\,LSB.

Furthermore, the intrinsic noise of the \gls{adc} was obtained from the standard deviation of the output, for an input 
signal corresponding to the middle of the dynamic range, i.e. 2047\,ADC counts, and was found to be 0.85\,LSB. This 
measurement can also be used to estimate the noise of the analog block, by quadratically subtracting it from the 
noise level of the full ASIC, realistically assuming that they are not correlated. Using the noise measurements,
derived with the 24 prototype chips (see Section\,\ref{subsec:ped}), the estimated noise of the analog block is 
2.34\,LSB (significantly dominant) for the high-gain, 1.23\,LSB for the medium-gain and 0.89\,LSB channel for the low-gain 
channel.

\begin{figure}[t]
\centering
  \subfloat[][]{\includegraphics[width=0.48\textwidth]{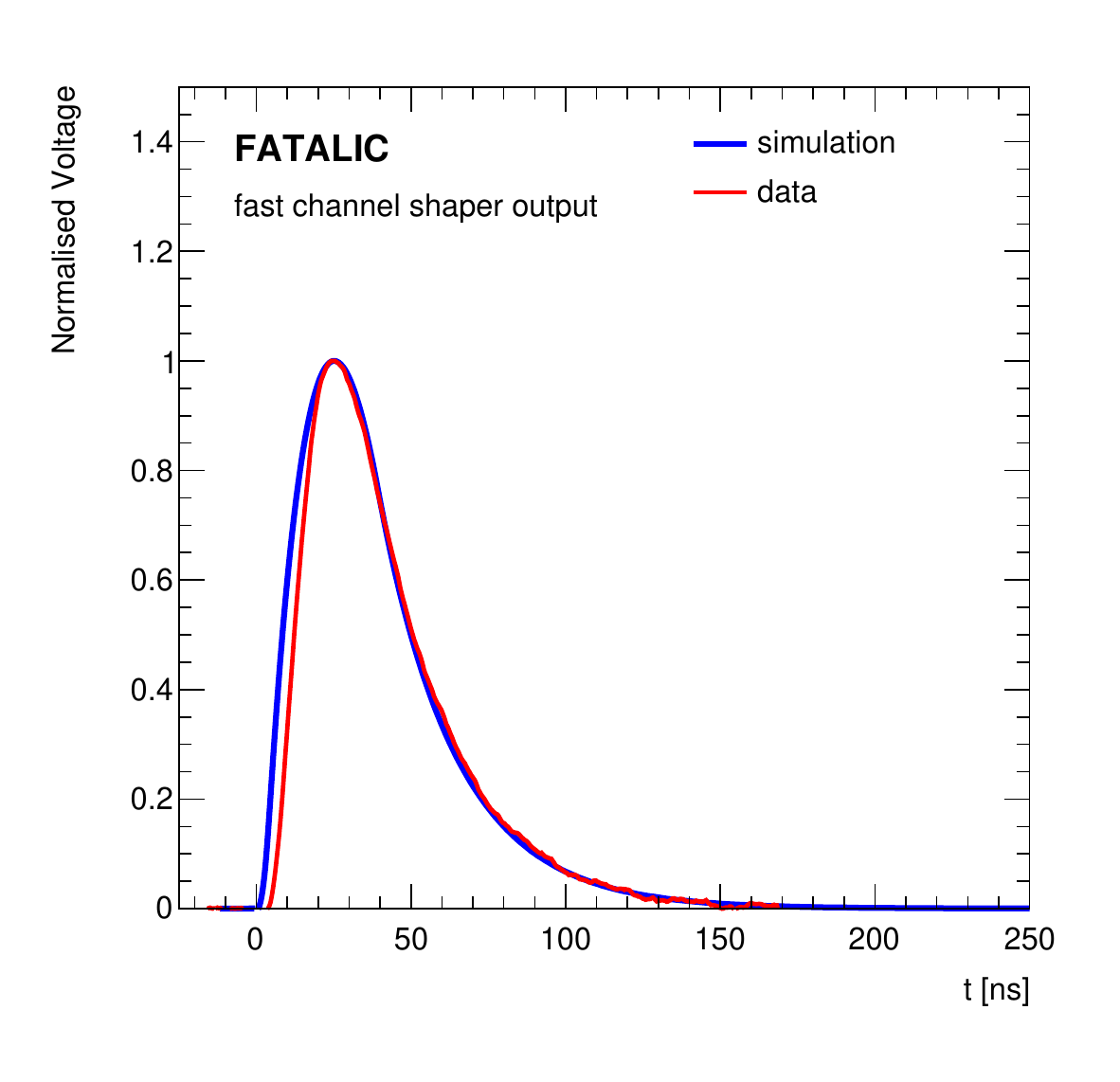}}
  \subfloat[][]{\includegraphics[width=0.48\textwidth]{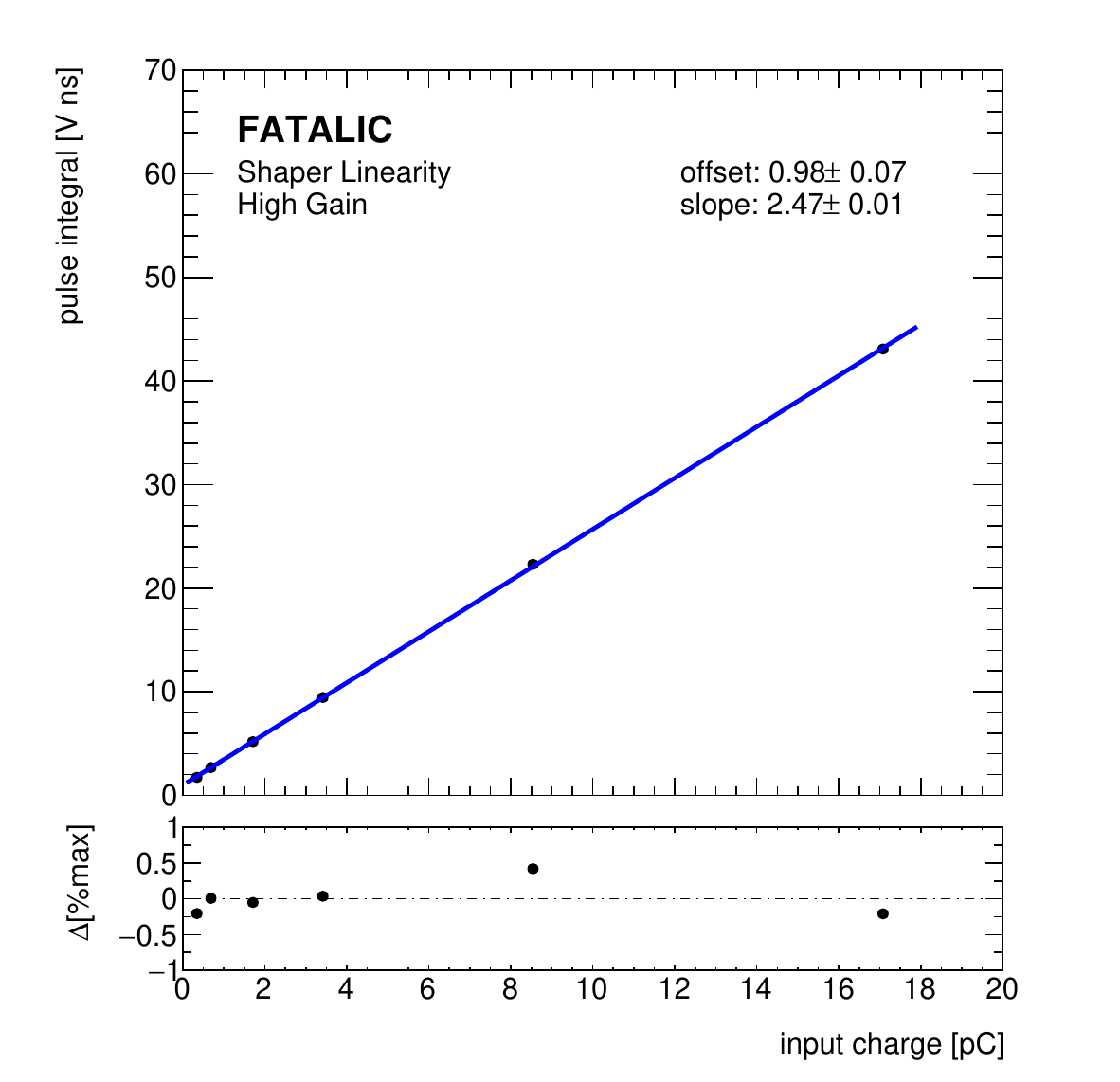}}\\
  \subfloat[][]{\includegraphics[width=0.48\textwidth]{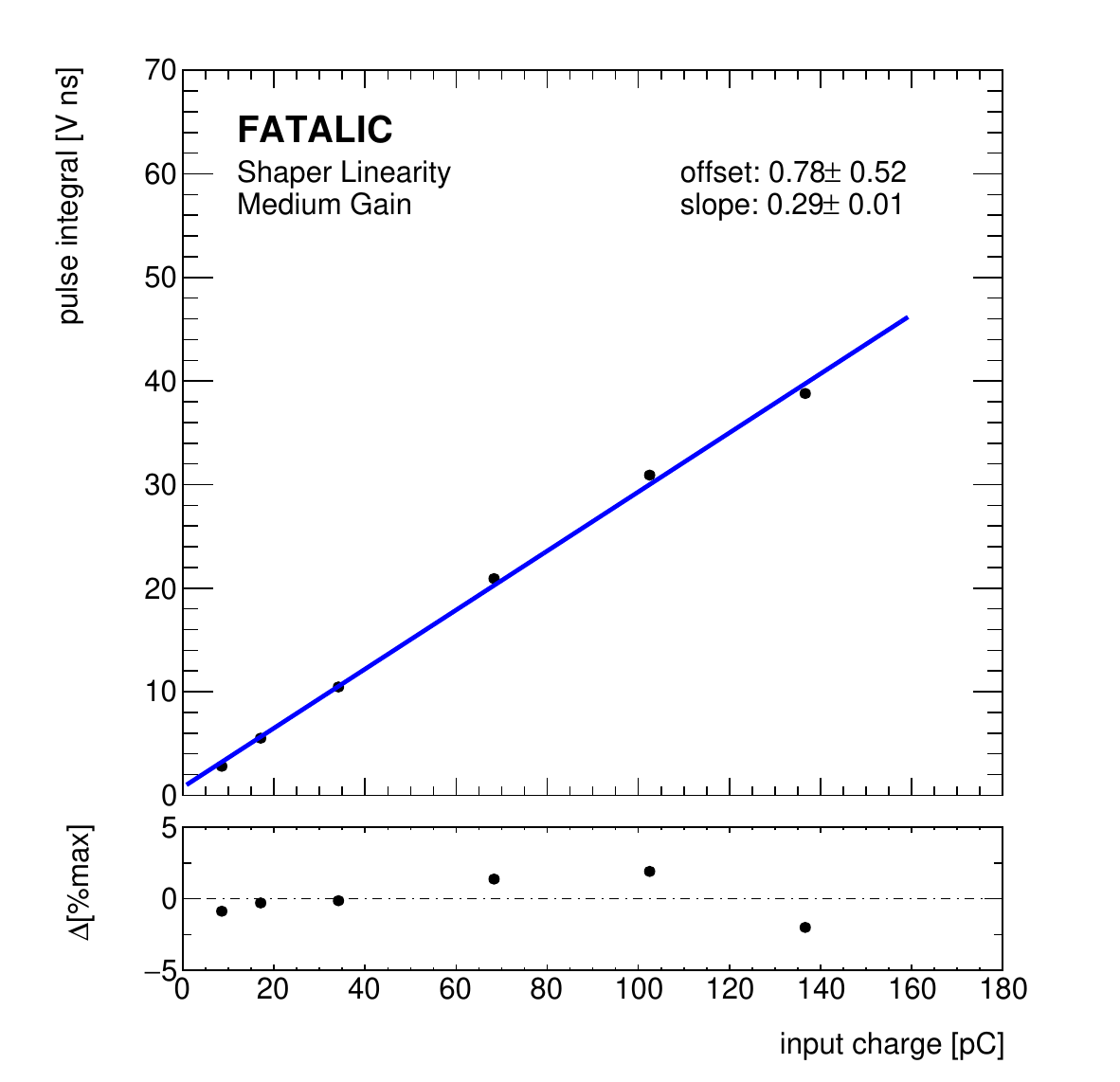}}
  \subfloat[][]{\includegraphics[width=0.48\textwidth]{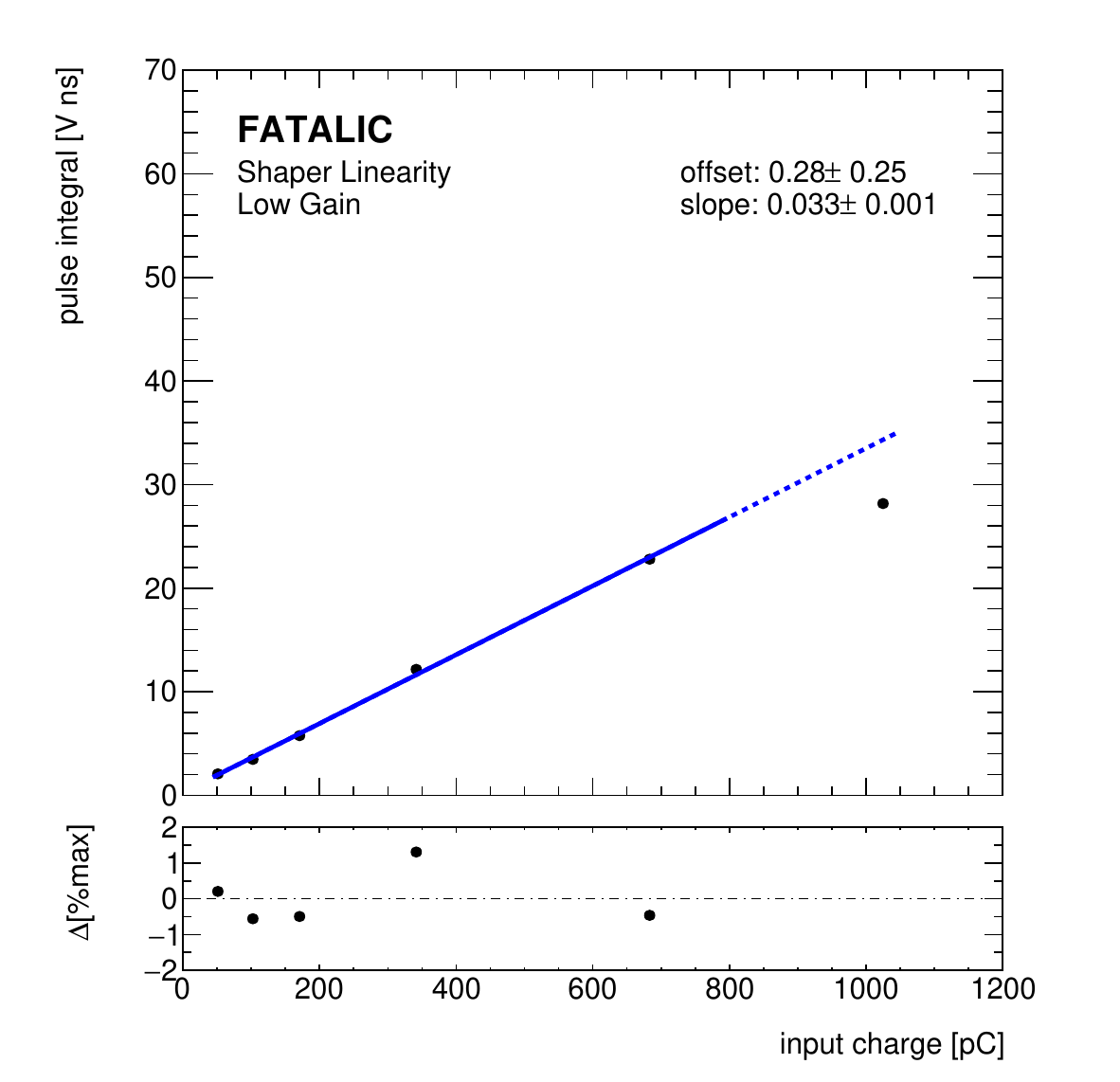}}
  \caption{Comparison of the observed analog shaper pulse to simulation (a) and linearity of the analog block in the
  high-gain (a), medium-gain (b) and low-gain (c) channels.\label{fig:shaper_perf}}
\end{figure}

Figure\,\ref{fig:shaper_perf} compares the observed analog pulse shape, obtained by illumination of the connected \gls{pmt}
with a Light-Emitting Diode (LED), to the simulated pulse with which it shows good agreement. The integral of the analog 
pulse as a function of the injected charge (using the \gls{cis}) is also presented in figure\,\ref{fig:shaper_perf} for
the three gain channels. The maximum deviation from linearity is found to be 0.4\% in the high-gain, 1.9\% in the medium-gain 
and 1.3\% in the high-gain channel, up to approximately \SI{700}{pC}. In this case as well, the observed linearity is 
significantly affected by imperfections of the \gls{cis}. The linearity for higher input charges is not comparable to that of 
the present version of \gls{fatalic} due to intermediate revisions in the input stage. Finally, although the dedicated output 
could be used to directly measure the noise of the analog block, this measurement would be significantly corellated to the 
noise of the scope. Therefore, the indirect estimation presented above is preferred.

%---------------------------------
\subsection{Radiation tolerance}
\label{subsec:rad}
%---------------------------------

The \SI{130}{nm} GF CMOS technology is recommended by ATLAS due to its high radiation tolerance in terms 
of Total Ionising Dose (TID). It is suitable for the development of radiation-hard chips up to at least \SI{100}{Mrad} with 
a peak of leakage current at $\sim$\SI{1}{Mrad}. Therefore, since the expected radiation level at the HL-LHC, based on both 
Monte Carlo and in-situ measurements, is \SI{50}{krad} (including conservative safety factors), well below the \SI{1}{Mrad} 
peak, no further assessment for TID or Non Ionising Energy Loss (NIEL) is deemed necessary for \gls{fatalic}. Single Event 
Effects (SEE) would need to be tested though for hadron fluxes up to $\SI{8.06e11}{particles/cm^2}$ (typically with \SI{200}{MeV} 
protons), including safety factors. Finally, the associated All-in-One card and Mainboard contain Commercial Off-The-Shelf (COTS) 
components, which have been tested with a different \gls{fe} electronic option to be tolerant for the radiation level expected at 
the HL-LHC. Since \gls{fatalic} was not the selected option for the \gls{TileCal} upgrade, no further tests are currently planned. 
Table\,\ref{tab:rad} lists the anticipated radiation levels for the ASIC and COTS.

\begin{table}[!h]
  \centering
  \medskip
  {\footnotesize
  \begin{tabular}{l c c c c c}
    \toprule
    \multirow{2}{*}{Radiation} & \multirow{2}{*}{\makecell[c]{Simulation\\\&\,Experiment}} & \multicolumn{3}{c}{Safety Factors} & \multirow{2}{*}{\makecell{Total\\ASIC (COTS)}} \\
    \cline{3-5}
                               &  & {\scriptsize Simulation} & \makecell{\scriptsize Batch variation\\\scriptsize ASIC (COTS)} & \makecell{\scriptsize Low Dose \\\scriptsize Rate Effect} & \\
    \midrule
    \makecell[l]{NIEL\\{\scriptsize {[}1\,MeV eq. neutron/cm$^2${]}}} & $2.69\times 10^{12}$ & 2 & 1(4) & 1 & $5.38\times 10^{12}$ ($2.15\times 10^{13}$) \\
    \midrule[\cmidrulewidth]
    \makecell[l]{TID\\ {\scriptsize {[}Gray{]}}} & 67.3 & 1.5 & 1(4) & 5 & 505 (2019) \\
    \midrule[\cmidrulewidth]
    \makecell[l]{SEE\\ {\scriptsize {[}>20\,MeV hadron/cm$^2${]}}}    & $4.03\times 10^{11}$ & 2 & 1(4) & 1 & $8.06\times 10^{11}$ ($3.22\times 10^{12}$) \\
  \bottomrule
  \end{tabular}}
  \caption{\label{tab:rad}Radiation levels in the worst location of the \gls{fe} electronics for an integrated luminosity of 4000 $\si{fb^{-1}}$.}
\end{table}

\clearpage

%%%%%%%%%%%%%%%%%%%%%%%%%%%%%%%%%%
\section{Energy reconstruction}
\label{sec:EnergyReco}
%%%%%%%%%%%%%%%%%%%%%%%%%%%%%%%%%%

As described in Section\,\ref{sec:strategy}, the fast channel shapers deliver an asymmetric pulse, the amplitude $A$ 
of which has to be reconstructed from the set of digitised samples (typically using seven samples), collected after
trigger decision. At the \SI{40}{MHz} sampling rate, the second sample is expected to coincide with the pulse peak. 
Small time-shifts $\tau$, of the order of a few nanoseconds, are however possible. Both $A$ and $\tau$ are reconstructed 
by Optimal Filtering~\cite{Fullana:2005dwa}, a weighted sum of the digitised samples with minimum sensitivity to both
correlated (e.g. pile-up) and uncorrelated noise.

\begin{figure}[!tb]
  \centering
  \subfloat[][]{\includegraphics[width=0.48\textwidth]{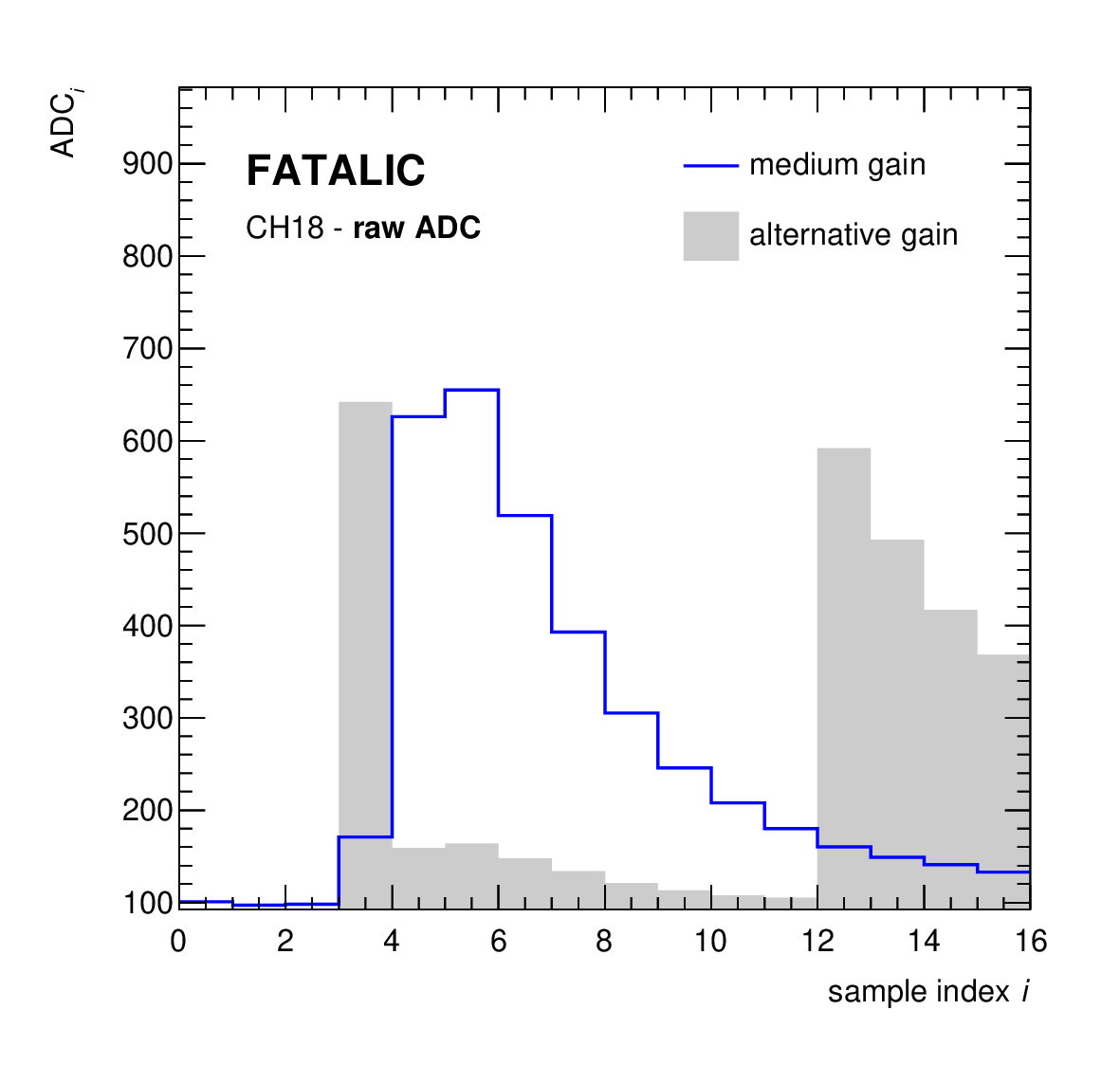}\label{fig:pulses_b}}
  \subfloat[][]{\includegraphics[width=0.48\textwidth]{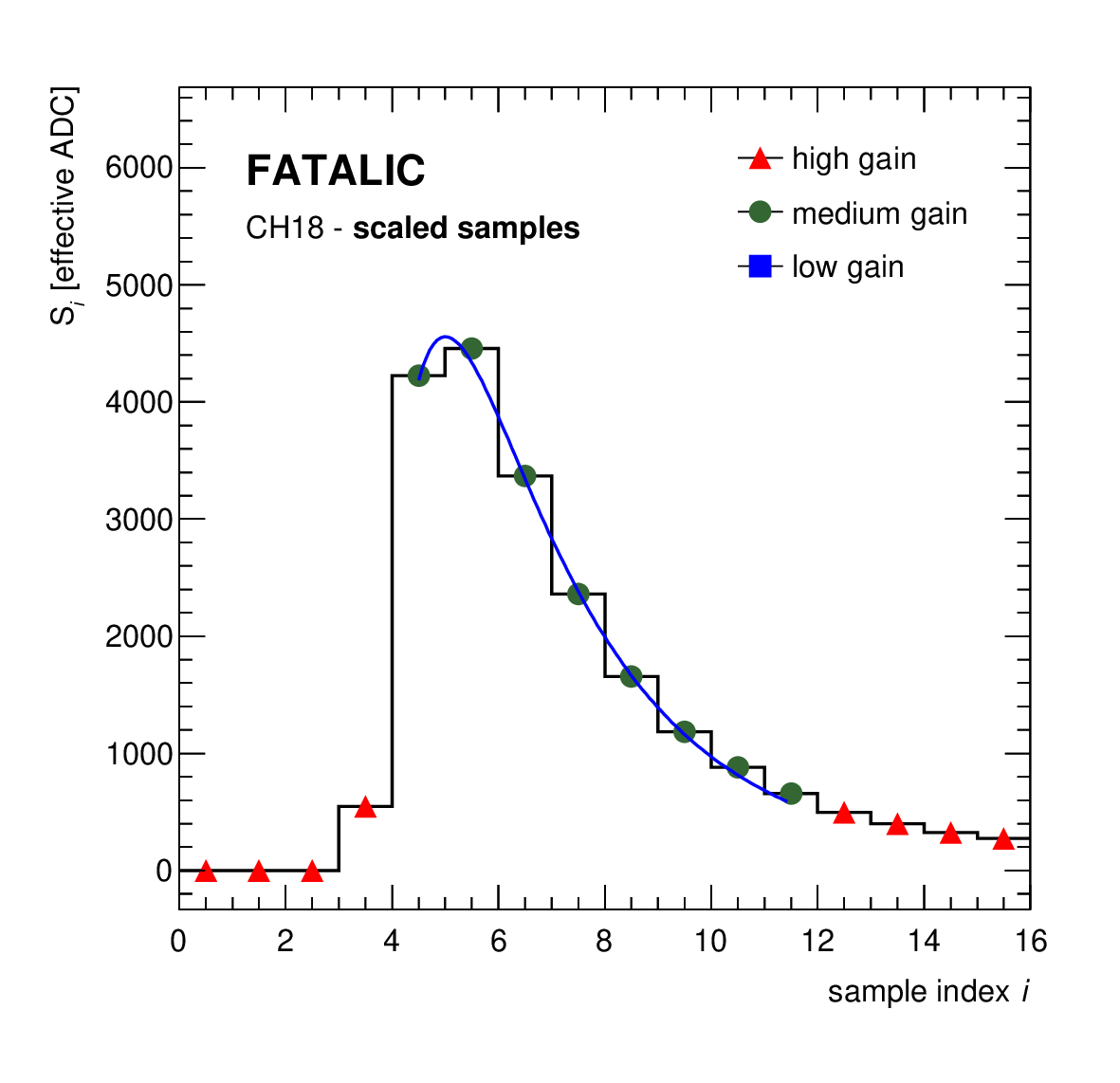}\label{fig:pulses_c}}
  \caption{(a) \gls{adc} measurements from a \SI{100}{GeV} electron beam event. (b) Selected samples, normalised 
  to the high-gain scale, are used to reconstruct the analog pulse by Optimal Filtering.
  \label{fig:pulses}}
\end{figure}

%-------------------------------------------------
\subsection{Selection of digitised samples}
\label{sec:samples}
%-------------------------------------------------

The dynamic gain switch of \gls{fatalic} is exploited in order to ensure the best possible resolution for the acquisition of 
each digitised sample $S_i$. If the high-gain channel does not saturate, the \gls{adc} measurement is acquired from the alternative 
gain output. Otherwise, the switch turns to low gain, in which case the medium gain output is preferred. If, however, the medium-gain 
channel also saturates, then the alternative, low-gain measurement is used. It is noted that, once the high-gain channel saturates, 
the switch remains to low gain for the next seven samples (low-gain block). This is imposed through the Mainboard FPGAs to allow the 
high-gain channel to recover from the saturation state.
%The same treatment is followed by the reconstruction algorithm for the medium gain channel; in case it saturates, the 
%alternative (low) gain readout is exclusively used for the next seven samples.
Once the \gls{adc} value has been acquired, the pedestal $p_i$ of the selected channel is subtracted and the sample is normalised to 
the high-gain scale. Figure\,\ref{fig:pulses} demonstrates a characteristic case in which digitised samples from different channels are 
selected.

%-------------------------------------------------
\subsection{Simulation of experimental effects}
\label{sec:simeffects}
%-------------------------------------------------

\begin{figure}[t]
  \centering
  \subfloat[][]{\includegraphics[width=0.48\textwidth]{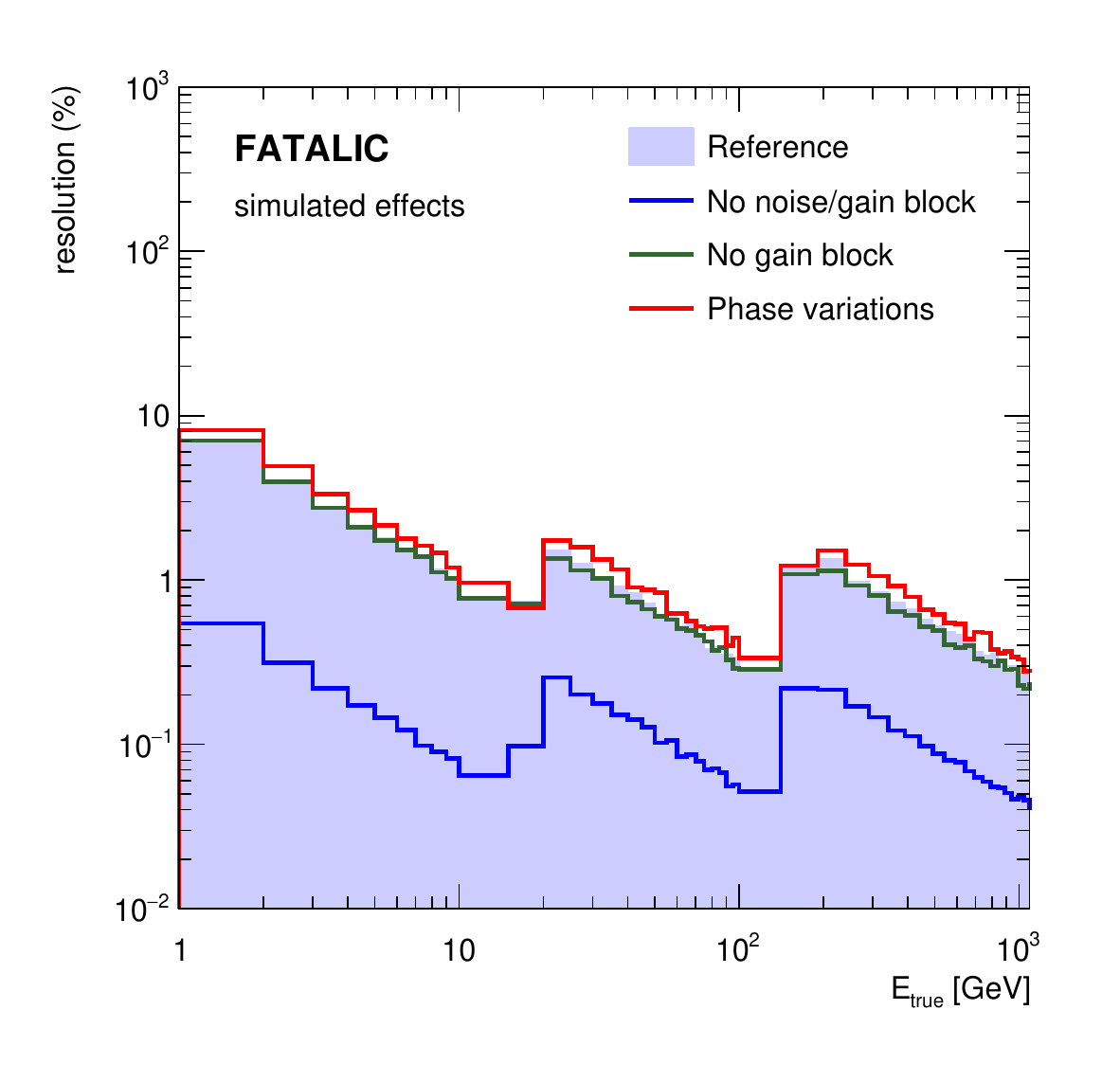}\label{fig:simu_a}}
  \subfloat[][]{\includegraphics[width=0.48\textwidth]{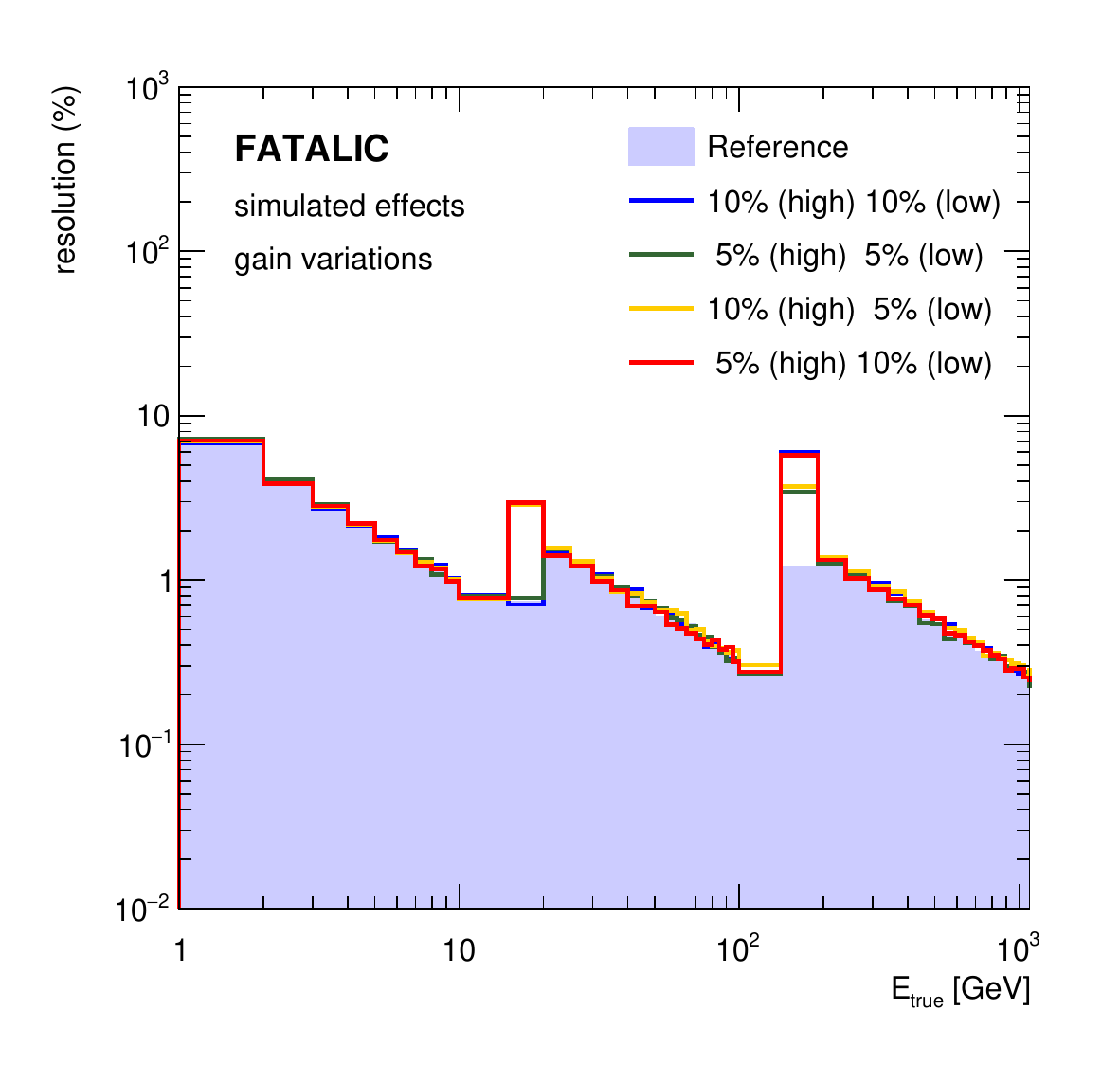}\label{fig:simu_b}}\\
  \subfloat[][]{\includegraphics[width=0.48\textwidth]{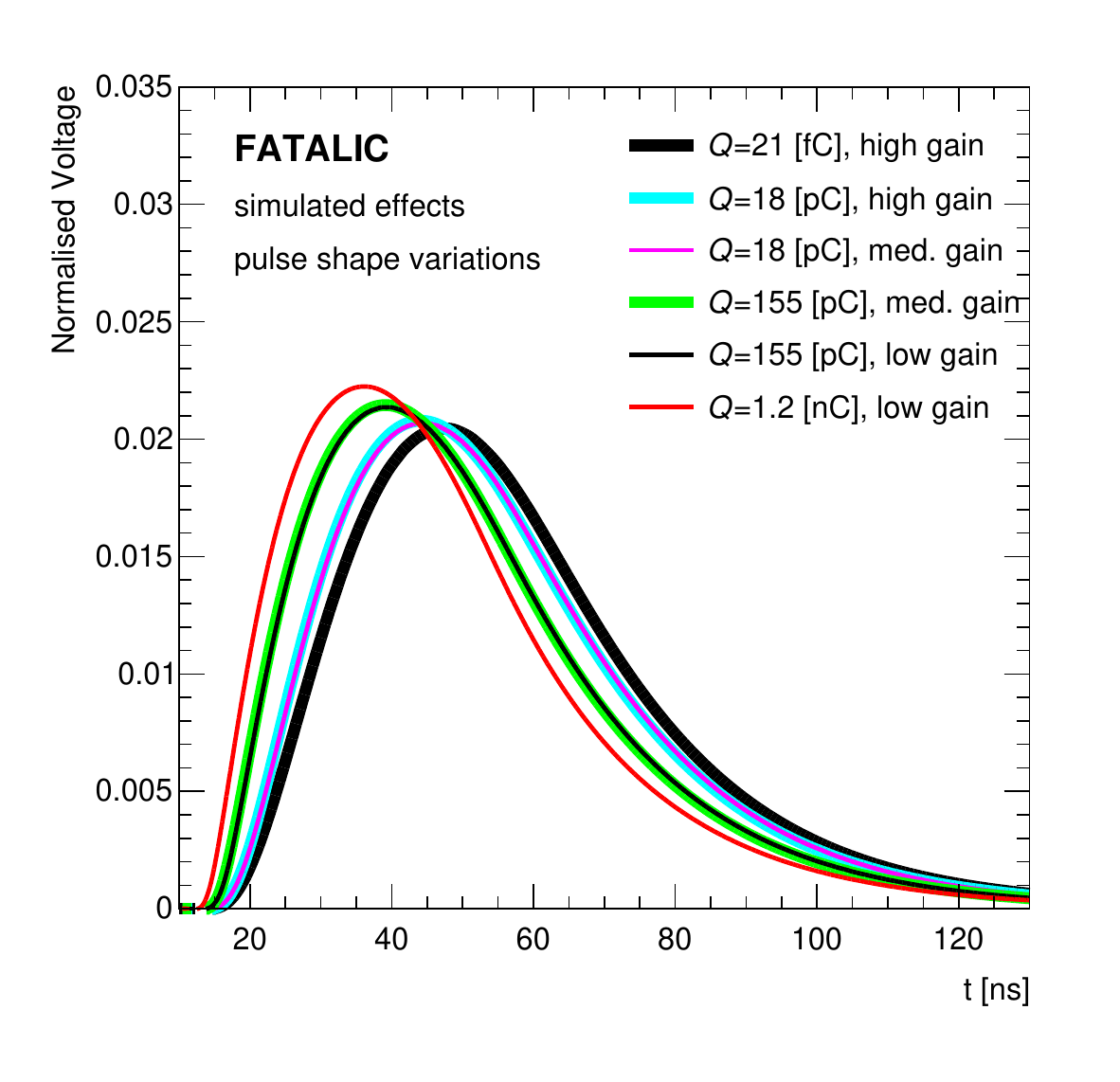}\label{fig:simu_pulses}}
  \subfloat[][]{\includegraphics[width=0.48\textwidth]{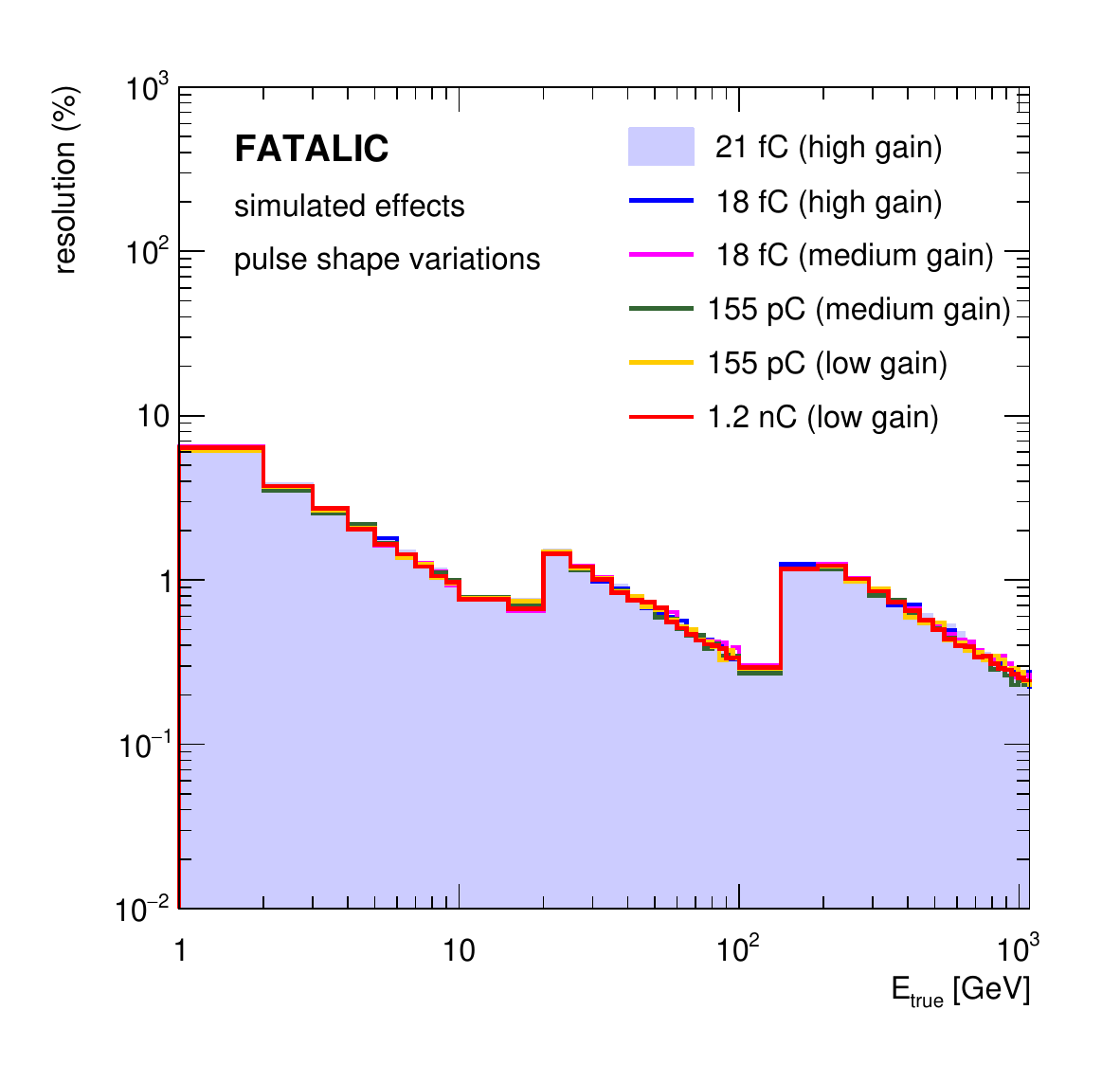}\label{fig:simu_c}}
  \caption{(a,b) Energy resolution as a function of the true energy in different scenarios probing the impact of various
           experimental effects. (c) Variation of the fast channel shaper output with the input charge and (d) its impact
           on the energy resolution.\label{fig:simu1}}
\end{figure}

Simulation studies have been carried out in order to test the impact of experimental effects on the energy resolution.
The latter is defined as the standard deviation of the $(E_{\text{reco}}-E_{\text{true}})/E_{\text{true}}$ distribution, where $E$ 
refers to the deposited energy read by one of the two \glspl{pmt} in a single Tile cell and the label ``true'' (``reco'') refers to the input 
(reconstructed) energy. In these studies, the Optimal Filter is calibrated with respect to the measured output pulse shape of one 
prototype chip. To simulate the charge-injection, this reference pulse is given random amplitudes 
and is subsequently sampled and digitised according to the procedure described in Section~\ref{sec:samples}. Random noise is
then added to each digitised sample, based on measurements with the same chip; \SI{1.5}{ADC} counts for the low- and medium-gain 
channels (corresponding to \SI{8.4}{fC} and \SI{32}{fC}, respectively), and \SI{3.5}{ADC} counts for the high-gain channel 
(\SI{256}{fC}). As shown in figure\,\ref{fig:simu1}, the expected resolution is lower than 2\% in the input range above \SI{2}{pC}, 
while for lower values it increases up to $\sim$7\%.

\paragraph{Electronic noise and gain saturation.} The reference scenario described above is first compared
     to the ideal case where no electronic noise and/or no low-gain block is applied. The results are presented in 
     figure\,\ref{fig:simu_a}. The impact of noise on the intrinsic 18-bit resolution is about one order of
     magnitude, whereas the cost of the low gain block is less than 1\% over the entire input range.

\paragraph{Phase variations.} The arrival time of the pulse, with respect to the digitising clock, may vary 
      by a few nanoseconds. To simulate this effect, the generated pulses are shifted by a random phase 
      $\tau\in[-8,8]$\,ns. The results (figure\,\ref{fig:simu_a}) show that such time-shifts are accounted for 
      by Optimal Filtering, affecting the resolution by $\sim$1\%.

%\paragraph*{Pedestal variations:} Pileup interactions introduce a continuous signal to the \glspl{pmt}, which 
%      is added to the pedestals, altering the relative gain of the fast channels. To simulate this effect, each pedestal
%      is convoluted with a gaussian distribution of mean 0 and standard deviation 0.20. The observed impact on the 
%      resolution is $\sim$3\% (figure\,\ref{fig:simu_a}).

\paragraph{Gain variations.} The impact of gain variations is tested by shifting the gain of each channel by 5\% 
      or 10\%. Such variations do not affect the resolution (figure\,\ref{fig:simu_b}) but rather introduce a shift to 
      the reconstructed energy, which can be recovered by calibration using the \gls{cis}.

\paragraph{Pulse shape variations.} Imperfections of the electronics have been found to distort the output pulse shape
      depending on the input charge. Figure\,\ref{fig:simu_pulses} compares pulses, obtained from the simulation of the 
      ASIC for different injected charges. These variations of the pulse shape have less than 0.5\% impact on the 
      resolution (figure\,\ref{fig:simu_c}). However, they introduce a shift to the reconstructed energy (of less than 1\%), 
      which can be accounted for by applying a scale factor to the gain as a function of the input charge.

\paragraph{Pile-up.} Inelastic $pp$ interactions, taking place in the same (in-time pile-up) and adjacent (out-of-time pile-up) 
      bunch crossings, introduce parasitic pulses, which contaminate the signal from the actual hard scattering. In the years
      2015-2017 the average number of inelastic interactions per bunch-crossing was measured to be $\langle\mu\rangle=32$,
      while in the HL-LHC, the nominal expected rate is $\langle\mu\rangle=140$ and it is foreseen to increase as much as 
      $\langle\mu\rangle\approx 200$.

      To test the performance of \gls{fatalic} in the presence of pile-up background, the description of the \gls{fatalic}
      readout is implemented into the official simulation software of ATLAS. Random in-time and out-of-time pile-up pulses 
      are added to each generated signal pulse, according to the energy distribution of minimum-bias events in a given 
      Tile cell. These distributions are available from physics simulation with different values of $\langle\mu\rangle$.
      The estimated impact on the resolution is presented in figure\,\ref{fig:simu_pu} as a function of the true deposited 
      energy for Tile cell A13 (the most exposed to radiation from $pp$ collisions) and cell D1 (the least exposed). It 
      is note though that the impact can be reduced by calibrating the Optimal Filter against the correlation matrix of 
      the pile-up background~\cite{Fullana:2005dwa}.

%-------------------------------------------------
\subsection{Two-gain scenario}
\label{sec:2gain}
%-------------------------------------------------

\begin{figure}[h]
  \centering
  \subfloat[][]{\includegraphics[width=0.48\textwidth]{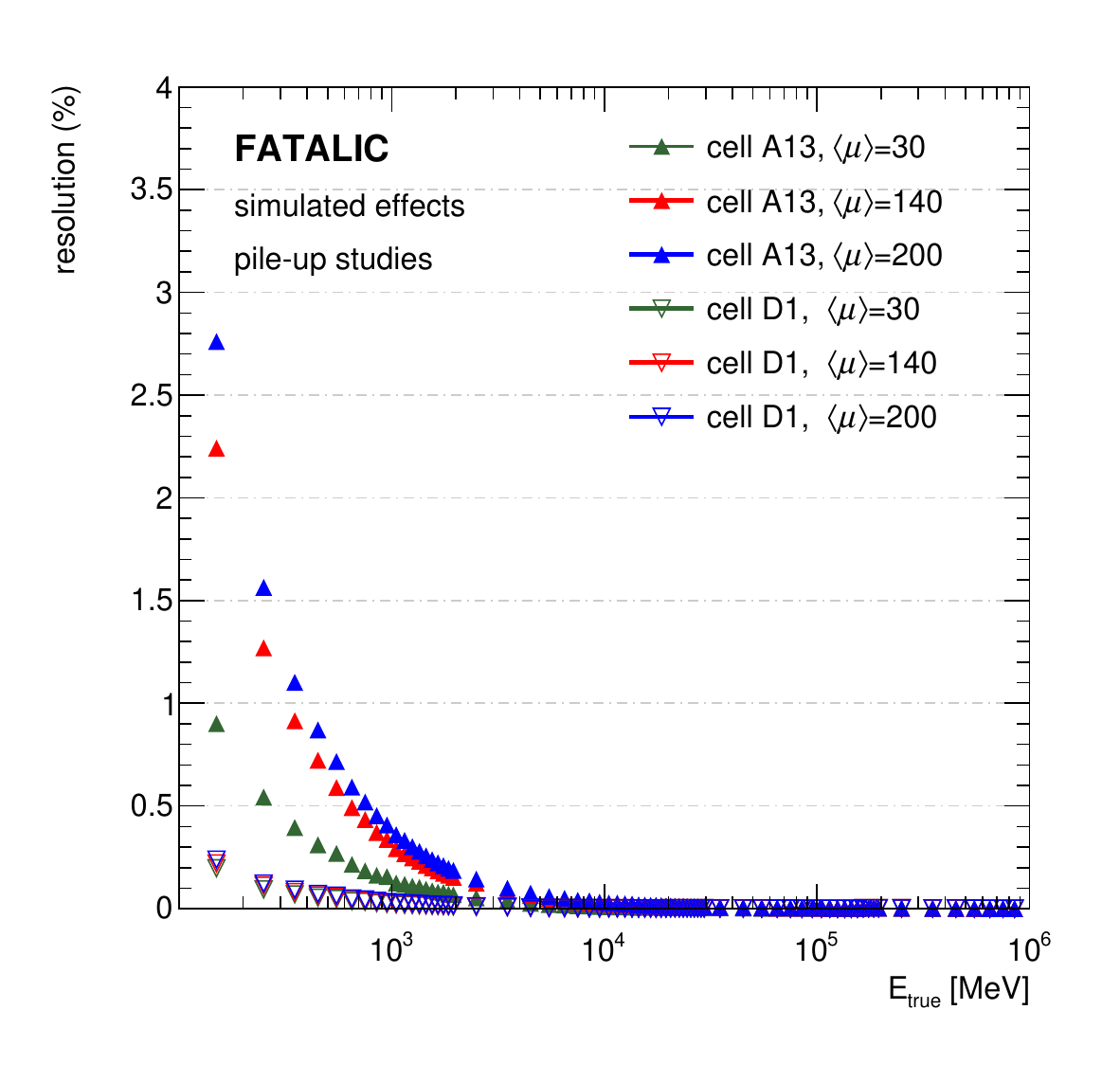}\label{fig:simu_pu}}
  \subfloat[][]{\includegraphics[width=0.48\textwidth]{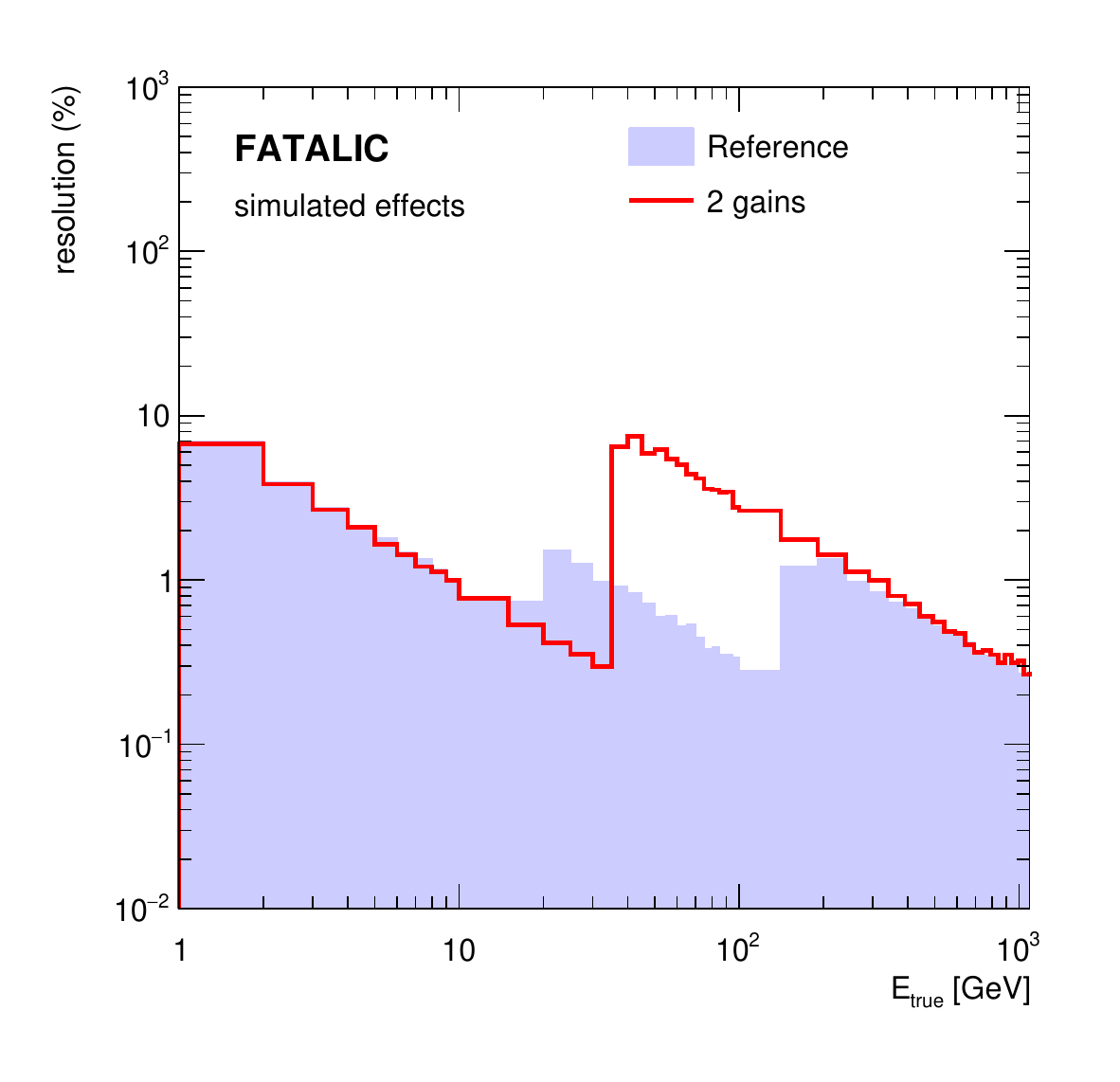}\label{fig:simu_2gain}}
  \caption{(a) Energy resolution as a function of the true energy for $\langle\mu\rangle=30$,140 and 200, in Tile cells 
           D1 and A13. (b) Energy resolution achieved with \gls{fatalic}, as a function of the true energy, compared to 
           the scenario of using only two fast channels with a gain ratio of 32.\label{fig:simu2}}
\end{figure}

In order to quantify the benefit of having three gains, the scenario of using two gains with a gain ratio of 32
is explored, based on the simulation described above. Considering digitisation with 12-bit \glspl{adc}, the effective 
output range in this case is 17-bits. The noise is assumed to be \SI{1.8}{ADC} counts for both channels. As seen in 
figure\,\ref{fig:simu_2gain} the resolution drops by $\sim$8\% in the intermediate range (\SI{40}{MeV}-\SI{180}{MeV})
which, in the case of \gls{fatalic}, is recovered by the medium-gain channel.

%% file: TeX/part5_tbeam.tex
%%%%%%%%%%%%%%%%%%%%%%%%%%%%%%%%%%%%%%%%%%%%%%%%%%%%%%%%%%%%%%%%%%%%%%
\section{Performance with particle beams}
\label{sec:TestBeamResults}
%%%%%%%%%%%%%%%%%%%%%%%%%%%%%%%%%%%%%%%%%%%%%%%%%%%%%%%%%%%%%%%%%%%%%%

The 24 prototype FATALIC were tested in the reconstruction of real energy deposits of hadrons, electrons and muons,
provided by the H8 secondary particle beam of CERN. Two mini-drawers, equipped with 12 \gls{fatalic} \gls{fe} electronics 
each, were inserted into a demonstrator Tile module, providing full readout of Tile cells A1-5, BC1-5, D1 and partial readout
of Tile cells A6 and D0. First, the detection chains were calibrated by running \gls{cs} scans, using the slow channel. The fast channels 
performance was then probed with particle beams of different energies and compositions.

%figure\,\ref{fig:tb:setup} demonstrates the mini-drawer assembly, before insertion, as well as a three 
%of the TileCal modules at the beam hall.

%\begin{figure}[!b]
%  \centering
%  \subfloat[][]{\includegraphics[width=0.48\textwidth]{figures/testbeam/tb_minidrawer.jpg}\label{fig:tb:minidrawer}}
 % \hspace{4pt}
%  \subfloat[][]{\includegraphics[width=0.48\textwidth]{figures/testbeam/tb_module.jpg}\label{fig:tb:module}}
%  \caption{(a) A TileCal mini-drawer, equipped with 12 FATALIC FEs, before insertion into the TileCal module. (b) Three of the TileCal demonstrator modules in the beam hall. }
%  \label{fig:tb:setup}
%\end{figure}

%~~~~~~~~~~~~~~~~~~~~~~~~~~~~~~~~~~~~~~~~~~~~~~~~~
\subsection{Inter-calibration of readout channels}
\label{subsec:Calibration}
%~~~~~~~~~~~~~~~~~~~~~~~~~~~~~~~~~~~~~~~~~~~~~~~~~

\begin{figure}[!tb]
  \centering
  \subfloat[][]{\includegraphics[width=0.48\textwidth]{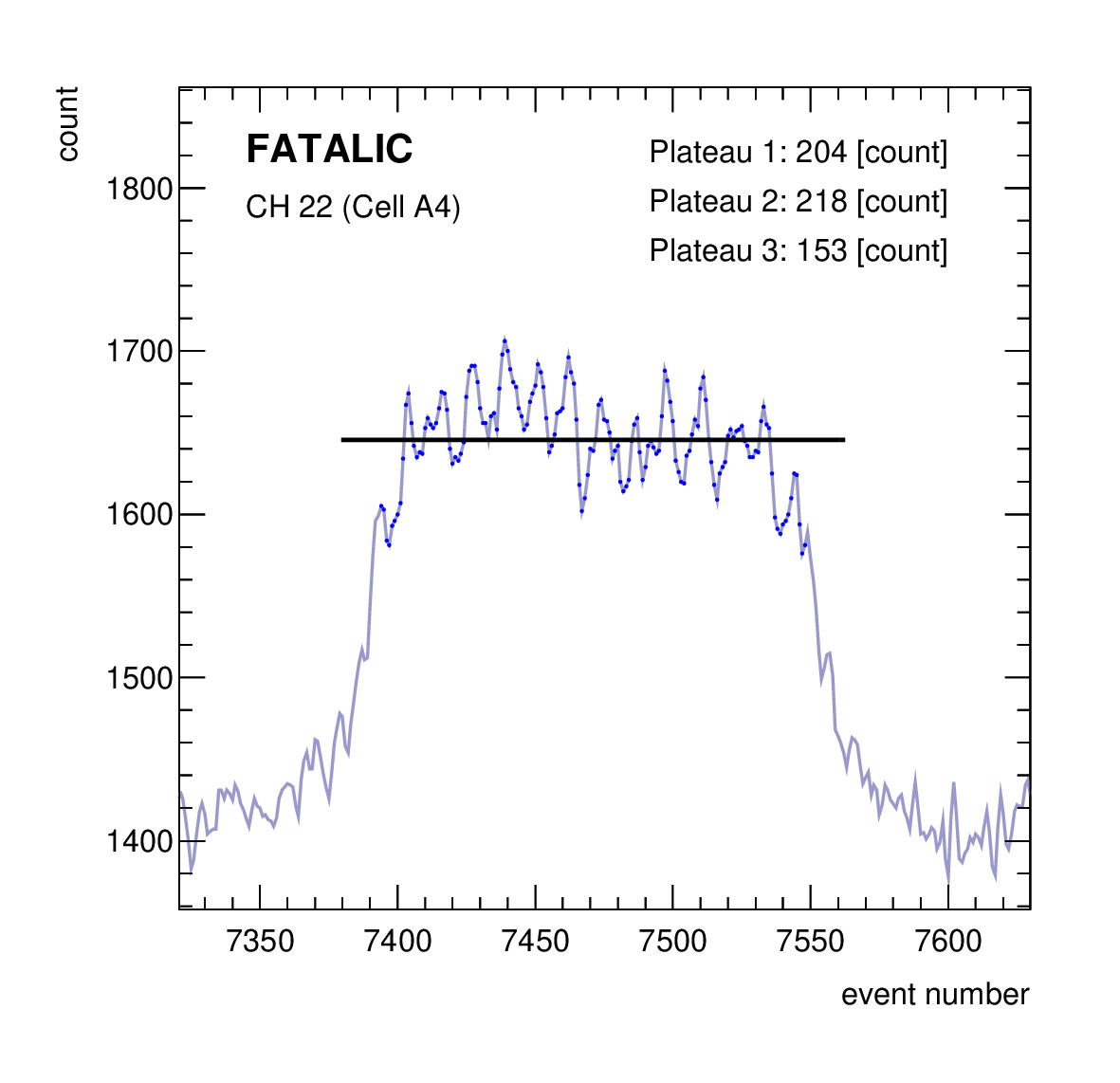}\label{fig:pl_a}}
  \subfloat[][]{\includegraphics[width=0.48\textwidth]{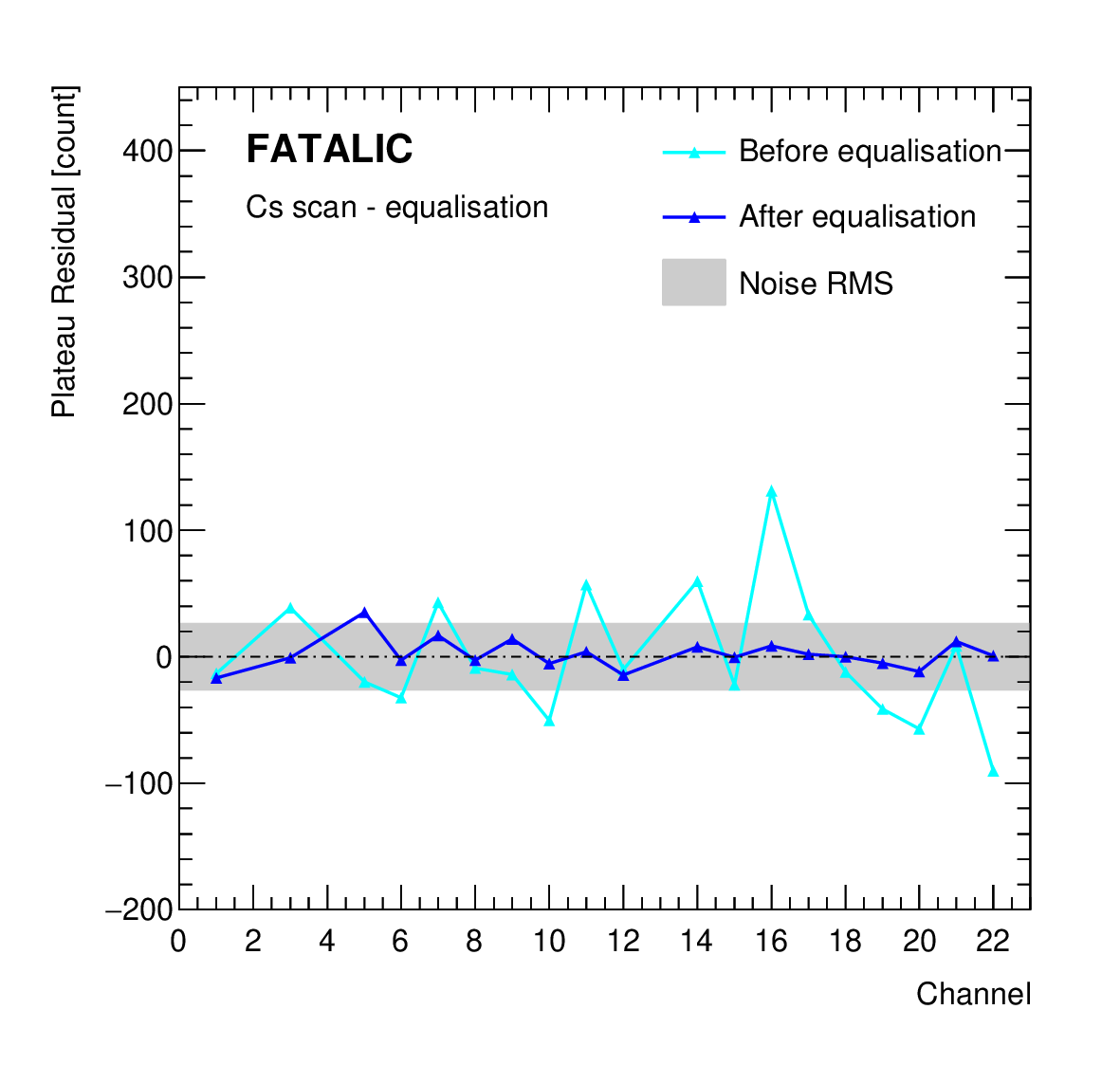}\label{fig:pl_b}}
  \caption{(a) Characteristic plateau, showing the response of one \gls{pmt} to the \gls{cs} source signal.
  (b) Plateau variations before and after adjustment of the high voltage of each \gls{pmt}.}
  \label{fig:pl}
\end{figure}

As the \gls{cs} source traverses a Tile cell, the response of the respective \glspl{pmt} exhibits a characteristic
plateau (figure\,\ref{fig:pl_a}), which reflects the sequential excitation of the tiles, with local maxima (minima), 
generated when the source traverses scintillator tiles (steel plates).
Since the energy deposited by the source is uniform, variations of the measured responses can be used to inter-calibrate 
the detection chains. This is performed by equalising each plateau $p_{i}$ to the overall mean $\langle p \rangle$. Since 
the dependence of the \gls{pmt} gain on the applied high voltage is $\sim$$V^{\beta}$, where $\beta\simeq 7$ for the 
particular \glspl{pmt}, the high voltage is corrected to $V^{'}_{i} = V_{i}\cdot(\langle p \rangle/p_{i})^{-7}$. 
Figure\,\ref{fig:pl_b} demonstrates the plateaus before and after equalisation. As shown, residual variations are 
successfully reduced to the noise level and can be used to derive correction factors for each channel's response.

%~~~~~~~~~~~~~~~~~~~~~~~~~~~~~~~~~~~~~~~~~~~~~~~~~
\subsection{Measurement of particle deposits}
\label{subsec:TBparticles}
%~~~~~~~~~~~~~~~~~~~~~~~~~~~~~~~~~~~~~~~~~~~~~~~~~

The following paragraphs present the results of data analysis, using electron, muon and hadron beams, targeting the center of
each A-cell at 20$^{\circ}$ incidence. The deposited energy is reconstructed using Optimal Filtering, as described in 
Section\,\ref{sec:EnergyReco}, and is expressed in units of input charge by applying the fC/ADC conversion factors, derived using 
the \gls{cis}. Unless specified otherwise, the deposited energy is estimated from the sum of the measurements in the targeted 
A and BC cell. Adjacent cells are also taken into account for containment. The average energy, deposited by a specific beam 
constituent, is obtained from the mean of a gaussian line-shape interpolated around the respective characteristic peak.

%~~~~~~~~~~~~~~~~~~~~~~~~~~~~~~~~~~~~~~~~~~~~~~~~~
\paragraph{Electrons.}

\begin{figure}[tb]
  \centering
  \subfloat[][]{\includegraphics[width=0.48\textwidth]{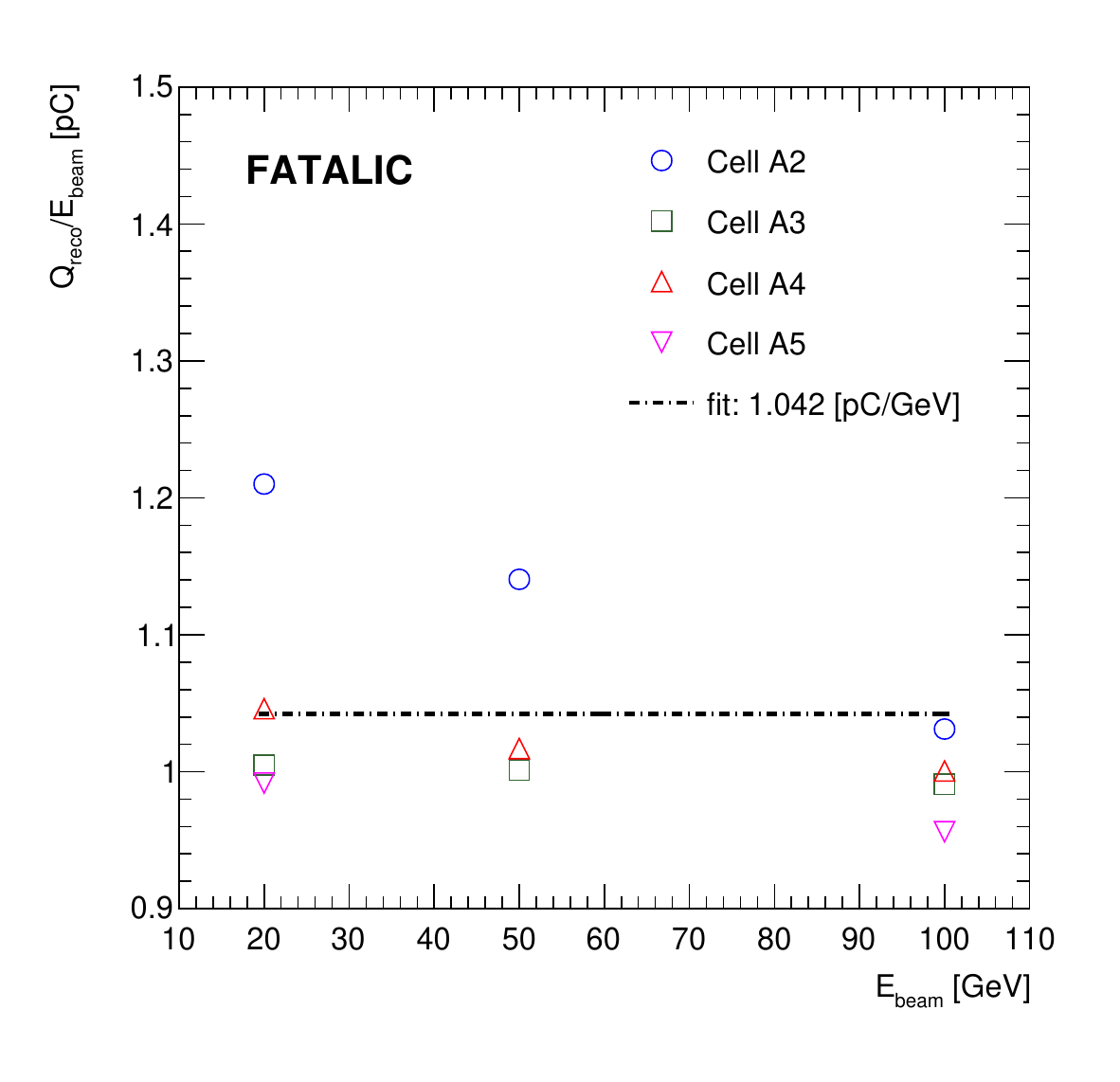}}
  \subfloat[][]{\includegraphics[width=0.48\textwidth]{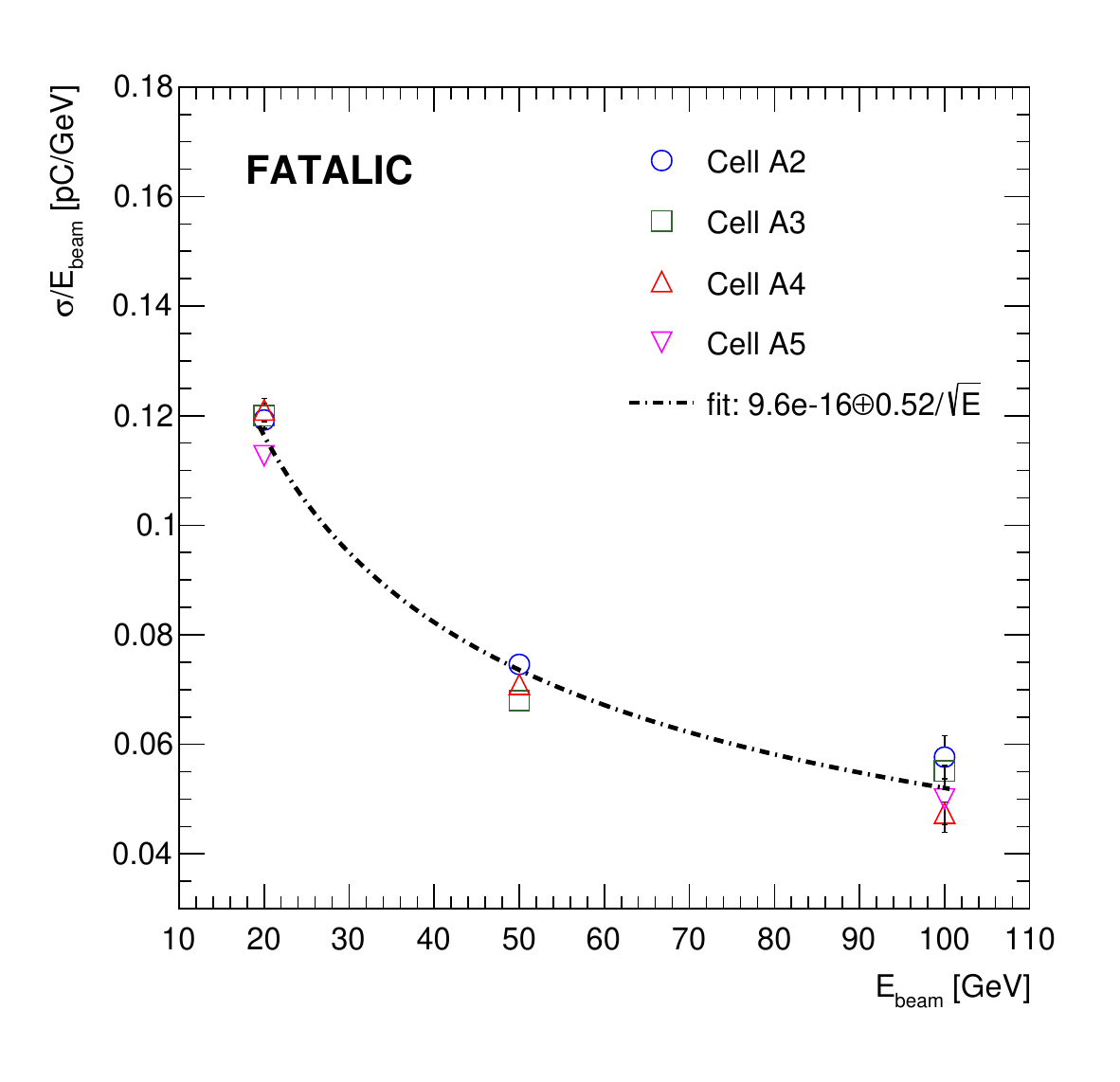}}
  \caption{Relative mean (a) and standard deviation (b) of the reconstructed electron energy in A-cells as a function of 
  the beam energy. Error bars represent statistical uncertainties from the gaussian fit to the energy distribution in each cell.}
  \label{fig:elec_reso}
\end{figure}

\begin{figure}[t]
  \centering
  \subfloat[][]{\includegraphics[width=0.48\textwidth]{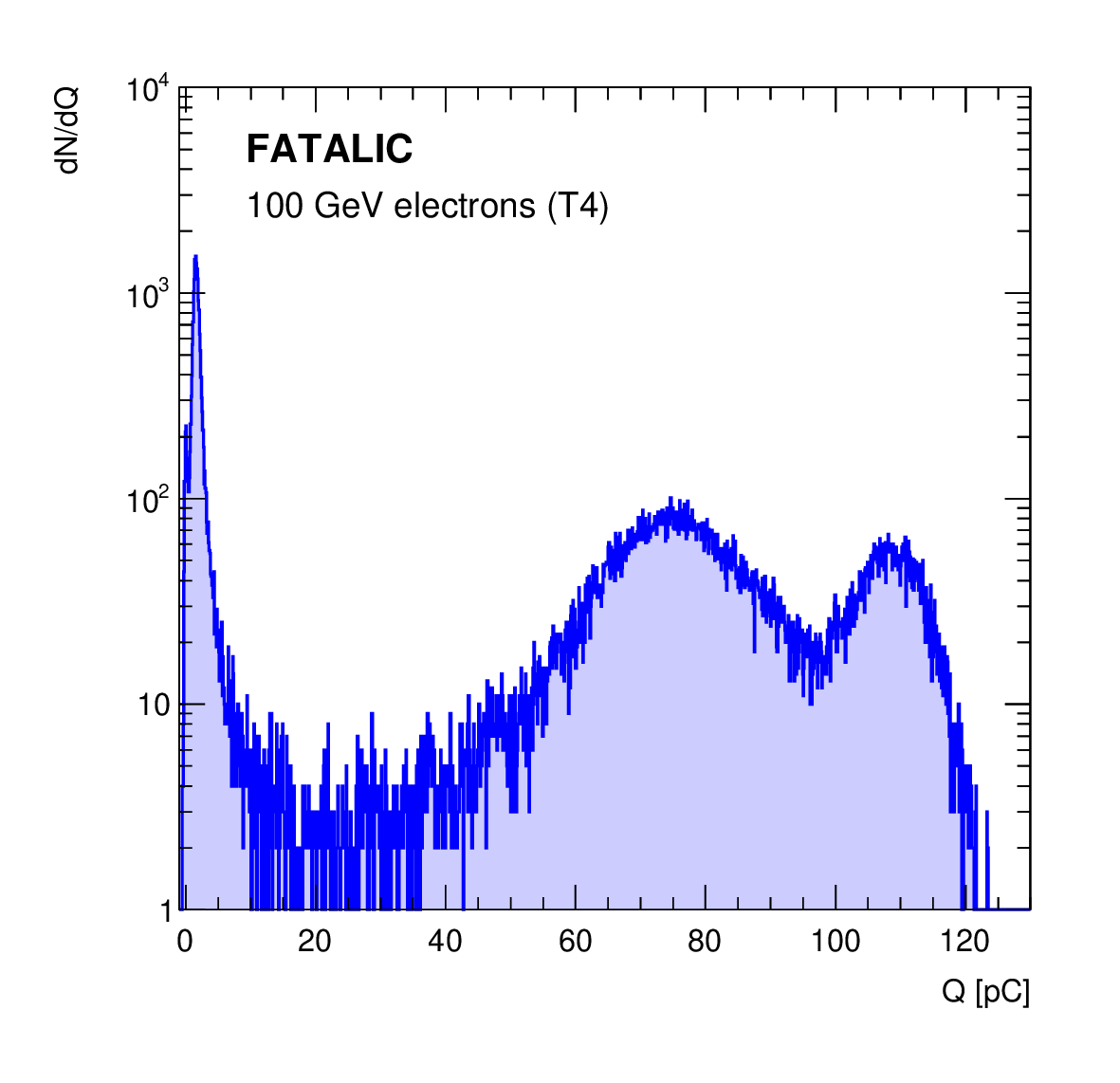}\label{fig:ereco_a}}
  \subfloat[][]{\includegraphics[width=0.508\textwidth]{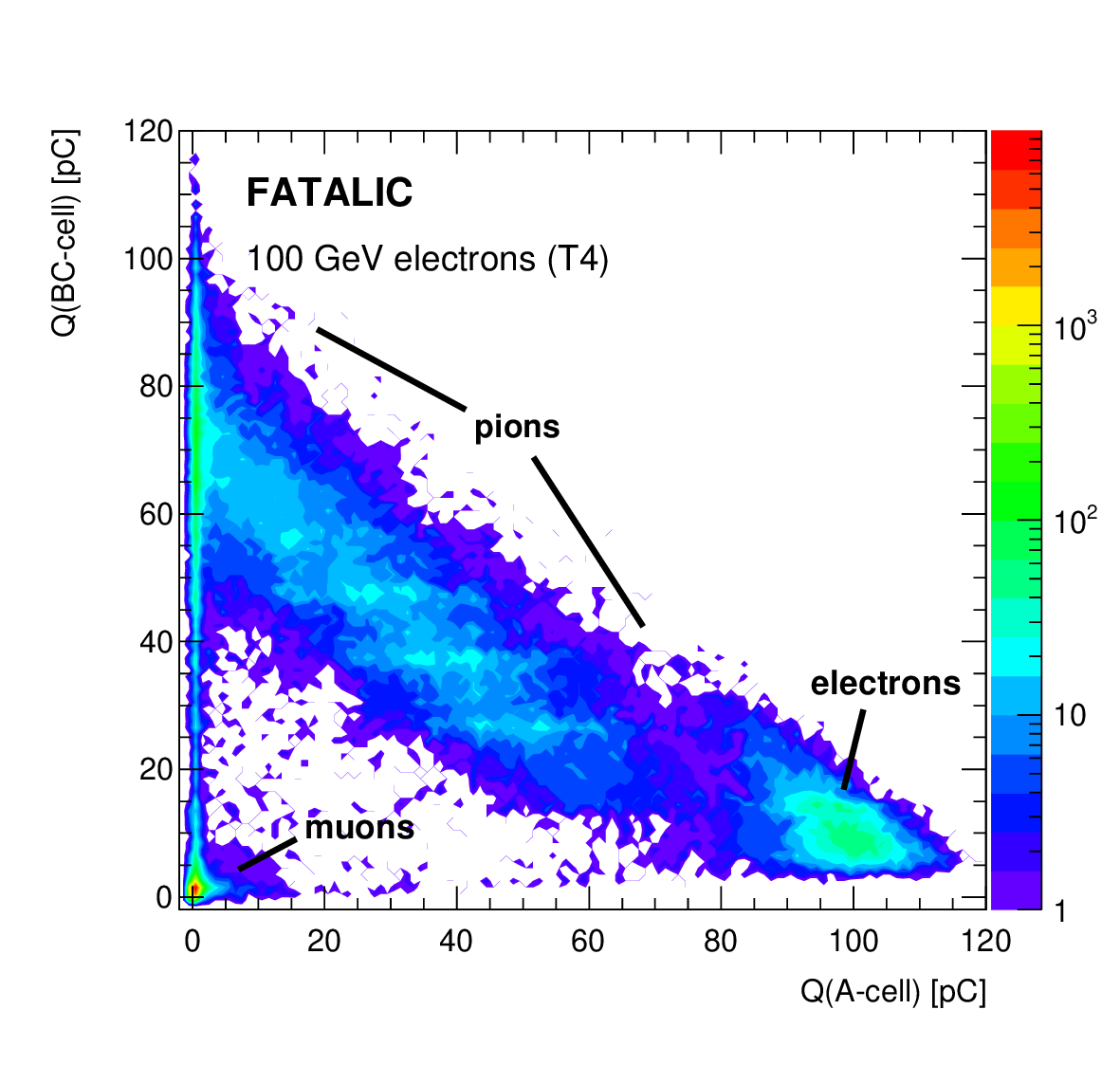}\label{fig:ereco_b}}\\
  \subfloat[][]{\includegraphics[width=0.48\textwidth]{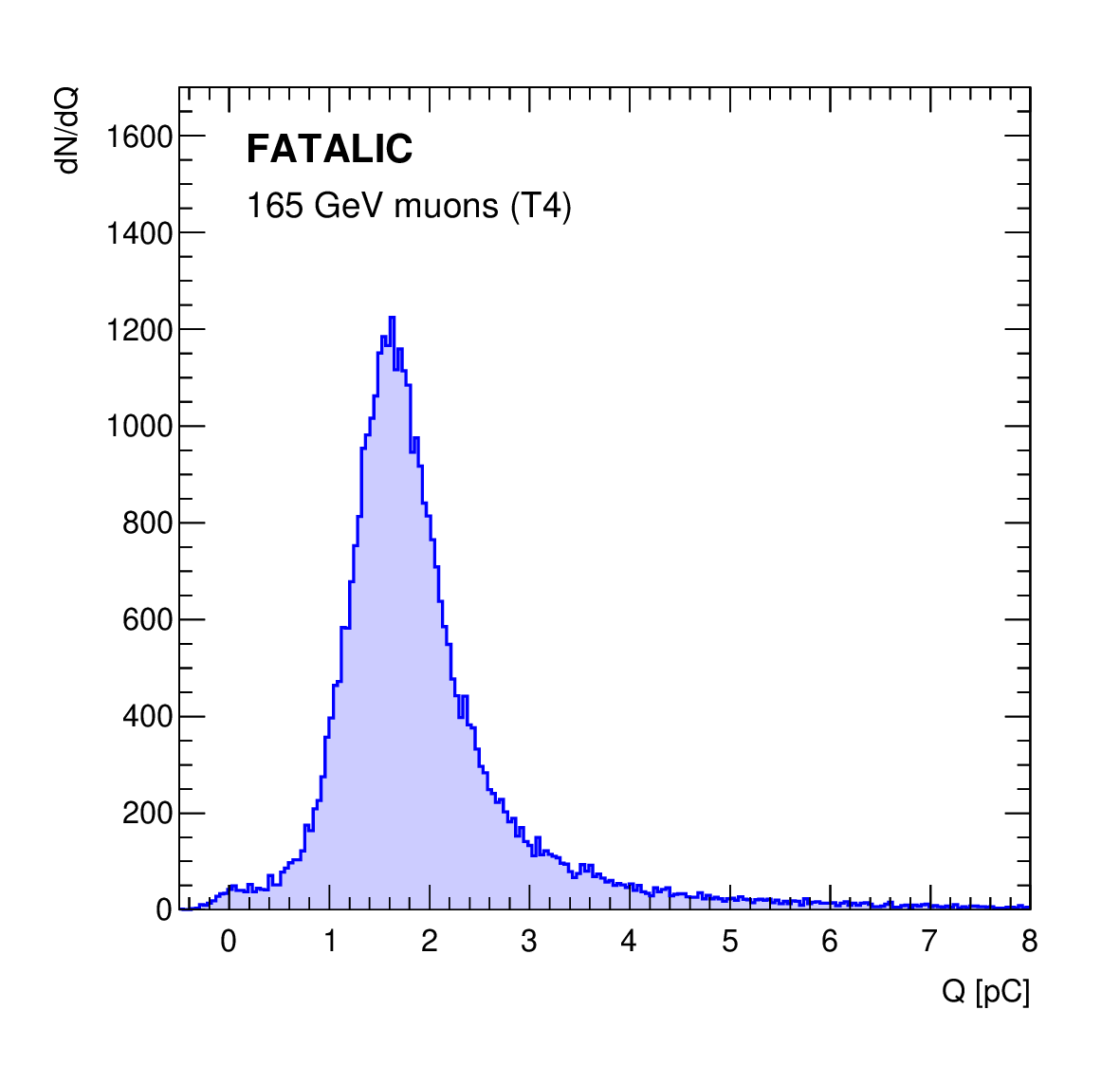}\label{fig:ereco_c}}
  \subfloat[][]{\includegraphics[width=0.48\textwidth]{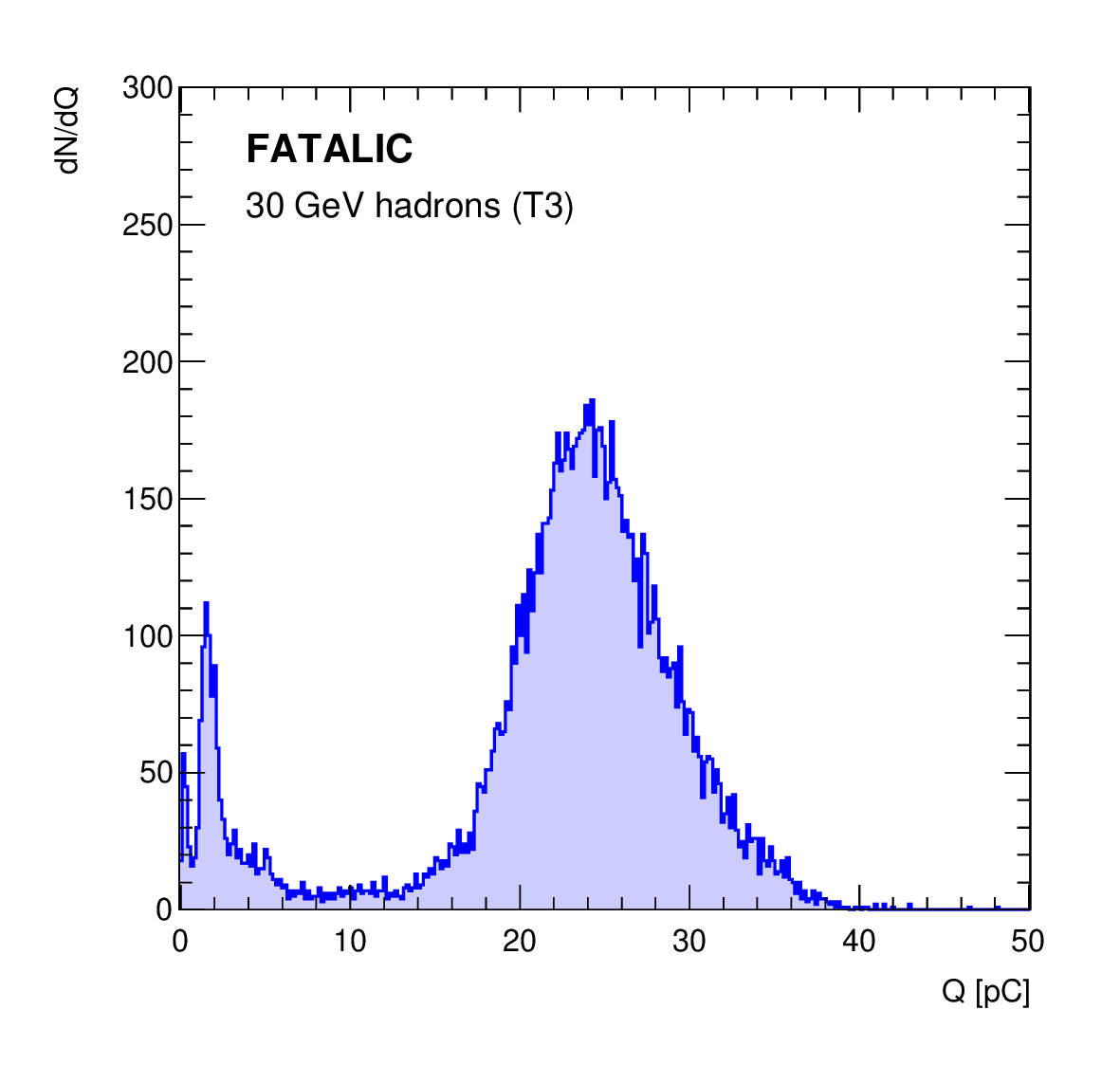}\label{fig:ereco_d}}
  \caption{(a) Reconstructed energy distribution with 100\,GeV electron beam targeting cell A4. (b) Two-dimensional distribution
  in the A/BC-cell plane, demonstrating the characteristic deposits of the different beam constituents. (c) 165 GeV muon beam 
  targeting cell A4. (d) 30 GeV\,hadron beam targeting cell A3 (adjacent towers are also included for containment).\label{fig:ereco}}
\end{figure}

The reconstruction of \SI{20}{GeV}, \SI{50}{GeV} and \SI{100}{GeV} electron signal is tested with the beam targeting Tile cells 
A2-A5. Since the electromagnetic shower is contained within a short distance in the Tile module, the electron energy is obtained
from the respective distribution in each targeted A-cell. The results are summarised in table\,\ref{tab:elecs} and displayed 
graphically in figure\,\ref{fig:elec_reso} as a function of the beam energy. Using these measurements (from 11\,Tile cells), the 
EM scale constant is estimated $1.04\pm 0.1$\si{pC/GeV}, which is consistent with the nominal value of $1.05\pm 0.1$\si{pC/GeV}, 
derived in precise test-beam studies\,\cite{Adragna:2009zz}, based on more than 200 Tile cells, using electrons of different
energies with $20^{\circ}$ incidence. Finally, an example showing the total reconstructed energy distribution (A and BC cells 
combined) is presented in figure\,\ref{fig:ereco_a}, while figure\,\ref{fig:ereco_b} displays the two-dimensional distribution in 
the A/BC cell plane, where the characteristic deposits of the different beam constituents (electrons, muons and pions) can be distinguished.

\begin{table}[h]
  \begin{center}{{\small
  \begin{tabular}{ r c c c c c c c c }
  {\bfseries Beam} & \multicolumn{2}{c}{\bfseries Cell A2} & \multicolumn{2}{c}{\bfseries Cell A3} & \multicolumn{2}{c}{\bfseries Cell A4} & \multicolumn{2}{c}{\bfseries Cell A5}\\
  \toprule
  GeV & $Q_\text{reco}$\,[pC] & $\sigma$\,[pC] & $Q_\text{reco}$\,[pC] & $\sigma$\,[pC] & $Q_\text{reco}$\,[pC] & $\sigma$\,[pC] & $Q_\text{reco}$\,[pC] & $\sigma$\,[pC]\\
   20   & 23.2 & 2.3  &  20.1 & 2.4  &  21.6 & 2.5  &  19.7 & 2.2\\
   50   & 54.3 & 3.5  &  50.1 & 3.5  &  52.5 & 3.7  &  -    & -  \\ 
  100   & 98.2 & 5.6  &  98.7 & 5.5  & 103.1 & 4.8  &  95.1 & 5.1\\ 
  \bottomrule
  \end{tabular}}
  \caption{\label{tab:elecs}Reconstructed electron energy in the targeted A-cells. }
  }\end{center}
\end{table}

%~~~~~~~~~~~~~~~~~~~~~~~~~~~~~~~~~~~~~~~~~~~~~~~~~
\paragraph{Muons.}

The reconstruction of muon signal is tested with \SI{165}{GeV} muon beams targeting cells A2-A5. The reconstructed energy distribution
for a characteristic case is shown in figure\,\ref{fig:ereco_c}. The most probable values (mpv), estimated for each targeted Tile tower 
(A- and BC-cells), are listed in table\,\ref{tab:muons}. The results are also expressed in terms of energy loss per unit distance (\dedx), 
using the track length in each Tile cell (\SI{31.925}{cm} for A-cells and \SI{89.391}{cm} for BC-cells, for 20$^\circ$ incidence). The 
overal energy loss is $14.2\pm 1.9$ \si{fC/cm}, which corresponds to $15.0\pm 2.0$ \si{MeV/cm} considering the EM calibration 
constant of \SI{1.04}{MeV/pC} estimated above and the $e/\mu$ response ratio of 0.91. The result is consistent with the estimate of
\SI{15.2}{MeV/cm}, reported in previous test-beam studies using \SI{180}{GeV} muons at projective angles.

\begin{table}[h]
  \begin{center}{{\small
  \begin{tabular}{ r c c c c  }
  {\bfseries Cell Type} & {\bfseries Tower 2(+3)} & {\bfseries Tower 3(+4)} & {\bfseries Tower 4(+5)} & {\bfseries Tower 5}\\
  \toprule
            & $Q_\text{reco}$\,[pC] & $Q_\text{reco}$\,[pC] & $Q_\text{reco}$\,[pC] & $Q_\text{reco}$\,[pC]\\
  A         &     0.49 &  0.41 & 0.47 & 0.41\\
  BC        &     1.66 &  1.17 & 1.11 & 1.22\\ 
  A+BC      &     2.21 &  1.60 & 1.60 & 1.67\\
  \midrule
            &   $dQ/dx$\,[fC/cm] & $dQ/dx$\,[fC/cm] & $dQ/dx$\,[fC/cm] & $dQ/dx$\,[fC/cm]\\
  A         &   15.3 & 12.8 & 14.7 & 12.8\\
  BC        &   18.6 & 13.1 & 12.4 & 13.6\\ 
  A+BC      &   18.2 & 13.2 & 13.2 & 13.8\\ 
  \bottomrule
  \end{tabular}}
  \caption{\label{tab:muons}Reconstructed energy and energy loss per unit distance of 165\,GeV muons targeting cells A2-A5.}
  }\end{center}
\end{table}

\begin{table}[h]
  \begin{center}{{\small
  \begin{tabular}{ r c c c }
  {\bfseries Beam} & {\bfseries Tower 2+3+4} & {\bfseries Tower 3} & {\bfseries Tower 2+4}\\
  \toprule
  GeV   & $Q_\text{reco}$\,[pC] & $Q_\text{reco}$\,[pC] & $Q_\text{reco}$\,[pC]\\
   30   &     24.4 &   20.1 &  2.8\\
  180   &    150.9 &  121.4 & 27.9\\ 
  \bottomrule
  \end{tabular}}
  \caption{\label{tab:hadrons}Reconstructed energy of 30\,GeV and 180\,GeV pions in the targeted tower\,3 and adjacent towers\,2,4.}
  }\end{center}
\end{table}

%~~~~~~~~~~~~~~~~~~~~~~~~~~~~~~~~~~~~~~~~~~~~~~~~~
\paragraph{Hadrons.}

Hadron signal is studied with beams of \SI{30}{GeV} and \SI{180}{GeV} pions targeting Tile cell A3. To account for energy leakage
towards adjacent Tile cells of the same module, deposits measured in Tile cells A2, BC2, A4, BC4 are also added to the measurement. 
The results are summarised in table\,\ref{tab:hadrons}, while the reconstructed energy distribution for the case of \SI{30}{GeV} pions is
shown in figure\,\ref{fig:ereco_d}. As expected, the measured energy is lower than the beam energy, since a single Tile module 
cannot provide full coverage, in solid angle, of the hadronic shower. The energy leakage towards neighbouring Tile cells of the 
same module is found approximately 11\% and 18\% in the cases of \SI{30}{GeV} and \SI{180}{GeV}, respectively.